\documentclass[aps,prd,nofootinbib,preprintnumbers,floatfix]{revtex4-2}

\usepackage{epsfig}
\usepackage{bm}
\usepackage{amssymb}
\usepackage{amsmath}
\usepackage{color}
\usepackage{dcolumn}
\usepackage[export]{adjustbox}
\usepackage[caption=false]{subfig}

\usepackage[utf8]{inputenc}

\usepackage{amsmath}
\usepackage{amssymb}
\usepackage{amsthm}
\usepackage{amscd, amsfonts, mathrsfs}
\usepackage{cases}
\usepackage{verbatim}
\usepackage{epsf}
\usepackage{xcolor}
\usepackage[bookmarksnumbered, pdfpagelabels=true, plainpages=false, colorlinks=true, linkcolor=blue, citecolor=blue, urlcolor=blue]{hyperref}

\usepackage{soul, color}

\theoremstyle{plain}

\theoremstyle{definition}

\allowdisplaybreaks

\newcommand{\e}{\epsilon}

\newcommand{\tr}{\operatorname{tr}}

\newcommand{\be}{\begin{equation}}
\newcommand{\ee}{\end{equation}}
\newcommand{\bea}{\begin{eqnarray}}
\newcommand{\eea}{\end{eqnarray}}

\newcommand{\bml}{\begin{subequations}}
\newcommand{\eml}{\end{subequations}}

\newcommand{\bbm}{\begin{bmatrix}}
\newcommand{\ebm}{\end{bmatrix}}
\newcommand{\bvm}{\begin{vmatrix}}
\newcommand{\evm}{\end{vmatrix}}

\newcommand{\mc}{\mathcal}
\newcommand{\p}{\n}

\newcommand{\ve}{\varepsilon}
\newcommand{\n}{\nabla}

\renewcommand{\k}{\bm k}

\newcommand{\ka}{|\bm k|}
\newcommand{\disc}{\operatorname{Disc}_3}
\newcommand{\AND}{\text{and}}
\newcommand{\OR}{\text{or}}

\newcommand{\ot}[1]{\{#1\}}
\newcommand{\td}[1]{\langle #1 \rangle}

\newcommand{\cf}{\tilde}
\newcommand{\cfe}{\tilde \ve}
\newcommand{\cfp}{\tilde \pi}
\newcommand{\cfn}{\tilde \nu}
\newcommand{\cfq}{\tilde \theta}
\newcommand{\cfj}{\tilde \gamma}

\newcommand{\lzero}{\lambda_0}
\newcommand{\lone}{\lambda_1}
\newcommand{\ltwo}{\lambda_2}
\newcommand{\lthree}{\lambda_3}

\newcommand{\blzero}{\bar \lambda_0}
\newcommand{\blone}{\bar \lambda_1}
\newcommand{\bltwo}{\bar \lambda_2}
\newcommand{\blthree}{\bar \lambda_3}

\makeatletter
\renewcommand*\env@matrix[1][\arraystretch]{%
  \edef\arraystretch{#1}%
  \hskip -\arraycolsep
  \let\@ifnextchar\new@ifnextchar
  \array{*\c@MaxMatrixCols c}}
\makeatother

\renewcommand{\Re}{\operatorname{Re}}
\renewcommand{\Im}{\operatorname{Im}}

\newcommand{\D}[1]{\Delta_{(#1)}}
\renewcommand{\bar}{\overline}

\begin{document}


\title{Causal and stable first-order chiral hydrodynamics}
\date{\today}

\author{Nick Abboud}
\email{nka2@illinois.edu}

\author{Enrico Speranza}
\email{espera@illinois.edu}

\author{Jorge Noronha}
\email{jn0508@illinois.edu}
\affiliation{Illinois Center for Advanced Studies of the Universe \& Department of Physics, \\
University of Illinois Urbana-Champaign, Urbana, Illinois 61801, USA}

\begin{abstract}
We derive the set of inequalities that is necessary and sufficient for nonlinear causality and linear stability of first-order relativistic hydrodynamics with either a $U(1)_V$ conserved current or a $U(1)_A$ current with a chiral anomaly or both. Our results apply to generic hydrodynamic frames in which no relations among the transport parameters are imposed. Furthermore, our analysis yields, to the best of our knowledge, the first theory of viscous chiral hydrodynamics proven to be causal and stable. We find that causality demands the absence of vorticity-induced heat flux, forcing a departure from the thermodynamic frame in the chiral case. The inequalities for causality and stability define a hypervolume in the space of transport parameters, wherein each point corresponds to a consistent formulation. Notably, causality is determined by just three combinations of transport parameters. We present our results in a form amenable to numerical hydrodynamic simulations.
\end{abstract}

\maketitle

\section{Introduction}

Relativistic fluid dynamics provides a powerful effective description for the long-wavelength, late-time dynamics of a wide variety of physical systems in, e.g., cosmology \cite{WeinbergCosmology}, astrophysics \cite{Rezzolla_Zanotti_book}, and high-energy nuclear collisions \cite{Romatschke:2017ejr}. In addition, microscopic quantum effects, such as those induced by quantum anomalies, can give rise to novel macroscopic transport phenomena which can also be reflected in the hydrodynamic description \cite{Kharzeev:2015znc,Huang:2015oca,Hosur:2013kxa}.
Although the formulation of the viscous theory of nonrelativistic hydrodynamics---the Navier-Stokes theory---is textbook material \cite{LandauLifshitzFluids}, its generalization to the relativistic regime remains an important and active topic of current research both with and without the effect of quantum anomalies. A theory of relativistic hydrodynamics which consistently includes dissipation is needed in particular to model the behavior of the quark-gluon plasma created in relativistic heavy-ion collisions \cite{Heinz:2013th} and the dynamics of ultradense matter in neutron star mergers \cite{Baiotti:2016qnr,Alford:2017rxf,Most:2021zvc,Most:2022yhe}.

For any theory of relativistic hydrodynamics to be suitable for numerical implementation and, thus, for predicting experimental outcomes, three fundamental requirements must be fulfilled: causality, stability, and local well-posedness. Causality demands that information does not propagate faster than the speed of light, in agreement with Einstein's relativity. Stability refers to the property that a system perturbed slightly away from global equilibrium will return to it. A locally well-posed theory is one for which a unique solution to the equations of motion exists in a neighborhood of any suitable hypersurface on which arbitrary initial data can be specified \cite{ChoquetBruhatGRBook}. One might na\"ively expect that these requirements are automatically satisfied once a covariant formulation of hydrodynamics is given, but this is not necessarily the case. For example, the standard first-order relativistic hydrodynamic theories proposed by Eckart \cite{EckartViscous} and by Landau and Lifshitz \cite{LandauLifshitzFluids} violate both causality and stability and are therefore unsuitable for numerical simulation.

The most widely used relativistic viscous hydrodynamic theories which can overcome the problems related to causality and stability originated in the pioneering works of M\"uller \cite{Muller:1967zza} and Israel and Stewart \cite{MIS-6,MIS-2}. In these so-called M\"uller-Israel-Stewart (MIS) theories, in contrast to first-order theories, new degrees of freedom in addition to the long-lived hydrodynamic modes are introduced, increasing the number of equations of motion needed to describe relativistic fluids. However, it is important to note that causality for MIS theories has only been derived in the linearized regime \cite{Hiscock_Lindblom_stability_1983,Olson:1989ey}, with the exception of some specific cases \cite{Floerchinger:2017cii,Bemfica:2019cop,Bemfica:2020xym} where nonlinear analyses were performed.

Recently, it has been shown by Bemfica, Disconzi, Noronha and Kovtun (BDNK) \cite{Bemfica:2017wps,Kovtun:2019hdm,Bemfica:2019knx,Hoult:2020eho,Bemfica:2020zjp} that one can construct causal, stable, and locally well-posed \emph{first-order} theories. As in the theories of Eckart and Landau and Lifshitz, BDNK is formulated with temperature, chemical potential, and flow velocity as the sole hydrodynamic degrees of freedom. More precisely, BDNK identified classes of hydrodynamic frames (i.e. definitions for the temperature, chemical potential, and flow velocity) for which the equations of motion are causal and locally well-posed in the fully nonlinear regime and for which perturbations of the hydrodynamic fields around global equilibria in Minkowski spacetime are linearly stable. Derivations from kinetic theory can be found in \cite{Bemfica:2017wps,Bemfica:2019knx,Rocha:2022ind,Hoult:2021gnb}, see also \cite{Biswas:2022cla, Biswas:2022hiv}. In Ref.~\cite{Noronha:2021syv}, a generalized second-order theory that unifies MIS and BDNK was proposed and proven to be causal and stable in the linearized regime (under certain conditions).

We note that the BDNK theory admits into the constitutive relations all symmetry-allowed terms proportional to first-order derivatives of the hydrodynamic fields. This includes time derivatives even in the local rest frame of the fluid, which are not present in the constitutive relations of standard first-order theories such as the one from Landau and Lifshitz \cite{LandauLifshitzFluids}. These additional time-derivative terms render the hydrodynamic equations of motion second-order with respect to time, even in the rest frame of the fluid, rather than first-order. This property leads to new gapped (i.e., nonhydrodynamic) modes that parametrize the behavior of the fluid in the ultraviolet regime. The BDNK formalism uses these nonhydrodynamic modes as ultraviolet regulators ensuring that the equations of motion are causal and stable even beyond the theory's infrared regime of applicability \cite{Bemfica:2017wps,Kovtun:2019hdm,Bemfica:2019knx,Hoult:2020eho,Bemfica:2020zjp}.

Although first-order BDNK hydrodynamics is known to be causal and stable in several specific classes of hydrodynamic frames \cite{Bemfica:2017wps,Kovtun:2019hdm,Bemfica:2019knx,Hoult:2020eho,Bemfica:2020zjp}, the complete set of necessary and sufficient conditions for causality and stability for the most general first-order theory has not yet been derived. In this paper, we derive such necessary and sufficient conditions for generic hydrodynamic frames in which no simplifying relations are imposed among the transport parameters. More precisely, the generic frames are those in which the principal part of the equations of motion is composed of terms proportional to second-order derivatives of the hydrodynamic fields. The inequalities we obtain carve out a 14-dimensional hypervolume in the space of transport parameters in which each point corresponds to a causal and stable formulation. We find that the values of just three combinations of transport parameters are needed to determine whether the theory is causal. An additional nine combinations determine whether the theory is stable. We provide these results in a form that can be immediately implemented in numerical simulations to check stability and causality given a set of transport parameters and an equation of state. The causality conditions should also be sufficient to establish local well-posedness in suitable Gevrey\footnote{A function is in the Gevrey class if all its derivatives obey certain bounds. For example, a one-dimensional function $f(x)$ is Gevrey of degree $s$ if there exists a constant $K$ such that $\left(\int_{-\infty}^\infty |f^n(x)|^2\right)^{1/(1+n)} < K (1+n)^s$. When $s=1$, this is the space of analytic functions (see \cite{RodinoGevreyBook}).} function spaces. However, the mathematical proof of such a statement is very technical and beyond the scope of this work. 

Moreover, there has been a growing interest in studying novel transport phenomena related to quantum anomalies in recent years. Two such phenomena are the chiral magnetic effect \cite{Kharzeev:2007jp, Fukushima:2008xe}, which is explicitly related to the axial anomaly, and the chiral vortical effect \cite{vilenkin1979macroscopic, vilenkin1980quantum, Erdmenger:2008rm, Banerjee:2008th, Son:2009tf, Landsteiner:2011cp, Landsteiner:2011iq,Amado:2011zx}, whose microscopic origin can be understood through various approaches \cite{Glorioso:2017lcn, Flachi:2017vlp, Avkhadiev:2017fxj, Stone:2018zel, Buzzegoli:2018wpy, Prokhorov:2020okl, Huang:2020kik}. While chiral (or anomalous) hydrodynamics \cite{Erdmenger:2008rm, Banerjee:2008th, Son:2009tf,Sadofyev:2010pr,Neiman:2010zi,Landsteiner:2012kd}, the hydrodynamic theory which incorporates the effect of quantum anomalies, has provided crucial understanding of how these anomalies manifest at the macroscopic level in fluids, there remain fundamental issues that must be considered when investigating the initial-value problem in such theories \cite{Speranza:2021bxf}. In particular, it was shown that ideal chiral hydrodynamics is causal in the Landau frame, but it is ill-posed in the most general frame as derived from kinetic theory \cite{Chen:2015gta,Son:2012wh,Stephanov:2012ki,Chen:2012ca,Son:2012zy,Manuel:2013zaa,Hidaka:2016yjf,Huang:2018wdl,Yang:2018lew}, effective actions \cite{Jensen:2012jy,Jensen:2012jh,Banerjee:2012iz,Ammon:2020rvg}, and quantum statistical mechanics \cite{Buzzegoli:2018wpy}. This demonstrates that the choice of hydrodynamic frame is important for chiral hydrodynamics even in the absence of dissipation \cite{Speranza:2021bxf}.

To date, it is not yet known whether chiral hydrodynamics can be causal and stable. In fact, previous works at first order \cite{Son:2009tf} have problems with causality \cite{Speranza:2021bxf}, while causality and stability analyses are still lacking for second-order approaches \cite{Gorbar:2017toh}. In this paper, we fill in this gap by presenting a first-order theory of viscous chiral hydrodynamics \`a la BDNK along with the necessary and sufficient conditions for this theory to be causal and stable.  We analyze two chiral theories: one in which the current is associated with an anomalous axial $U(1)_A$ symmetry, and a combined case with both vector and axial-vector currents, i.e., $U(1)_V\times U(1)_A$. As the classical $U(1)_A$ symmetry is exact only in the high-temperature or massless-fermion limits, such theories are often employed to study fluids obeying exact conformal symmetry; our results pertain also to the case where conformal symmetry is lifted.

The rest of the paper is organized as follows. In Sec.~\ref{theory}, we review the construction of first-order hydrodynamics and define the notation to be used in subsequent sections. In Sec.~\ref{causality-sec}, we derive the necessary and sufficient conditions for nonlinear causality in a generic frame, and in Sec.~\ref{stability-sec} we derive the necessary and sufficient conditions for linear stability about all homogeneous equilibrium states. Finally, in Sec.~\ref{chiral-sec} we define first-order chiral hydrodynamics in the $U(1)_A$ and $U(1)_V\times U(1)_A$ cases and derive the causality and stability conditions. A concise summary of our results can be found in the Conclusions (Sec.~\ref{conclusions}). \emph{Notation}: we use the mostly plus metric signature $(-\ +\ +\ +)$, with Greek indices running from $0$ to $3$. We work in natural units where $\hbar = c = k_B = 1$.

\section{Review of first-order hydrodynamics} \label{theory}

We begin by considering relativistic fluids described by a symmetric energy-momentum tensor $T^{\mu\nu}$ and a conserved vector $U(1)_V$ current $J^\mu$, e.g., baryon current. The hydrodynamic equations of motion are provided by the conservation laws
\begin{align} \label{conservation}
    \n_\mu T^{\mu\nu} = 0 \qquad\text{and} \qquad \n_\mu J^\mu = 0,
\end{align}
where $\n_\mu$ is the spacetime covariant derivative. To construct dissipative hydrodynamics in a derivative expansion, one first assumes that the relevant degrees of freedom are the same as those describing thermal equilibrium. These degrees of freedom, the hydrodynamic fields, can be taken as the fluid velocity $u^\mu$ ($u^\mu u_\mu = -1$), the temperature $T$, and the chemical potential $\mu$. One then expresses $T^{\mu\nu}$ and $J^\mu$ as an expansion in spacetime derivatives of these fields. In this way, Eqs.~\eqref{conservation} produce a closed system of equations for $u^\mu$, $T$, and $\mu$. Note that we are considering the most general case in which the system can also be coupled to dynamical gravity \cite{Bemfica:2017wps, Bemfica:2019knx, Bemfica:2020zjp}.

In BDNK first-order hydrodynamics \cite{Bemfica:2017wps,Kovtun:2019hdm,Bemfica:2019knx,Hoult:2020eho,Bemfica:2020zjp}, we truncate the derivative expansion of $T^{\mu\nu}$ and $J^\mu$ at first order and, following an effective field theory approach, we include all possible terms allowed by symmetries. The BDNK constitutive relations read, in the notation of \cite{Kovtun:2019hdm},
\begin{subequations}\label{decomp-kovtun}
\begin{align} 
    T^{\mu\nu} &= \mc Eu^\mu u^\nu + \mc P \Delta^{\mu\nu} + \mc Q^\mu u^\nu + \mc Q^\nu u^\mu + \mc T^{\mu\nu}, \\*
    J^\mu &= \mc N u^\mu + \mc J^\mu,
\end{align}
\end{subequations}
where
\begin{subequations} \label{constitutive-Talpha-kovtun}
\begin{align}
\mathcal{E} & =\varepsilon+\varepsilon_1 \frac{D T}{T}+\varepsilon_2 \nabla_\lambda u^\lambda+\varepsilon_3 D(\mu/T), \\
\mathcal{P} & =P+\pi_1 \frac{D T}{T}+\pi_2 \nabla_\lambda u^\lambda+\pi_3 D(\mu/T), \\
\mathcal{N} & =n+\nu_1 \frac{D T}{T}+\nu_2 \nabla_\lambda u^\lambda+\nu_3 D(\mu/T), \\
\mathcal{Q}^\mu & =\theta_1 \frac{\nabla_\perp^\mu T}{T}+\theta_2 D u^\mu+\theta_3 \nabla_\perp^\mu(\mu/T), \\
\mathcal{J}^\mu & =\gamma_1 \frac{\n_\perp^\mu T}{T}+\gamma_2 D u^\mu+\gamma_3 \n^\mu_\perp(\mu/T), \\
\label{taumunu}
\mathcal{T}^{\mu \nu} & =-2 \eta \sigma^{\mu \nu}.
\end{align}
\end{subequations}
We have defined the derivative operators $D = u^\mu \n_\mu$ and $\n_\perp^\mu = \Delta^{\mu\nu} \n_\nu$ with $\Delta^{\mu\nu} = g^{\mu\nu} + u^\mu u^\nu$ and the shear tensor $\sigma^{\mu\nu} = \Delta^{\mu\nu\alpha\beta} \n_\alpha u_\beta$ with $\Delta^{\mu\nu\alpha\beta} = (\Delta^{\mu\alpha}\Delta^{\nu\beta}+\Delta^{\mu\beta}\Delta^{\nu\alpha})/2-\Delta^{\mu\nu}\Delta^{\alpha\beta}/3$. The expansion coefficients $\ve$, $P$, $n$, $\eta$, $\ve_i$, $\pi_i$, $\nu_i$, $\theta_i$, and $\gamma_i$ ($i=1,2,3$) are functions of $T$ and $\mu$. We refer to shear viscosity $\eta$, $\ve_i$, $\pi_i$, $\nu_i$, $\theta_i$, and $\gamma_i$ ($i=1,2,3$) as transport parameters. The familiar transport coefficients $\zeta$ (bulk viscosity) and $\sigma$ (charge conductivity) can be expressed in terms of these transport parameters as \cite{Kovtun:2019hdm, Hoult:2020eho}
\begin{subequations} \label{coefficients}
\begin{align}
\zeta &= -\left[\pi_2 - \left(\frac{\partial P}{\partial \varepsilon}\right)_n \varepsilon_2 - \left(\frac{\partial P}{\partial n}\right)_\varepsilon \nu_2 \right]+\left(\frac{\partial P}{\partial \varepsilon}\right)_n\left[\pi_1 - \left(\frac{\partial P}{\partial \varepsilon}\right)_n \varepsilon_1 - \left(\frac{\partial P}{\partial n}\right)_\varepsilon \nu_1 \right] \\*
&\qquad \qquad + \frac{1}{T}\left(\frac{\partial P}{\partial n}\right)_\ve\left[\pi_3 - \left(\frac{\partial P}{\partial \varepsilon}\right)_n \varepsilon_3 - \left(\frac{\partial P}{\partial n}\right)_\varepsilon \nu_3 \right], \nonumber \\
\sigma & = \frac{n}{w} \left( \gamma_1 - \frac{n}{w}\theta_1\right) - \frac{1}{T} \left( \gamma_3 - \frac{n}{w}\theta_3 \right),
\end{align}
\end{subequations}
where we have introduced the notation $w\equiv \varepsilon + P$.

Equations \eqref{decomp-kovtun} and \eqref{constitutive-Talpha-kovtun} can be equivalently rearranged as
\begin{subequations} \label{decomp-Talpha}
\begin{align}
    T^{\mu\nu} &= T_0^{\mu\nu} + H^{\mu\nu\rho\sigma}\n_\rho (u_\sigma/T) + X^{\mu\nu\rho}\n_\rho (\mu/T), \\
    J^\mu &= J_0^\mu + Y^{\mu \rho \sigma}\n_\rho (u_\sigma/T) + Z^{\mu\rho}\n_\rho (\mu/T),
\end{align}
\end{subequations}
where
\begin{subequations}\label{constitutive-Talpha}
\begin{align} 
    \label{T0}
    T^{\mu\nu}_0 &= \ve u^\mu u^\nu + P \Delta^{\mu\nu}, \\
    \label{J0}
    J^\mu_0 &= n u^\mu, \\
    H^{\mu\nu\rho\sigma} &= T \bigg[ \ve_1 u^\mu u^\nu u^\rho u^\sigma + \ve_2u^\mu u^\nu \Delta^{\rho\sigma} + \pi_1 u^\rho u^\sigma\Delta^{\mu\nu} + \theta_1\left( u^\mu u^\sigma \Delta^{\nu\rho} + u^\nu u^\sigma \Delta^{\mu\rho} \right) \\
    &\quad + \theta_2 \left( u^\mu u^\rho \Delta^{\nu\sigma} + u^\nu u^\rho \Delta^{\mu\sigma} \right) + \pi_2 \Delta^{\mu\nu}\Delta^{\rho\sigma} -2 \eta \Delta^{\mu\nu\rho\sigma} \bigg], \nonumber \\
    Y^{\mu\rho\sigma} &= T \bigg[ \nu_1 u^\mu u^\rho u^\sigma + \nu_2 u^\mu \Delta^{\rho\sigma} + \gamma_1 u^\sigma \Delta^{\mu\rho} + \gamma_2 u^\rho \Delta^{\mu\sigma} \bigg], \\
    X^{\mu\nu\rho} &= \ve_3 u^\mu u^\nu u^\rho + \pi_3 u^\rho \Delta^{\mu\nu} + \theta_3 \left( u^\mu \Delta^{\nu \rho} + u^\nu \Delta^{\mu\rho} \right), \\
    Z^{\mu\rho} &= \nu_3 u^\mu u^\rho + \gamma_3 \Delta^{\mu\rho}.
\end{align}
\end{subequations}
Equations \eqref{constitutive-Talpha} are the most general combinations of the tensors at our disposal subject to the assumed symmetry of the energy-momentum tensor, i.e., $H^{\mu\nu\sigma\rho}=H^{\nu\mu\sigma\rho}$ and $X^{\mu\nu\rho}=X^{\nu\mu\rho}$. This way of organizing the constitutive relations allows the equations of motion \eqref{conservation} to be expressed compactly (see Sec.~\ref{causality-sec}).

Beyond the demands of symmetries, additional constraints on the constitutive relations emerge from the need to provide a consistent description of global equilibrium when external sources are present \cite{Jensen:2012jh, Banerjee:2012iz}. Reference \cite{Jensen:2012jh} obtained those consistency constraints in a specific class of hydrodynamic frames collectively termed the thermodynamic frame. As will be reviewed in Sec.~\ref{frame-sec}, a hydrodynamic frame is a choice of definitions for the fields $u^\mu$, $T$, and $\mu$. The thermodynamic frame is defined by requiring the fields to take a prescribed form in states of global thermal equilibrium. Namely, the fluid velocity is chosen to point along a timelike Killing vector $K^\mu$, i.e.~$u^\mu = K^\mu/\sqrt{-K^2}$, in equilibrium; the temperature satisfies $T= T_0/\sqrt{-K^2}$ in equilibrium, where $T_0$ is a constant that fixes the units of temperature; and the ratio $\mu/T$ is constant in equilibrium \cite{Kovtun:2022vas}. Equivalently, the thermodynamic frame can be defined by requiring the hydrodynamic fields $u^\mu$, $T$, and $\mu$ to satisfy the following Killing conditions in global equilibrium:
\begin{subequations} \label{killing}
\begin{align}
    \label{killing2}
    \nabla_{\mu}(u_{\nu}/T) + \nabla_\nu(u_\mu/T) &=0,\\
    \nabla_\alpha(\mu/T) &= 0.
\end{align}
\end{subequations}
We emphasize that identifying global equilibrium states with solutions of Eq.~\eqref{killing} constitutes an implicit definition of the thermodynamic frame. The thermodynamic frame is also known as the natural frame \cite{Van:2011yn}, the thermometer/J\"uttner frame \cite{Van:2013sma}, and the beta frame \cite{Becattini:2014yxa}. We defer to these works for more detailed discussions and definitions via kinetic theory and quantum statistical mechanics.

The methods of Ref. \cite{Jensen:2012jh} lead to the following equality-type constraints in the thermodynamic frame \cite{Kovtun:2019hdm}. First, the zeroth-order coefficients must be related by the standard thermodynamic identities $\ve + P = T(\partial P/\partial T)_\mu + \mu(\partial P/\partial \mu)_T$ and $n=(\partial P/\partial \mu)_T$. That is, $\ve$, $P$, and $n$ are interpreted in the usual way as the equilibrium energy density, pressure, and charge density, respectively. Furthermore, the first-order coefficients must obey $H^{\mu\nu\rho\sigma}=H^{\mu\nu\sigma\rho}$, or equivalently $\theta_1 = \theta_2$ and $\gamma_1 = \gamma_2$.

The equilibrium thermodynamic quantities are assumed to describe stable matter with a unique equilibrium state, so the Hessian matrix of the equilibrium entropy density $s=(\partial P/\partial T)_\mu$ with respect to any two independent thermodynamic variables is negative-definite (e.g., \cite{prigogine2014modern}). This is equivalent to the conditions
\begin{subequations} \label{hessian}
\begin{align}
    0 &< \left( \frac{\partial^2 s}{\partial \varepsilon^2} \right)_{n} = - \frac{1}{T^2c_V}, \\
    0 &< \left( \frac{\partial^2 s}{\partial \varepsilon}\right)_n\left( \frac{\partial^2 s}{\partial n}\right)_\varepsilon - \left[ \frac{\partial}{\partial \varepsilon}\bigg|_n\left( \frac{\partial s}{\partial n} \right)_\varepsilon \right]^2 = \frac{1}{T^2 c_V}\left(\frac{\partial (\mu/T)}{\partial n}\right)_\varepsilon - \left(\frac{\partial (\mu/T)}{\partial \varepsilon}\right)_n^2,
\end{align}
\end{subequations}
where $c_V = T(\partial s/\partial T)_n$ is the volumetric heat capacity. Equations \eqref{hessian} contain the well-known conditions for thermal and diffusive stability, respectively $c_V>0$ and $(\partial(\mu/T)/\partial n)_\varepsilon>0$.

Another set of constraints on the constitutive relations arises from the second law of thermodynamics. From the canonical entropy current $T S^\mu = Pu^\mu - u_\nu T^{\mu\nu} - \mu J^\mu$, one can compute the entropy production rate $\nabla_\mu S^\mu$ on-shell using the equations of motion \eqref{conservation} and constitutive relations \eqref{decomp-kovtun} and \eqref{constitutive-Talpha-kovtun}. Provided $\eta, \zeta, \sigma\ge 0$, the entropy production can be shown to be non-negative within the regime of validity of first-order theory \cite{Kovtun:2019hdm}. That is, $\nabla_\mu S^\mu = R + \mathcal O(\nabla^3)$, where $R\ge 0$ is second-order in derivatives using the solutions of the equations of motion. Therefore, the second law of thermodynamics holds for BDNK theory only within its regime of validity (i.e., when the constitutive relations are defined to first-order in derivatives). This situation is not uncommon in hydrodynamic theories constructed using systematic power-counting arguments; it also happens, for example, in DNMR theory \cite{Denicol:2012cn}. Although it is in principle possible to demand $\nabla_\mu S^\mu \ge 0$ even beyond the regime of validity of the theory, such a requirement goes beyond the regime of applicability of the first-order effective theory, resulting in a more restrictive set of constraints that excludes causality and stability, as pointed out in Refs.~\cite{Bemfica:2017wps,Kovtun:2019hdm,Bemfica:2019knx,Hoult:2020eho,Bemfica:2020zjp}.

\subsection{Hydrodynamic frames} \label{frame-sec}

We provide here a brief review of hydrodynamic frames. The fields $u^\mu$, $T$, and $\mu$ are auxiliary variables that parametrize the evolution of the fluid, but they do not have unambiguous first-principles definitions out of equilibrium \cite{Israel:1976tn,MIS-6,Kovtun:2012rj}. Therefore, one is free to perform field redefinitions valid at first order, provided the new fields coincide with the old ones when restricted to homogeneous equilibrium states. In the context of a derivative expansion truncated to first order, the most general redefinition one needs to consider is \cite{Kovtun:2019hdm}
\begin{subequations} \label{redef}
\begin{align}
    &u^\alpha \rightarrow u'^\alpha = u^\alpha + r_{u1} \frac{\n_\perp^\alpha T}{T} + r_{u2} D u^\alpha + r_{u3} \n_\perp^\alpha (\mu/T), \\
    &T \rightarrow T' = T + r_{T1} \frac{DT}{T} + r_{T 2}\n_\lambda u^\lambda + r_{T 3}D(\mu/T), \\
    &\mu \rightarrow \mu' = \mu + r_{\mu1} \frac{DT}{T} + r_{\mu2}\n_\lambda u^\lambda + r_{\mu3}D(\mu/T),
\end{align}
\end{subequations}
where $r_{ui}$, $r_{T i}$ and $r_{\mu i}$ are arbitrary functions of $T$ and $\mu$. When Eqs.~\eqref{redef} are inserted into Eqs.~\eqref{decomp-Talpha} and \eqref{constitutive-Talpha} and the latter are truncated to first order in derivatives, the resulting constitutive relations are formally identical in the primed variables but with new coefficients given by
\begin{subequations}\label{frame-transf}
\begin{align}
    \ve_i' &= \ve_i - \left(\frac{\partial \ve}{\partial T}\right)_\mu r_{T i} - \left(\frac{\partial \ve}{\partial \mu}\right)_T r_{\mu i}, \\
    \pi_i' &= \pi_i - \left( \frac{\partial P}{\partial T} \right)_\mu r_{T i} - \left( \frac{\partial P}{\partial \mu} \right)_T r_{\mu i}, \\
    \nu_i' &= \nu_i -\left( \frac{\partial n}{\partial T} \right)_\mu r_{T i} - \left( \frac{\partial n}{\partial \mu} \right)_T r_{\mu i}, \\
    \theta_i' &= \theta_i - wr_{ui}, \\
    \gamma_i' &= \gamma_i - nr_{ui}, \\
    \label{eta-transf}
    \eta' &= \eta,
\end{align}
\end{subequations}
where $i=1,2,3$. We continue to use the notation $w = \varepsilon + P$ throughout the paper.

A given definition of $u^\mu$, $T$, and $\mu$ is known as a hydrodynamic frame, and Eqs.~\eqref{redef} and \eqref{frame-transf} describe the frame transformations of first-order hydrodynamics. There is a wide variety of frames to choose from, each seemingly equally legitimate. However, the equations of motion following from Eqs.~\eqref{conservation} can have qualitatively different causality and stability properties in different hydrodynamic frames, especially when Eqs.~\eqref{frame-transf} are used to set one or more of the transport parameters to zero. The subset of hydrodynamic frames in which the theory is causal and stable defines physically acceptable theories that are suitable for numerical simulation.

Furthermore, it is important to note that the transport parameters cannot all be changed to arbitrary values by frame transformations, as reflected by the existence of seven frame-invariant quantities\footnote{ The bulk viscosity and charge conductivity in Eqs.~\eqref{coefficients} are frame invariant and can be expressed in terms of $f_i$ and $\ell_i$.}:
\begin{subequations}\label{frame-invariants}
\begin{align}
    f_i &= \pi_i - \ve_i (\partial P/\partial \ve)_n - \nu_i(\partial P/\partial n)_\ve,\\
    \ell_i &= \gamma_i - n\theta_i/w,
\end{align}
\end{subequations}
and $\eta$ itself \cite{Kovtun:2019hdm}. For example, by appropriately choosing $r_{u3}$, $r_{T3}$, and $r_{\mu 3}$, one can set $\ve'_3=\pi_3'=\theta_3'=0$, but then $\nu_3'$ and $\gamma_3'$ are fixed in terms of the frame invariants, namely $\nu_3'=-f_3/(\partial P/\partial n)_\ve$ and $\gamma_3'=\ell_3$. In particular, it is possible to obtain a theory in which derivatives of $\mu/T$ are absent from the constitutive relations only if the frame invariants $f_3$ and $\ell_3$ already vanish, either by an act of fine-tuning or by way of some additional constraints not already incorporated into the effective theory. If a desired set of transport parameter values is overconstrained in this way, as is the set $\ve_3=\pi_3=\nu_3=\theta_3=\gamma_3=0$, then this set of values is inaccessible by frame transformations.

Finally, although the thermodynamic consistency constraints described above, namely $\theta_1 = \theta_2$ and $\gamma_1 = \gamma_2$, were derived within the thermodynamic frame, they imply the frame-invariant constraint $\ell_1 = \ell_2$. The thermodynamic frame is actually a wide class of frames; indeed, any frame transformation \eqref{redef} satisfying $r_{u1}=r_{u2}$ preserves the Killing conditions \eqref{killing} that define the thermodynamic frame. On the other hand, if one is willing to part with the identification between global equilibrium states and solutions of the Killing conditions, then one can permit frame transformations with $r_{u1}\neq r_{u2}$. After such a transformation, one lands in a nonthermodynamic frame in which $\theta_1'\neq \theta_2'$ and $\gamma_1'\neq \gamma_2'$, but the content of the consistency constraints lives on in the invariant statement $\ell_1 = \ell_2$.

\subsection{Change of variables}

Following \cite{Bemfica:2020zjp}, we preemptively change variables from $T$ and $\mu$ to $\ve$ and $n$, as doing so will simplify the analysis of stability (see Sec.~\ref{stability-sec}). Equations \eqref{constitutive-Talpha-kovtun} become\footnote{ We emphasize that Eqs.~\eqref{constitutive-kovtun} are obtained from Eqs.~\eqref{constitutive-Talpha-kovtun} via an exact change of variables from $T=T(\varepsilon, n)$ to $\mu=\mu(\varepsilon, n)$ to $\varepsilon$ and $n$ using standard thermodynamic relations. Equations \eqref{xi13} show how this change of variables affects the transport coefficients and are exact.
}
\begin{subequations} \label{constitutive-kovtun}
\begin{align}
\label{E}
\mathcal{E} &= \ve + \cfe_1 D\ve+ \ve_2 \nabla_\lambda u^\lambda+ \cfe_3 Dn, \\
\mathcal{P} &= P+\cfp_1 D\ve+\pi_2 \nabla_\lambda u^\lambda+\cfp_3 Dn, \\
\mathcal{N} &= n+\cfn_1 D\ve+\nu_2 \nabla_\lambda u^\lambda+\cfn_3 Dn, \\
\mathcal{Q}^\mu &= \cfq_1 \n_\perp^\mu \ve + \theta_2 D u^\mu + \cfq_3 \n_\perp^\mu n, \\
\mathcal{J}^\mu &= \cfj_1 \n_\perp^\mu \ve + \gamma_2 D u^\mu + \cfj_3 \n_\perp^\mu n,
\end{align}
\end{subequations}
while $\mc T^{\mu\nu}$ is still given as in Eq.~\eqref{taumunu}. The new transport parameters in Eq.~\eqref{constitutive-kovtun} are given in terms of the old ones by
\begin{subequations} \label{xi13}
\begin{align} 
    \cf \psi_1 &= \frac{\psi_1}{w} \left( \frac{\partial P}{\partial \ve} \right)_n + \left( \psi_3 - \psi_1 \frac{nT}{w} \right) \left( \frac{\partial (\mu/T)}{\partial \ve} \right)_n, \\
    \cf \psi_3 &= \frac{\psi_1}{w} \left( \frac{\partial P}{\partial n} \right)_\ve + \left( \psi_3 - \psi_1 \frac{nT}{w} \right) \left( \frac{\partial (\mu/T)}{\partial n} \right)_\ve,
\end{align}
\end{subequations}
for every $\cf\psi_i\in \{ \cf\ve_i, \cf\pi_i, \cf\theta_i, \cf\nu_i, \cf\gamma_i \}$ ($i=1,3$), and an analogous notation applies for $\psi_i$. The thermodynamic consistency constraint $\theta_1 = \theta_2$, described above, now becomes an equality relating the parameters $\cfq_1$, $\theta_2$ and $\cfq_3$; similarly, the constraint $\gamma_1=\gamma_2$ relates $\cfj_1$, $\gamma_2$ and $\cfj_3$. For simplicity of notation, and to leave open the possibility of transforming to a nonthermodynamic frame, nowhere in the rest of the paper will we solve these constraints to eliminate any of the transport parameters.

The alternative organization of the constitutive relations in Eqs.~\eqref{decomp-Talpha} becomes
\begin{subequations} \label{decomp}
\begin{align}
    \label{decompT}
    T^{\mu\nu} &= T_0^{\mu\nu} + \tilde H^{\mu\nu\rho\sigma}\n_\rho u_\sigma + \tilde X^{\mu\nu\rho}_1 \n_\rho \ve + \tilde X^{\mu\nu\rho}_3 \n_\rho n, \\
    J^\mu &= J_0^\mu + \tilde Y^{\mu\rho\sigma}\n_\rho u_\sigma + \tilde Z^{\mu\rho}_1\n_\rho \ve + \tilde Z^{\mu\rho}_3 \n_\rho n,
\end{align}
\end{subequations}
where
\begin{subequations} \label{constitutive}
\begin{align}
    \label{H}
    \tilde H^{\mu\nu\rho\sigma} &= \ve_2 u^\mu u^\nu \Delta^{\rho\sigma} + \theta_2\left( u^\mu u^\rho \Delta^{\nu\sigma} + u^\nu u^\rho \Delta^{\mu\sigma} \right) + \pi_2 \Delta^{\mu\nu}\Delta^{\rho\sigma} -2\eta \Delta^{\mu\nu\sigma\rho}, \\
    \label{Y}
    \tilde Y^{\mu\rho\sigma} &= \nu_2 u^\mu \Delta^{\rho\sigma} + \gamma_2 u^\rho \Delta^{\mu\sigma}, \\
    \tilde X_1^{\mu\nu\rho} &= \cfe_1 u^\mu u^\nu u^\rho + \cfp_1 u^\rho \Delta^{\mu\nu} + \cfq_1 ( u^\mu \Delta^{\nu\rho} + u^\nu \Delta^{\mu\rho}), \\
    \tilde X_3^{\mu\nu\rho} &= \cfe_3 u^\mu u^\nu u^\rho + \cfp_3 u^\rho \Delta^{\mu\nu} + \cfq_3 (u^\mu \Delta^{\nu\rho} + u^\nu \Delta^{\mu\rho}), \\
    \tilde Z^{\mu\rho}_1 &= \cfn_1 u^\mu u^\rho + \cfj_1 \Delta^{\mu\rho}, \\
    \tilde Z^{\mu\rho}_3 &= \cfn_3 u^\mu u^\rho + \cfj_3 \Delta^{\mu\rho},
\end{align}
\end{subequations}
and $T^{\mu\nu}_0$ and $J^\mu_0$ remain as in Eqs.~\eqref{T0} and \eqref{J0}.

\section{Causality} \label{causality-sec}

We now derive the set of necessary and sufficient conditions under which the general BDNK hydrodynamic equations are causal in the fully nonlinear regime. The results we obtain are a set of inequalities constraining the transport parameters. The analysis is based on standard techniques from the theory of partial differential equations (PDEs) \cite{Courant_and_Hilbert_book_2,ChoquetBruhatGRBook}.

The equations of motion for $u^\mu$, $\ve$, and $n$ are obtained by inserting the constitutive relations \eqref{decomp} and \eqref{constitutive} into the conservation laws \eqref{conservation}. We must be mindful of the constraint $u^\mu u_\mu = -1$, which can be implemented in several equivalent ways. One way is to decompose $\n_\mu T^{\mu\nu} = 0$ into two equations, $\Delta^{\alpha}{}_\nu \n_\mu T^{\mu\nu} = 0$ and $u_\nu \n_\mu T^{\mu\nu} = 0$. Here, instead, following Ref.~\cite{Bemfica:2019knx}, it is more convenient to leave $\n_\mu T^{\mu\nu}=0$ alone and to apply the operator $u^\lambda \n_\lambda$ twice to the constraint equation $u^\mu u_\mu = -1$, yielding 
\begin{align}
    u^\mu u^\rho u^\sigma  \p_\mu \p_\rho u_\sigma + u^\rho[\n_\rho (u^\sigma u^\mu)] \n_\mu u_\sigma = 0
\end{align}
as an additional equation of motion. We thus obtain a system of quasilinear PDEs for $\Psi = (u_\sigma, \ve, n)$,
\begin{align} \label{eom}
    [A^{\mu\rho}]^{IJ} \n_\mu \n_\rho \Psi_J + [B^{\mu\rho}]^{IJK}(\nabla_\mu \Psi_J)(\nabla_\rho \Psi_K) + [C^\mu]^{IJ} \n_\mu \Psi_J &= 0,
\end{align}
where the indices $I$, $J$ and $K$ run from $0$ to $5$ and the coefficient matrices are given by
\begin{subequations} \label{matrices}
\begin{align}
    \label{principal}
    [A^{\mu\rho}]^{IJ}
    &=
    \begin{pmatrix}[1.5]
       \tilde H^{\mu\nu\rho\sigma} & \tilde X_1^{\mu\nu\rho} & \tilde X_3^{\mu\nu\rho} \\
       u^\mu u^\rho u^\sigma & 0 & 0 \\
       \tilde Y^{\mu\rho\sigma} & \tilde Z_1^{\mu\rho} & \tilde Z_3^{\mu\rho}
    \end{pmatrix}, \\
    [C^{\mu}]^{IJ}
    &=
    \begin{pmatrix}[1.5]
        w(g^{\mu\sigma} u^\nu + g^{\nu\sigma} u^\mu)\qquad~ & u^\mu u^\nu + \Delta^{\mu\nu} \left( \frac{\partial P}{\partial \ve} \right)_n\qquad~ & \Delta^{\mu\nu}\left( \frac{\partial P}{\partial n} \right)_\ve \\
        0 & 0 & 0 \\
        n g^{\mu\sigma} & 0 & u^\mu
    \end{pmatrix}.
\end{align}
\end{subequations}
The $[B^{\mu\rho}]^{IJK}$ depend on $\Psi$ but not its derivatives; their explicit form will not be important for the rest of the paper. As mentioned earlier, gravity can be coupled to the fluid by inserting Eq.~\eqref{decompT} into Einstein's equations. As shown in Refs.~\cite{Bemfica:2017wps, Bemfica:2019knx, Bemfica:2020zjp}, the causality conditions derived in this section remain the same whether $g^{\mu\nu}$ is dynamically coupled to the fluid or treated as an external source. The reason is that, upon including Einstein's equations, the characteristic determinant to be discussed below would factorize into two pieces, one corresponding to the metrical degrees of freedom and one corresponding to the hydrodynamic fields.

Equations \eqref{eom} are causal if, and only if, their characteristics, the hypersurfaces along which initial data are propagated, are everywhere nonspacelike \cite{Courant_and_Hilbert_book_2, ChoquetBruhatGRBook}. The characteristics are determined by the principal part of Eq.~\eqref{eom}, i.e., the terms containing the highest-order derivatives of each unknown (the highest order can be different for different unknowns). As mentioned in the Introduction, in this paper we treat the hydrodynamic frames in which the transport parameters have generic values and are not subject to finely tuned relations.  We, therefore, call this the ``generic case'' and speak of ``generic hydrodynamic frames.'' Then, the principal part is exactly $[A^{\mu\rho}]^{IJ} \n_\mu \n_\rho \Psi_J$ (see Sec.~\ref{causality-subsec} for further discussion and examples of nongeneric cases).

The system is causal if, and only if, the normal vectors to the characteristics, $\varphi^\mu$, which are obtained in generic frames as the solutions of the characteristic equation
\begin{align} \label{detAzero}
    \det[A^{\mu\rho}\varphi_\mu\varphi_\rho] = 0,
\end{align}
all satisfy
\begin{align} \label{causality-abstract}
    \varphi^\mu \text{ is real and } \varphi^\mu\varphi_\mu \ge 0.
\end{align}
Upon inserting \eqref{principal}, we obtain the characteristic determinant
\begin{align} \label{characteristic-det}
     \det[A^{\mu\rho}\varphi_\mu \varphi_\rho] = -b^2\left( \theta_2 b^2 - \eta v^2 \right)^2 q(b^2, v^2)
\end{align}
with
\begin{align} \label{q}
q(b^2, v^2) = \lzero b^6 - \lone b^4 v^2 + \ltwo b^2 v^4 - \lthree v^6,
\end{align}
where $b= u^\mu \varphi_\mu$, $v^\mu = \Delta^{\mu\nu} \varphi_\nu$, and
\begin{subequations} \label{lambdai}
\begin{align}
    \lzero &= \theta_2 \ot{\cfe,\cfn}, \\
    \lone &= \left(\frac{4}{3}\eta - \pi_2\right)\ot{\cfe,\cfn} - (\ve_2+\theta_2)\ot{\cfn,\cfp + \cfq} + (\nu_2+\gamma_2)\ot{\cfe,\cfp + \cfq} + \theta_2(\ot{\cfj,\cfe}+\ot{\cfn,\cfq}), \\
    \ltwo &= \left(\frac{4}{3}\eta - \pi_2\right)(\ot{\cfn,\cfq}+\ot{\cfj,\cfe}) + (\ve_2+\theta_2)\ot{\cfj,\cfp + \cfq} + (\nu_2+\gamma_2)\ot{\cfp,\cfq} + \theta_2\ot{\cfq,\cfj}, \\
    \lthree &= \left(\frac{4}{3}\eta - \pi_2\right)\ot{\cfq,\cfj}.
\end{align}
\end{subequations}
Here we have introduced a compact notation $\ot{\cf\psi, \cf\chi}= \cf\psi_1\cf \chi_3-\cf \psi_3\cf \chi_1$, where $\tilde \psi_i$ and $\cf \chi_i$ can be any of the transport parameters \eqref{xi13}, e.g., $\{\cf \ve, \cf \nu\} = \cf \ve_1 \cf \nu_3 - \cf \ve_3 \cf \nu_1$ and $\ot{\cf\nu, \cf \pi + \cf \theta} = \cf\nu_1(\cf\pi_3+\cf\theta_3) - \cf\nu_3(\cf\pi_1+\cf\theta_1)$. 

\begin{figure}
    \captionsetup[subfigure]{labelformat=empty}
    \subfloat[\label{subfig:a}]{
        \includegraphics[valign=c,width=0.48\textwidth]{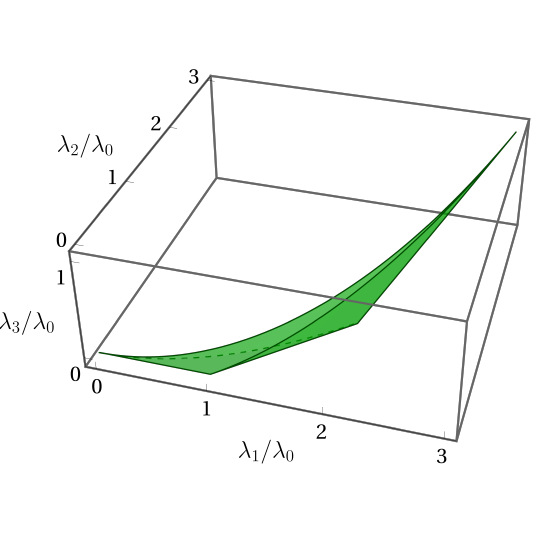}
    }
    \hfill
    \subfloat[\label{subfig:b}]{
        \includegraphics[valign=c,width=0.48\textwidth]{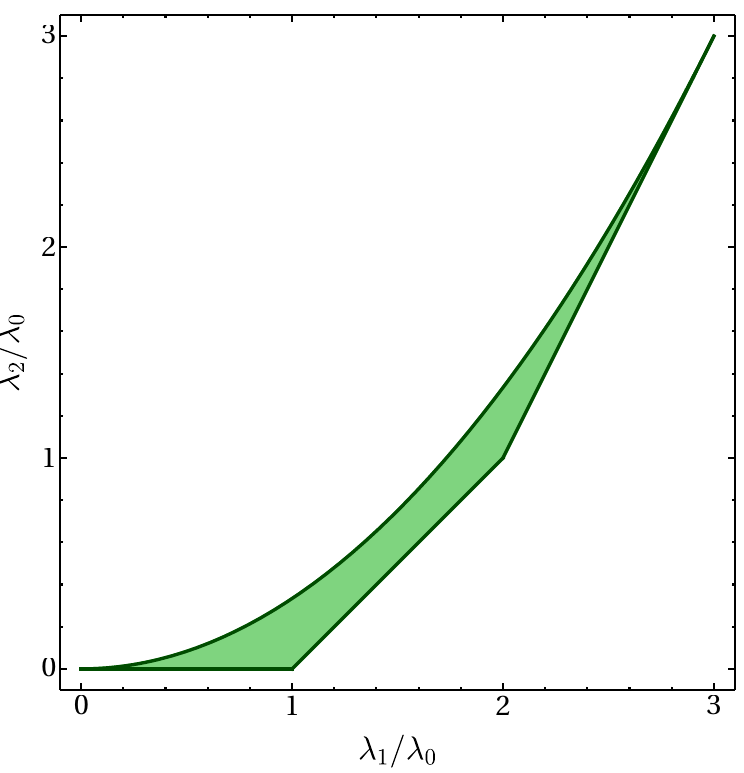}
    } \\[-3ex]
    
    \subfloat[\label{subfig:c}]{
        \includegraphics[valign=c,width=0.48\textwidth]{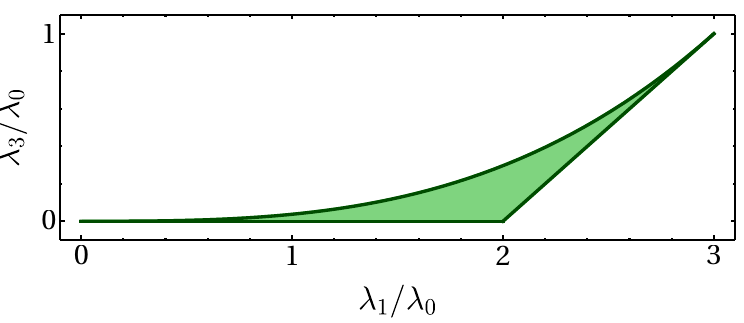}
    }
    \hfill
    \subfloat[\label{subfig:d}]{
        \includegraphics[valign=c,width=0.48\textwidth]{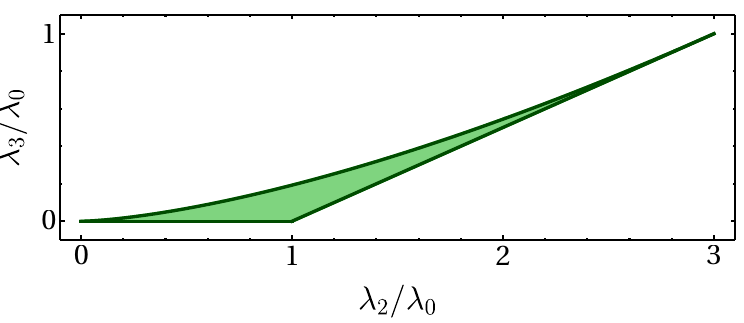}
    }
    \caption{The shaded region in the upper-left panel is the intersection of the causality conditions \eqref{first-causality-cond}-\eqref{last-causality-cond}, and the other panels are its projections onto the three coordinate planes. As described in the text, the causal hydrodynamic frames are those for which $0 \le \eta/\theta_2 \le 1$ is satisfied and $\lambda_i/\lambda_0$ ($i=1,2,3$) lie within the shaded region in the upper-left panel.}
    \label{fig:amplituhedron}
\end{figure}

For the system to be causal, $\lambda_0$ cannot vanish. If it did, then Eq.~\eqref{q} would have an overall factor of $v^2$, so Eq.~\eqref{detAzero} would have solutions with $v^\mu = 0$ and $b\neq 0$ and the conditions \eqref{causality-abstract} could not be satisfied. Assuming $\lzero\neq 0$, Eq.~\eqref{characteristic-det} can be factorized as
\begin{align} \label{characteristic-speeds}
     \det[A^{\mu\rho}\varphi_\mu \varphi_\rho] = -\theta_2^{\;2}\lzero b^2\left( b^2 - \frac{\eta}{\theta_2} v^2 \right)^2\prod_{i=1}^3(b^2-c_i^{\;2} v^2),
\end{align}
where $c_1^{\;2}$, $c_2^{\;2}$ and $c_3^{\;2}$ are the roots of the polynomial
\begin{align} \label{qx}
    \lzero x^3 - \lone x^2 +\ltwo x - \lthree.
\end{align}
The quantities $\eta/\theta_2$, $c_1^{\;2}$, $c_2^{\;2}$, and $c_3^{\;2}$ are the squares of the so-called characteristic speeds at which information propagates.\footnote{To understand this statement in the context of a simple example, take the wave equation ${u^\mu u^\nu \n_\mu \n_\nu \psi - c^2 \Delta^{\mu\nu}\n_\mu \n_\nu \psi = 0}$ for a scalar function $\psi$. Its characteristic equation reads $b^2 - c^2 v^2 = 0$. The solutions $\varphi^\mu$ of the characteristic equation therefore have $\varphi^\mu \varphi_\mu = -b^2 + v^2 = (1-c^2)v^2$, and the causality conditions \eqref{causality-abstract} become $0 \le c^2 \le 1$.} The conditions \eqref{causality-abstract} for causality are equivalent to the condition that the squared characteristic speeds all lie within the interval $[0,1]$. By evaluating the latter, we find that the theory is causal in a generic frame if, and only if, all of the conditions\footnote{To write these conditions in a simple way, we have used $\lambda_0 > 0$, which will arise in the next section as a necessary condition for stability. We also assume throughout the paper that the equation of state is such that $T>0$ and $w>0$.}
\begin{subequations}  \label{causality-conditions}
\begin{align}
\label{cc1}
&\theta_2\neq 0, \\
\label{cc2}
&\ot{\cfe, \cfn} \neq 0, \\
&0 \le \eta / \theta_2 \le 1, \\
\label{first-causality-cond}
&0 \le \bar\lambda_1 \le 3 \bar\lambda_0, \\
&\bar\lambda_2 \ge \max\{ 0, -3\bar\lambda_0+2\bar\lambda_1 \}, \\
&0 \le \bar \lambda_3 \le \bar\lambda_0 - \bar\lambda_1 + \bar\lambda_2, \\
\label{last-causality-cond}
&\bar\lambda _2^{\, 2} \bar\lambda _1^{\, 2}-4 \bar\lambda _0 \bar\lambda _2^{\, 3}-4 \bar\lambda _1^{\, 3}\bar\lambda _3 -27 \bar\lambda _0^{\, 2} \bar\lambda _3^{\, 2}+18 \bar\lambda _0 \bar\lambda _1 \bar\lambda _2 \bar\lambda _3 \ge 0
\end{align}
\end{subequations}
are simultaneously satisfied, where we have defined dimensionless quantities $\bar\lambda_i= (T^3/w)\lambda_i$ ($i=0,1,2,3$). These conditions are visualized in Fig. \ref{fig:amplituhedron}. Note that conditions \eqref{cc1} and \eqref{cc2} are a restatement of the requirement $\lambda_0 \neq 0$, and they imply, for example, that the Landau frame, which is defined by setting to zero the parameters $\tilde \ve_1$, $\ve_2$, $\cf \ve_3$, $\tilde \nu_1$, $\nu_2$, $\cf \nu_3$, $\tilde \theta_1$, $\theta_2$, $\cf \theta_3$ and having at least one of $\cf \gamma_1$, $\gamma_2$, $\cf \gamma_3$ nonzero, cannot be causal \cite{PichonViscous, Hiscock_Lindblom_instability_1985, Denicol:2008ha}. In particular, Eq.~\eqref{cc2} requires at least one of $\tilde \varepsilon_1$ or $\tilde \nu_1$ to be nonzero. Therefore, as mentioned in the Introduction, the constitutive relations of first-order BDNK theory also contain time derivatives in the local rest frame of the fluid [see Eq.~\eqref{constitutive-kovtun}].

Given specific choices for the transport parameters as functions of $\ve$ and $n$, the conditions \eqref{causality-conditions} can be checked throughout a domain of values $\ve, n \in U \subset \mathbb R^2$ prior to hydrodynamic simulation. If they are met, then any solution of the equations of motion \eqref{eom} for which $\ve(x)$ and $n(x)$ lie within $U$ at every spacetime event $x$ is guaranteed to obey causality.

It should be possible to show that the conditions \eqref{causality-conditions} imply local well-posedness in suitable Gevrey spaces, using the results of Ref.~\cite{DisconziFollowupBemficaNoronha}. However, proving such a result goes beyond the scope of this paper (for details, we refer the reader to  Appendix A of Ref.~\cite{DisconziFollowupBemficaNoronha} and the recent review \cite{Disconzi:2023rtt}). Therefore, we leave detailed statements concerning the well-posedness of solutions to future work.

Let us now consider two classes of frames in which the full causality conditions are simpler. The first is defined by $\cfe_3=\cfp_3=\cfq_3=0$. The second is defined by $\cfj_1=\gamma_2=\cf\gamma_3=\cfn_1=\nu_2=0$ (with $\cf\nu_3 \neq 0$; see Sec.~\ref{causality-subsec}). In both classes of frames, the principal part $[A^{\mu\rho}\varphi_\mu\varphi_\rho]^{IJ}$ becomes block triangular in $I$ and $J$ (lower triangular in the first class and upper triangular in the second). Therefore, both have exactly the same characteristic determinant,
\begin{subequations}
\begin{align}
    \det[A^{\mu\rho}\varphi_\mu \varphi_\rho] = -\theta_2^{\;2} \tilde\nu_3 b^2 \left( b^2 - \frac{\eta}{\theta_2} v^2 \right)^2\left(b^2 + \frac{\tilde\gamma_3}{\tilde \nu_3} v^2 \right)(\Lambda_0 b^4 - \Lambda_1 b^2 v^2 + \Lambda_2 v^4),
\end{align}
\end{subequations}
where
\begin{subequations} \label{Lambdai}
\begin{align}
    \Lambda_0 &= \cfe_1\theta_2, \\
    \Lambda_1 &= \left( \frac{4}{3}\eta - \pi_2 \right) \cfe_1 + \cfp_1(\ve_2+\theta_2) + \ve_2 \tilde \theta_1, \\
    \Lambda_2 &= -\left( \frac{4}{3}\eta - \pi_2 \right)\cfq_1.
\end{align}
\end{subequations}
From condition \eqref{causality-abstract}, causality holds if, and only if, all of the conditions
\begin{subequations} \label{causality-conditions-simpler}
\begin{align}
    \label{3cond1}
    &\tilde\nu_3 \neq 0,\\
    \label{3cond2}
    &0 \le -\tilde\gamma_3/\tilde\nu_3 \le 1,\\
    \label{Lfirst}
    &\theta_2 \neq 0,\\
    &\tilde \ve_1\neq 0,\\
    &0 \le \eta/\theta_2 \le 1,\\
    &0 \le \bar\Lambda_1 \le \bar\Lambda_0 + \bar\Lambda_2, \\
    &0 \le \bar\Lambda_2 \le \bar\Lambda_0,\\
    \label{Llast}
    &\bar\Lambda_1^{\;2} \ge 4 \bar\Lambda_2 \bar\Lambda_0
\end{align}
\end{subequations}
are simultaneously satisfied, where $\bar\Lambda_i = (T^2/w)\Lambda_i$ ($i=0,1,2$). The first class of frames, where $\cf\ve_3 = \cf \pi_3 = \cf \theta_3=0$, is closely analogous to the class $\ve_3 = \pi_3 = \theta_3 = 0$ already studied in detail by Ref. \cite{Hoult:2020eho}, because the determinant factorizes in the same way. Indeed, their causality conditions in these frames are formally identical to \eqref{causality-conditions-simpler}.

\subsection{Comment on linear vs. nonlinear causality} \label{causality-subsec}

We have treated the generic case in which $[A^{\mu\rho}]^{IJ}\n_\mu \n_\rho \Psi_J$ defines the principal part of \eqref{eom}. Equivalently, this is the case in which $\det[A^{\mu\rho}\varphi_\mu\varphi_\rho]$ does not vanish identically (for all values of $u^\mu$, $\ve$, $n$, and $\varphi^\mu$). In this case, it can be shown that the causality conditions \eqref{causality-conditions}, which were derived from the full nonlinear equations of motion \eqref{eom}, are exactly the same as the causality conditions for the \emph{linearization} of the equations of motion [see Eq.~\eqref{lineom}], as was observed by Ref.~\cite{Hoult:2020eho}.

A nontrivial example falling outside the generic case occurs when
\begin{align}\label{example-outside}
    \frac{\cf \psi_1}{\cf\psi_3} = \frac{\left(\frac{\partial P}{\partial \ve}\right)_n - n T \left(\frac{\partial (\mu/T)}{\partial \ve}\right)_n}{\left(\frac{\partial P}{\partial n}\right)_\ve - n T \left(\frac{\partial (\mu/T)}{\partial n}\right)_\ve}
\end{align}
for every $\cf\psi_i\in \{ \cf\ve_i, \cf\pi_i, \cf\theta_i, \cf\nu_i, \cf\gamma_i \}$ ($i=1,3$), since it can be directly verified that $\det[A^{\mu\rho}\varphi_\mu\varphi_\rho] = 0$ identically. The reason this is outside our scope is more directly seen by returning to $T$ and $\mu$ as the independent variables. By inverting Eqs.~\eqref{xi13} with Eqs.~\eqref{example-outside} inserted, we find $\ve_3=\pi_3=\nu_3=\theta_3=\gamma_3=0$, so there are no second-order derivatives of $\mu/T$ appearing in the equations of motion. Consequently, the principal part contains, in addition to $[A^{\mu\rho}]^{IJ}\n_\mu \n_\rho \Psi_J$, some terms from $[B^{\mu\rho}]^{IJK}(\nabla_\mu \Psi_J)(\nabla_\rho \Psi_K)$, and \eqref{detAzero} must be modified accordingly. However, these additional terms drop out when taking the linearization of \eqref{eom}, so this provides an example of a first-order hydrodynamic theory whose nonlinear causality conditions do not coincide with those of its linearization. It is interesting to note that Ref.~\cite{Rocha:2023hts} has realized this example from a microscopic theory of weakly self-interacting classical ultrarelativistic scalar particles. From the effective field theory perspective outlined in Sec.~\ref{frame-sec}, we note that this example is not accessible by frame transformations.

Another example which is outside our scope is $\cf \nu_1 = \nu_2 = \cf \nu_3 = \cf \gamma_1 = \gamma_2 = \cf \gamma_3 = 0$. In this class of frames, the bottom row of \eqref{principal} vanishes, so $\det[A^{\mu\rho}\varphi_\mu\varphi_\rho]$ vanishes identically. The charge conservation equation (Eq.~\eqref{eom} with $I=5$) then contains only first-order derivatives of the hydrodynamic fields and plays the role of a constraint equation. These frames have already been analyzed in detail by Ref.~\cite{Bemfica:2020zjp}.

\section{Linear stability} \label{stability-sec}

In this section, we obtain the necessary and sufficient conditions for the linear stability of Eqs.~\eqref{eom} by considering the fate of small perturbations around a nonrotating state of global equilibrium in Minkowski spacetime. Such a state is characterized by constant $u^\mu$, $\ve$, and $n$, and we write perturbations around this background state as ${u^\mu \rightarrow u^\mu + \delta u^\mu}$ (with $u^\mu \delta u_\mu = 0$), $\ve \rightarrow \ve + \delta \ve$, and $n \rightarrow n + \delta n$. The perturbations $\delta \Psi = (\delta u_\sigma, \delta \ve, \delta n)$ evolve according to the linearization of Eqs.~\eqref{eom}, which is
\begin{align} \label{lineom}
     [A^{\mu\rho}]^{IJ} \n_\mu \n_\rho \delta \Psi_J + [C^{\mu}]^{IJ}\n_\mu \delta \Psi_J &= 0.
\end{align}
Inserting the Fourier transform $\delta \Psi_J(t, \bm x) = \int d^4k/(2\pi)^4 \exp\left[T(\Gamma t + i \k \cdot \bm x)\right] \delta \Psi_J(\Gamma, \bm k)$, where $T$ is the constant background temperature and $k^\mu = (i\Gamma, \k)$ is dimensionless, Eqs.~\eqref{lineom} become
\begin{align}
     \left(T[A^{\mu\rho}]^{IJ} k_\mu k_\rho - i[C^{\mu}]^{IJ}k_\mu\right) \delta\Psi_J(\Gamma, \k) &= 0,
\end{align}
and the dispersion relations $\Gamma(\k) = \Re[\Gamma(\k)] + i \Im[\Gamma(\k)]$ are the solutions of the corresponding secular equation
\begin{align}\label{secular}
    \det[T A^{\mu\rho} k_\mu k_\rho - i C^\mu k_\mu] = 0.
\end{align}
The theory is said to be modally linearly stable when, for any constants $u^\mu$, $\ve$, and $n$, all of the dispersion relations satisfy
\begin{align} \label{stability}
    \Re [\Gamma(\k)] \le 0
\end{align}
for all $\k \in \mathbb R^3$. Despite the Lorentz-covariant form of \eqref{lineom}, stability for one value of $u^\mu$ does not automatically imply stability for all other values of $u^\mu$. However, it was recently proven that if the theory is causal, this is indeed the case \cite{Bemfica:2020zjp, Gavassino:2021owo} (for more recent developments on the connection between stability and causality, see e.g. Refs.~\cite{Gavassino:2021kjm, Gavassino:2023myj, Gavassino:2023mad}). Furthermore, causality is necessary for the stability of a dissipative theory \cite{Gavassino:2021owo}. Therefore, it is necessary and sufficient to impose the causality conditions \eqref{causality-conditions} and only check stability for a given choice of the background $u^\mu$, which we take to be $u^\mu=(1, \bm 0)$ for convenience. Equation \eqref{secular} splits into shear and sound channels, given, respectively, by
\begin{subequations}\label{stability-polynomials}
\begin{align} 
    \label{stability-polynomials-shear}
    &T\theta_2 \Gamma^2 + w\Gamma + T\eta|\k|^2 = 0, \\
    \label{sound}
    &a_0\Gamma^6 + a_1\Gamma^5 + a_2\Gamma^4 + a_3 \Gamma^3 + a_4 \Gamma^2 + a_5 \Gamma + a_6 = 0.
\end{align}
\end{subequations}
The coefficients $a_i=a_i(\k)$ appearing in the sound polynomial \eqref{sound} read
\begin{subequations} \label{ai}
\begin{align}
    \label{a0}
    a_0 &=\blzero, \\
    \label{a1}
    a_1 &= \bar A, \\
    a_2 &= \bar B+\blone |\k|^2, \\
    a_3 &= 1 + \bar C|\k|^2, \\
    a_4 &= \bar D|\k|^2 + \bltwo|\k|^4, \\
    a_5 &= c_s^2|\k|^2 + \bar E|\k|^4, \\
    \label{a6}
    a_6 &= \bar F|\k|^4 + \blthree|\k|^6,
\end{align}
\end{subequations}
where we have denoted by $c_s$ the speed of sound given by
\begin{align}
    c_s^2 = \left( \frac{\partial P}{\partial \ve} \right)_{n} + \frac{n}{w} \left( \frac{\partial P}{\partial n} \right)_{\ve}.
\end{align}
The coefficients $\bar\lambda_i$ are the same as those appearing in Sec.~\ref{causality-sec}, and we have defined the dimensionless quantities $\bar A = (T^2/w) A$, $\bar B = (T/w)B$, $\bar C = (T^2/w) C$, $\bar D = (T/w) D$, $\bar E = (T^2/w) E$ and $\bar F = (T/w) F$, with
\begin{subequations} \label{ABCDEF}
\begin{align}
    A &= w\ot{\cfe,\cfn} + \theta_2(\cfe_1+\cfn_3), \\
    \label{barbie}
    B &= \theta_2 + w(\cfe_1+\cfn_3), \\
    C &= \left(\frac{4}{3}\eta - \pi_2\right)(\cfe_1+\cfn_3) + (\ve_2 + \theta_2)(\cfp_1 + \cfq_1 - \td{\cfn}) + (\nu_2+\gamma_2)(\cfp_3 + \cfq_3 + \td{\cfe}) \\
    &\quad + w(\ot{\cfp,\cfn}+\ot{\cfj,\cfe}) + n\ot{\cfe,\cfp + \cfq} - \theta_2(\cfq_1+\cfj_3),\nonumber \\
    \label{bardie}
    D &= \frac{4}{3}\eta - \pi_2 + w(\cfp_1-\cfj_3 - \td{\cfn}) + n(\cfp_3 + \cfq_3 + \td{\cfe}) + \left( \frac{\partial P}{\partial \ve} \right)_n(\ve_2+\theta_2) + \left( \frac{\partial P}{\partial n}\right)_\ve(\nu_2+\gamma_2), \\
    E &= -\left(\frac{4}{3}\eta - \pi_2\right)(\cfq_1+\cfj_3) +w\ot{\cfj,\cfp} + n \ot{\cfp,\cfq} + (\ve_2+\theta_2)\td{\cfj} - (\nu_2 + \gamma_2)\td{\cfq}, \\
    \label{F}
    F &= w\td{\cfj} - n\td{\cfq}.
\end{align}
\end{subequations}
Here, we have introduced another shorthand notation
\begin{align}
    \td{\cf \xi} = \cf\xi_1\left( \frac{\partial P}{\partial n}\right)_\ve - \cf\xi_3 \left( \frac{\partial P}{\partial \ve} \right )_n.
\end{align}

In the shear channel \eqref{stability-polynomials-shear}, the dispersion relations can be calculated analytically. We find
\begin{align} \label{shear-dispersion}
    \Gamma_\pm(\k) = -\frac{w}{2T\theta_2}\left(1 \pm \sqrt{1-4T^2\eta\theta_2|\k|^2/w^2} \right).
\end{align}
Then, by inspection of Eq.~\eqref{shear-dispersion}, one can see that the stability condition \eqref{stability} in the shear channel is equivalent to $\theta_2 \ge 0$ and $\eta \ge 0$.

The conditions for stability in the sound channel are considerably more involved. A necessary and sufficient set of conditions for modal stability is provided by the Routh-Hurwitz criterion. The Routh-Hurwitz criterion \cite{gradshteyn2007} states that \eqref{stability} holds for all sound channel dispersion relations if, and only if,\footnote{There is a relatively minor exception to the ``only if" part of this statement. That is, there are edge cases in which the Routh-Hurwitz conditions \eqref{RH} are violated but the theory is still stable. In all such cases, the theory is only marginally stable, i.e., at least one mode does not dissipate ($\Re[\Gamma]=0$). More importantly, while Eqs.~\eqref{causality-conditions} and \eqref{RH} define a hyper\emph{volume} in the space of transport parameters, these edge cases occupy a set of measure zero (a union of hyper\emph{surfaces} of various co-dimensions in the space of transport parameters). Therefore, achieving these edge cases would require the transport parameters to be finely tuned, with any small deviation rendering the theory unstable. To understand this in the context of a simple example, suppose \eqref{sound} were instead the quadratic $a \Gamma^2 + b \Gamma + c = 0$. The Routh-Hurwitz conditions replacing \eqref{RH} would be $a \ge 0$, $b \ge 0$ and $c \ge 0$ \cite{gradshteyn2007}. However, it is easy to verify that if $b=0$, $a \le 0$, and $c \le 0$, then the roots $\Gamma = \pm i\sqrt{|c|/|a|}$ are still stable. (This situation does not arise in the shear channel because of the assumption $w>0$.)}

\begin{align} \label{RH}
    \quad a_0,a_1,a_6,\Delta_1,\Delta_2,\Delta_3,\Delta_4 \ge 0
\end{align}
for all $\k \in \mathbb R^3$, where
\begin{subequations} \label{Delta-ai}
\begin{align}
    \Delta_1 &= a_1a_2-a_0a_3, \\
    \Delta_2 &= a_3\Delta_1-a_1(a_1a_4-a_0a_5), \\
    \Delta_3 &= a_4\Delta_2 + (a_1 a_6-a_2a_5)\Delta_1 + a_0a_5(a_1a_4-a_0a_5), \\
    \Delta_4 &= a_5\Delta_3 -a_6a_3\Delta_2 +a_1a_5a_6\Delta_1-a_1^3a_6^2.
\end{align}
\end{subequations}
Inserting Eqs.~\eqref{ai} into Eqs.~\eqref{Delta-ai} gives
\begin{subequations}\label{Deltai}
\begin{align} 
    \label{Delta1}
    \Delta_1 &= \Delta_{(1, 0)} + \Delta_{(1,2)} \ka^2, \\
    \label{Delta2}
    \Delta_2 &= \Delta_{(2, 0)} + \Delta_{(2, 2)} \ka^2 + \Delta_{(2, 4)} \ka^4, \\
    \label{Delta3}
    \Delta_3 &= \Delta_{(3, 2)} \ka^2 + \Delta_{(3, 4)} \ka^4 + \Delta_{(3, 6)} \ka^6 + \Delta_{(3, 8)} \ka^8, \\
    \label{Delta4}
    \Delta_4 &= \Delta_{(4, 4)} \ka^4 + \Delta_{(4, 6)} \ka^6 + \Delta_{(4, 8)} \ka^8 + \Delta_{(4, 10)} \ka^{10} + \Delta_{(4, 12)} \ka^{12},
\end{align}
\end{subequations}
where full expressions for the $\D{i, j}$ are given in Eqs.~\eqref{Deltaij}. Then the conditions \eqref{RH} can be translated into an equivalent set of $\k$-independent conditions. For example, in view of Eq.~\eqref{Delta1}, the condition $\Delta_1(\k) \ge 0$ for all $\k$ is equivalent to $\Delta_{(1,0)}\ge 0$ and $\Delta_{(1,2)} \ge 0$. To make an analogous statement for the condition $\Delta_2(\k)\ge 0$, we define the set
\begin{align}
    S_2 &= \{(a, b, c)\in\mathbb R^3 | a\ge 0,c \ge 0, b \ge - 2 \sqrt{a c} \}.
\end{align}
As is easily verified, $\Delta_2(\k) \ge 0$ for all $\k \in \mathbb R^3$ if, and only if, $(\D{2, 4}, \D{2, 2}, \D{2,0})\in S_2$.
Similarly, we define $S_3$ to be the set of $(a, b, c, d)\in \mathbb R^4$ satisfying one of the four alternatives
\begin{subequations}
\begin{align}
    \bullet\quad& a \ge 0\quad\AND\quad b \ge 0\quad\AND\quad c \ge 0\quad\AND\quad d \ge 0,  \\
    \bullet\quad& a > 0\quad\AND\quad d > 0\quad\AND\quad \disc(a, b, c, d)\le 0, \\
    \bullet\quad& a = 0\quad\AND\quad (b, c, d)\in S_2, \\
    \bullet\quad& d = 0\quad\AND\quad (a, b, c)\in S_2,
\end{align}
\end{subequations}
where the function $\disc(a, b, c, d) = 18 a b c d + b^2 c^2 - 4 a c^3 - 4 d b^3 - 27 a^2 d^2$ is the discriminant of a cubic polynomial $a x^3 + b x^2 + c x + d$. Finally, following \cite{doi:10.1137/0915035}, we define $S_4$ to be the set of $(a, b, c, d, e)\in \mathbb R^5$ for which one of the following five (mutually exclusive) alternatives holds:
\begin{subequations}
\begin{align}
    \bullet\quad& a>0 \quad\AND\quad e>0 \quad\AND\quad \chi_2<-2 \quad\AND\quad L \le 0 \quad\AND\quad \chi_1 + \chi_3 > 0, \\
    \bullet\quad& a>0\quad\AND\quad e>0 \quad\AND\quad -2 \le \chi_2 \le 6 \quad\AND\quad
    \begin{Bmatrix}[1.5]
        L \le 0 \quad \AND \quad \chi_1 + \chi_3 > 0  \\
        \OR \\
        L \ge 0 \quad \AND \quad K_1 \le 0
    \end{Bmatrix}, \\
    \bullet\quad&a>0 \quad \AND \quad e>0 \quad \AND \quad \chi_2 > 6 \quad \AND \quad
    \begin{Bmatrix}[1.5]
        L \le 0 \quad \AND \quad \chi_1 + \chi_3 > 0 \\
        \OR \\
        \chi_1 > 0 \quad \AND \quad \chi_3 > 0 \\
        \OR \\
        L \ge 0 \quad \AND \quad K_2 \le 0
    \end{Bmatrix}, \\
    \bullet\quad&a=0 \quad\AND\quad (b, c, d, e)\in S_3, \\
    \bullet\quad&e=0 \quad\AND\quad (a, b, c, d)\in S_3,
\end{align}
\end{subequations}
where
\begin{gather}
\chi_1= ba^{-3/4} e^{-1/4}, \quad \chi_2 = c a^{-1/2} e^{-1/2}, \quad \chi_3= d a^{-1/4}e^{-3/4}, \nonumber \\
L=\left(\chi_2^{\,2}-3 \chi_1 \chi_3+12\right)^3-\left(72 \chi_2+9 \chi_1 \chi_2 \chi_3-2 \chi_2^{\,3}-27 \chi_1^{\,2}-27 \chi_3^{\,2}\right)^2, \\
K_1 =(\chi_1-\chi_3)^2-16(\chi_1+\chi_2+\chi_3+2), \quad K_2 =(\chi_1-\chi_3)^2-\frac{4(\chi_2+2)}{\sqrt{\chi_2-2}}(\chi_1+\chi_3+4 \sqrt{\chi_2-2}). \nonumber
\end{gather}

We can now state the full result. In a generic frame, the hydrodynamic  equations \eqref{eom} are linearly modally stable if, and only if, the causality conditions \eqref{causality-conditions} are satisfied and all of the following conditions hold:
\begin{subequations} \label{stability-conditions}
\begin{align}
    \label{stability-shear}
    &\theta_2 > 0 \quad\AND\quad \eta \ge 0, \\
    \label{stability-a}
    &\bar\lambda_0 > 0 \quad\AND\quad \bar A \ge 0 \quad \AND \quad \bar F \ge 0 \quad \AND \quad \bar \lambda_3 \ge 0,\\
    \label{stability-Delta1}
    &\D{1,0} \ge 0 \quad\AND\quad \D{1,2} \ge 0, \\
    \label{stability-Delta2}
    &(\D{2, 4},\D{2,2},\D{2,0})\in S_2, \\
    \label{stability-Delta3}
    &(\D{3, 8},\D{3, 6},\D{3, 4},\D{3,2}) \in S_3, \\
    \label{stability-Delta4}
    &(\D{4, 12},\D{4, 10},\D{4, 8},\D{4,6},\D{4,4}) \in S_4.
\end{align}
\end{subequations}
Contained within this statement is the fact that causality conditions are necessary conditions for stability.

As mentioned above, the conditions \eqref{stability-shear} correspond to stability in the shear channel, while the conditions \eqref{stability-Delta1} correspond to the non-negativity of Eq.~\eqref{Delta1}. Conditions \eqref{stability-a} correspond to non-negativity of Eqs.~\eqref{a0}, \eqref{a1} and \eqref{a6}. Conditions \eqref{stability-Delta2}, \eqref{stability-Delta3} and \eqref{stability-Delta4} correspond to non-negativity of Eqs.~\eqref{Delta2}, \eqref{Delta3} and \eqref{Delta4}, respectively, thus completing Eq.~\eqref{RH}. We have verified that these conditions are nonempty by finding an explicit causal and stable example using the gamma-law equation of state (the equation of state of a relativistic ideal gas \cite{Rezzolla_Zanotti_book}). Further study is needed along the lines of Ref.~\cite{Pandya:2022sff} to develop systematic methods for constructing causal and stable hydrodynamic frames given an equation of state.

In some circumstances, it may be useful to have at hand a simpler set of sufficient (but not necessary) or necessary (but not sufficient) conditions for stability. If the causality conditions \eqref{causality-conditions} are satisfied, then
\begin{align} \label{sufficient}
    \bar \lambda_0> 0~\quad\AND\quad \bar A \ge 0 \quad\AND\quad \bar F \ge 0 \quad\AND\quad \bar \lambda_3 \ge 0 \quad\AND\quad \Delta_{(i, j)} \ge 0,
\end{align}
for every $\D{i,j}$ appearing in Eqs.~\eqref{Deltai}, are sufficient but not necessary for stability. These conditions are the simplest way to guarantee \eqref{RH}, as they impose non-negativity of all the coefficients in the expansions of $a_0$, $a_1$, $a_6$, $\Delta_1$, $\Delta_2$, $\Delta_3$ and $\Delta_4$ in powers of $|\k|^2$. The conditions
\begin{align}
    \bar A, \bar B, \bar C, \bar D, \bar E, \bar F, c_s^2 \ge 0 \quad\AND\quad \bar\lambda_0 > 0 \quad\AND\quad \bar\lambda_i \ge 0
\end{align}
for $i=1,2,3$, together with the causality conditions, are necessary but not sufficient for stability. They arise from the fact that all the $a_i(\k)$ must have the same sign in order for Eq.~\eqref{stability} to hold for all solutions of \eqref{sound}.

Finally, consider again the special class of frames defined by $\cf \ve_3 = \cf \pi_3 = \cf \theta_3 = 0$, as discussed in Sec.~\ref{causality-sec}. If we assume that the equation of state satisfies $(\partial P/\partial n)_\ve = 0$ (as, e.g., in a conformal fluid\footnote{Full conformal symmetry would impose further constraints on the constitutive relations, namely $\ve_i = 3 \pi_i$ ($i=1,2,3$), $\pi_1 = 3 \pi_2$, and $\nu_1 = 3 \nu_2$. See e.g.~Appendix B of \cite{Kovtun:2019hdm}.}), then the sound polynomial \eqref{sound} factorizes further in these frames. The necessary and sufficient conditions for stability are therefore much simpler; the theory is stable if, and only if, Eqs.~\eqref{causality-conditions-simpler} hold and
\begin{subequations} \label{stability-simpler}
\begin{align}
    \label{nugamma}
    &\cf \nu_3 > 0 \quad\AND\quad \cf\gamma_3 \le 0, \\
    \label{Cfirst}
    &\theta_2 > 0 \quad\AND\quad \eta \ge 0, \\
    &\bar\Lambda_0 > 0 \quad\AND\quad \bar B \ge 0 \quad\AND\quad \left( \frac{\partial P}{\partial \ve}\right)_n \ge 0 \quad\AND\quad \bar \Lambda_2 \ge 0, \\
    &\bar B\, \bar\Lambda_1 - \bar D\,\bar \Lambda_0 \ge 0 \quad\AND\quad \bar D- \bar B \left( \frac{\partial P}{\partial \ve}\right)_n \ge 0, \\
    \label{Clast}
    &\bar D^2 \bar \Lambda_0 - \bar B\, \bar D\, \bar \Lambda_1 + \bar B^2 \bar\Lambda_2 \le  0,
\end{align}
\end{subequations}
where $\bar\Lambda_i=(T^2/w)\Lambda_i$ with the $\Lambda_i$ defined in Eqs.~\eqref{Lambdai}. The quantities $\bar B=(T/w)B$ and $\bar D=(T/w)D$ are obtained from Eqs.~\eqref{barbie} and \eqref{bardie} by setting $\cf \ve_3 = \cf \pi_3 = \cf \theta_3 = (\partial P/\partial n)_\ve = 0$, which we write here for convenience:
\begin{subequations} \label{BD}
\begin{align}
    &B = \theta_2 + w\cf \ve_1,\\
    &D = \frac{4}{3}\eta - \pi_2 + w \cf \pi_1 + \left( \frac{\partial P}{\partial \ve}\right)_n(\ve_2 + \theta_2).
\end{align}
\end{subequations}

\section{Chiral BDNK  hydrodynamics} \label{chiral-sec}

We now turn to an analysis of chiral hydrodynamics \cite{Erdmenger:2008rm, Banerjee:2008th, Son:2009tf,Sadofyev:2010pr,Neiman:2010zi,Landsteiner:2012kd}. We first consider the case in which the current $J^\mu$ is associated with an axial $U(1)_A$ symmetry that can be broken by a chiral anomaly, as opposed to the preserved vector $U(1)_V$ symmetry contemplated in the previous sections. Although it is typical in discussions of chirality, for the sake of generality we do not necessarily work in the conformal limit, i.e., we allow the terms in the constitutive relations which violate conformal symmetry. Furthermore, we allow for coupling to a nondynamical external $U(1)_A$ gauge field $A^\mu$. This theory can be obtained from the one set out in Sec.~\ref{theory} by making the following modifications (generalizing Ref.~\cite{Son:2009tf} to a generic hydrodynamic frame).

First, the conservation laws \eqref{conservation} are replaced by
\begin{align}\label{Aeom}
    \n_\mu T^{\mu\nu} = F^{\nu\lambda} J_\lambda \qquad \text{and} \qquad \n_\mu J^\mu = \mc C E_\mu B^\mu,
\end{align}
where $F_{\mu\nu} = \n_\mu A_\nu - \n_\nu A_\mu$ is the background field strength, $E^\mu = F^{\mu\nu} u_\nu$ is a covariant $U(1)_A$ electric field, and $B^\mu = \frac{1}{2}\e^{\mu\nu\rho\sigma}u_\nu F_{\rho\sigma}$ is a covariant $U(1)_A$ magnetic field, with Levi-Civita tensor $\e^{\mu\nu\rho\sigma}$ (we use the convention $\sqrt{-g}\e^{0123}=1$ where $g=\det[g_{\mu\nu}]$). The right-hand sides of Eqs.~\eqref{Aeom} account, respectively, for work done on the fluid by the external fields and for the chiral anomaly with anomaly coefficient $\mc C$.

The general first-order constitutive relations now contain additional terms. We adopt a weak-field power-counting scheme in which $A^\mu$ is treated as zeroth order in derivatives (hence $E^\mu$ and $B^\mu$ are first order). Equations \eqref{constitutive-kovtun} are replaced with
\begin{subequations} \label{Aconstitutive-kovtun}
\begin{align}
\label{AE}
\mathcal{E} &= \ve + \cfe_1 D\ve+ \ve_2 \nabla_\lambda u^\lambda+ \cfe_3 Dn, \\
\mathcal{P} &= P+\cfp_1 D\ve+\pi_2 \nabla_\lambda u^\lambda+\cfp_3 Dn, \\
\mathcal{N} &= n+\cfn_1 D\ve+\nu_2 \nabla_\lambda u^\lambda+\cfn_3 Dn, \\
\mathcal{Q}^\mu &= \cfq_1 \n_\perp^\mu \ve + \theta_2 D u^\mu + \cfq_3 \n_\perp^\mu n + \theta_E E^\mu + \theta_B B^\mu + \theta_\omega \omega^\mu, \\
\mathcal{J}^\mu &= \cfj_1 \n_\perp^\mu \ve + \gamma_2 D u^\mu + \cfj_3 \n_\perp^\mu n + \gamma_E E^\mu + \gamma_B B^\mu + \gamma_\omega \omega^\mu.
\end{align}
\end{subequations}
Here $\omega^\mu = \frac{1}{2}\e^{\mu\nu\rho\sigma}u_\nu\n_\rho u_\sigma$ is the fluid vorticity, and the new transport parameters $\theta_B$, $\theta_\omega$, $\gamma_B$ and $\gamma_\omega$ are the chiral conductivities (more conventionally denoted by $\xi$ in the literature).

All of the thermodynamic consistency constraints described in Sec.~\ref{theory} translate verbatim to the chiral case. Furthermore, we must have $\theta_E = -\theta_3/T$ and $\gamma_E = -\gamma_3/T$, and the chiral conductivities are tightly constrained in terms of the anomaly \cite{Jensen:2012jy}. The most general functional forms for the chiral conductivities in the thermodynamic frame (neglecting $CPT$-violating terms \cite{Bhattacharya:2011tra}) were obtained from thermodynamic consistency by Ref. \cite{Jensen:2012jy},
\begin{subequations} \label{xi}
\begin{align}
    \gamma_B &= \mc C \mu, \\
    \label{gop}
    \gamma_\omega &= \mc C \mu^2 + \tilde {\mc C} T^2, \\
    \theta_B &= \frac{1}{2}\mc C \mu^2 + \frac{1}{2}\tilde {\mc C} T^2, \\
    \label{xiT}
    \theta_\omega &= \frac{2}{3}\mc C \mu^3 + 2\tilde {\mc C} T^2\mu,
\end{align}
\end{subequations}
where $\tilde {\mc C}$ is an integration constant that can be related to the mixed-gravitational anomaly \cite{Jensen:2013kka}. The same constraints can be obtained from an entropy-current analysis \cite{Son:2009tf, Neiman:2010zi} and can be explicitly computed from kinetic theory \cite{Chen:2015gta,Yang:2018lew}.

\subsection{Causality and stability}

Assembling the modifications described above, we find that the generic-frame principal part for BDNK chiral hydrodynamics is
\begin{align} \label{Acharacteristic-det}
     \det[A^{\mu\rho}\varphi_\mu \varphi_\rho] = -b^2\left[ \theta_2^{\;2} b^4 - 2\left(\eta\theta_2 - \frac{\theta_\omega^2}{8}\right)b^2 v^2 + \eta^2 v^4 \right]q(b^2, v^2),
\end{align}
where $q(b^2, v^2)$ is given by Eq.~\eqref{q} in terms of the very same $\lambda_i$ defined in \eqref{lambdai}. As before, if $\theta_2=0$, then \eqref{Acharacteristic-det} vanishes when $v^\mu = 0$ for any $b\neq0$, and the theory is acausal. Imposing $\theta_2 \neq 0$, the factor in Eq.~\eqref{Acharacteristic-det} in square brackets factorizes as
\begin{align}
    \theta_2\left( b^2 - c_+^2 v^2 \right) \left( b^2 - c_-^2 v^2 \right),
\end{align}
where $c_\pm^2$ are two squared characteristic speeds associated with the propagation of shear waves. The $c_\pm^2$ are the roots of the polynomial
\begin{align}
    p(x) = x^2 +2\left (z^2 - h\right)x + h^2,
\end{align}
where we have defined dimensionless quantities $h=\eta/\theta_2$ and $z=\theta_\omega/(2\sqrt{2}\theta_2)$. The polynomial $p(x)$ takes its minimum value $p(x^*)=z^2(2h-z^2)$ when $x^*=h-z^2$. If the theory is to be causal, then the equations of motion must be hyperbolic PDEs; i.e., $c_\pm$ must be real. This in turn implies $x^* \ge 0$ and $p(x^*) \le 0$ which, by inspection, cannot happen unless $z=0$. We conclude that generic-frame BDNK chiral hydrodynamics cannot be causal unless $\theta_\omega = 0$, i.e., unless there is no anomalous vorticity-induced heat flux. If $\theta_\omega=0$, then the necessary and sufficient conditions for causality are completely identical to Eqs.~\eqref{causality-conditions}. There are no conditions involving the parameters $\theta_B$, $\gamma_B$, or $\gamma_\omega$.

The frame transformations in Eq.~\eqref{frame-transf} are insufficient to cure the pathology associated with $\theta_\omega \neq 0$. However, in the presence of $B^\mu$, $\omega^\mu$, and the pseudoscalar degree of freedom $n$, the collection of first-order frame transformations is enlarged to include those of the form
\begin{align}
    u^\alpha \rightarrow u'^\alpha = u^\alpha + r_{uB} B^\alpha + r_{u\omega} \omega^\alpha,
\end{align}
which induce the following transformations on the transport parameters:
\begin{subequations}
\begin{align} \label{Aframe-transf}
    &\theta_B \rightarrow \theta_B' = \theta_B - w r_{uB}, \\
    &\theta_\omega \rightarrow \theta_\omega' = \theta_\omega - w r_{u\omega}, \\
    &\gamma_B \rightarrow \gamma_B' = \gamma_B - n r_{uB}, \\
    &\gamma_\omega \rightarrow \gamma_\omega' = \gamma_\omega - n r_{u\omega}.
\end{align}
\end{subequations}
In particular, by setting $r_{u\omega} = \theta_\omega/w$ we obtain
\begin{subequations}
\begin{align}
    \theta_\omega' &= 0, \\
    \label{gammaomegaprime}
    \gamma_\omega' &= \mc C \mu^2\left( 1 - \frac{2}{3} \frac{n\mu}{w}\right) + \tilde {\mc C} T^2 \left( 1 - \frac{4\mu n}{w} \right).
\end{align}
\end{subequations}
However, all transformations with $r_{u\omega} \neq 0$ take the theory out of the thermodynamic frame, as they do not preserve the Killing condition \eqref{killing2} for global equilibrium states with nonzero $\omega^\mu$. This is reflected in the fact that Eq.~\eqref{gammaomegaprime} is not of the form \eqref{gop}. In other words, demanding causality in generic-frame BDNK first-order chiral hydrodynamics forces one to give up on the Killing conditions Eq.~\eqref{killing} as characteristics of global equilibrium. An analogous result was previously obtained for ideal chiral hydrodynamics \cite{Speranza:2021bxf}.

The importance of hydrodynamic frame choices in chiral hydrodynamics has also been discussed in other contexts. For example, the no-drag frame \cite{Stephanov:2015roa, Rajagopal:2015roa} is the frame in which an object at rest experiences no drag force, a natural notion of ``rest frame'' for the fluid. This frame coincides with the thermodynamic frame for the $U(1)_A$ model under consideration and is, hence, excluded by causality.

Modal linear stability is defined as in Sec.~\ref{stability-sec} with vanishing external fields $E^\mu=B^\mu=0$. The conditions for stability in the chiral case are precisely the same as Eqs.~\eqref{stability-conditions} because the term proportional to $\gamma_\omega$ does not survive the process of linearization.\footnote{ If the electromagnetic fields were dynamical, then homogeneous configurations with $\mu \neq 0$ and $B^\mu = 0$ would be unstable to the formation of helical electromagnetic structures \cite{Akamatsu:2013pjd, Hirono:2015rla, Manuel:2015zpa}, so such states are not thermal equilibria. When interpreting the stability conditions \eqref{stability-conditions} for the $U(1)_A$ case, all parameters should be evaluated at $\mu=0$.} It is reasonable to expect that this should be the case, as the frame transformations are already exhausted by causality; any further conditions would constrain the coefficients $\mc C$ and $\tilde {\mc C}$ which are, however, already determined microscopically by anomalies.

\subsection{The $U(1)_V\times U(1)_A$ theory} \label{VA}

Realistic hydrodynamic studies of chiral effects in heavy-ion collisions should include both $U(1)_V$ and $U(1)_A$ currents simultaneously, with an ordinary $U(1)_V$ gauge field instead of the $U(1)_A$ analog. The equations of motion for this theory are
\begin{align}
    \n_\mu T^{\mu\nu} = F^{\nu}_{\;\;\,\lambda} J_{V}^{\lambda} \qquad \text{and} \qquad \n_\mu J_V^\mu = 0 \qquad \text{and} \qquad \n_\mu J_A^\mu = \mc C E^\mu B_\mu,
\end{align}
where $J_V^\mu$ is a vector current; $J_A^{\mu}$ is an axial current with a chiral anomaly; and $F^{\mu\nu}$, $E^\mu$, and $B^\mu$ are now associated with a $U(1)_V$ gauge field. The most general constitutive relations are, in a straightforward extension of the previous notation,
\begin{subequations} 
\begin{align}
\label{VAE}
&\mathcal{E} = \ve + \cfe_1 D\ve+ \ve_2 \nabla_\lambda u^\lambda+ \cfe_{3V} Dn_V + \cf \ve_{3A} Dn_A, \\
&\mathcal{P} = P+\cfp_1 D\ve+\pi_2 \nabla_\lambda u^\lambda+\cfp_{3V} Dn_V + \cf \pi_{3A} Dn_A, \\
&\mathcal{N}_V = n_V+\cfn_{V1} D\ve+\nu_{V2} \nabla_\lambda u^\lambda+\cfn_{V3V} Dn_V + \cf \nu_{V3A} Dn_A, \\
&\mathcal{N}_A = n_A+\cfn_{A1} D\ve+\nu_{A2} \nabla_\lambda u^\lambda+\cfn_{A3V} Dn_V + \cf \nu_{A3A} Dn_A, \\
&\mathcal{Q}^\mu = \cfq_1 \n_\perp^\mu \ve + \theta_2 D u^\mu + \cfq_{3V} \n_\perp^\mu n_V + \cf \theta_{3A}\n_\perp^\mu n_A + \theta_E E^\mu + \theta_B B^\mu + \theta_\omega \omega^\mu, \\
&\mathcal{J}^\mu_V = \cfj_{V1} \n_\perp^\mu \ve + \gamma_{V2} D u^\mu + \cfj_{V3V} \n_\perp^\mu n_V + \cf \gamma_{V3A}\n_\perp^\mu n_A + \gamma_{VE} E^\mu + \gamma_{VB} B^\mu + \gamma_{V\omega} \omega^\mu \\
&\mathcal{J}^\mu_A = \cfj_{A1} \n_\perp^\mu \ve + \gamma_{A2} D u^\mu + \cfj_{A3V} \n_\perp^\mu n_V + \cf \gamma_{A3A}\n_\perp^\mu n_A + \gamma_{AE} E^\mu + \gamma_{AB} B^\mu + \gamma_{A\omega} \omega^\mu.
\end{align}
\end{subequations}

The conditions for causality in this theory are unwieldy in a generic frame [since Eq.~\eqref{qx} is replaced by a quartic polynomial], but they take a simpler form in the class of frames defined by $\cf\ve_{3V}=\cf\ve_{3A}=\cf\pi_{3V}=\cf\pi_{3A}=\cf\theta_{3V}=\cf\theta_{3A} = 0$. To state them, we first define
\begin{subequations}
\begin{gather}
    N_3 = \begin{pmatrix}
        \cf \nu_{V3V} & \cf \nu_{V3A} \\
        \cf \nu_{A3V} & \cf \nu_{A3A}
    \end{pmatrix},
    \qquad
    G_3 = \begin{pmatrix}
        \cf \gamma_{V3V} & \cf \gamma_{V3A} \\
        \cf \gamma_{A3V} & \cf \gamma_{A3A}
    \end{pmatrix},
    \\
    f = - \cf\nu_{V3V}\cf\gamma_{A3A} - \cf\nu_{A3A}\cf\gamma_{V3V} + \cf\nu_{V3A}\cf\gamma_{A3V} + \cf\nu_{A3V}\cf\gamma_{V3A}.
\end{gather}
\end{subequations}
Then, in the considered class of frames, we find that the theory is causal if, and only if, $\theta_\omega = 0$ and
\begin{subequations} \label{new3conds}
\begin{align}
    \label{detN3}
    &\det  N_3 \neq 0, \\
    \label{GN1}
    &0 \le f \le \det N_3 + \det G_3, \\
    \label{GN2}
    &0 \le \det G_3 \le \det N_3, \\
    \label{GN3}
    &f^2 \ge 4 \det N_3 \det G_3,
\end{align}
\end{subequations}
and the conditions \eqref{Lfirst} through \eqref{Llast} are satisfied.

If, in addition to our choice of frame, the equation of state satisfies $(\partial P/\partial n_V)_{\ve,n_A} = (\partial P/\partial n_A)_{\ve,n_V} = 0$, then the theory is stable if, and only if, it is causal and
\begin{subequations}
\begin{align}
    \label{N3cond}
    \det N_3 > 0 \quad\AND\quad \tr N_3 > 0, \\
    \label{G3cond}
    \det G_3 \ge 0 \quad\AND\quad \tr G_3 \le 0, \\
    \det N_3 \tr G_3 + f \tr N_3 \ge 0,
\end{align}
\end{subequations}
and conditions \eqref{Cfirst} through \eqref{Clast} are satisfied [with $B$ and $D$ given by \eqref{BD}]. Condition \eqref{N3cond} states that $N_3$ is positive definite, while condition \eqref{G3cond} states that $G_3$ is negative semidefinite; together they generalize \eqref{nugamma}.

\section{Conclusions} \label{conclusions}

In this paper, we have fully characterized for the first time to the best of our knowledge all of the generic causal and stable frames of first-order relativistic hydrodynamics with a $U(1)$ current by providing necessary and sufficient conditions. We treated two cases exhaustively: the vector $U(1)_V$ case and the anomalous axial $U(1)_A$ case. In the combined $U(1)_V\times U(1)_A$ case, with both vector and axial-vector currents, we obtained necessary and sufficient conditions only within a convenient class of frames. We also noted that the conditions for causality in these theories should imply local well-posedness in Gevrey function spaces, leaving a detailed statement of this property for future work. For the sake of generality, our results do not require conformal symmetry, even in anomalous cases.

In the $U(1)_V$ case, we showed that first-order hydrodynamics in a generic frame is causal if, and only if, conditions \eqref{causality-conditions} are simultaneously satisfied. This result remains valid in the presence of dynamical gravity. Furthermore, the theory is linearly stable around all homogeneous equilibrium states if, and only if, it is causal and all conditions \eqref{stability-conditions} are simultaneously satisfied. As a special case, we also studied the class of frames in which $\mc E$, $\mc P$, and $\mc N$ do not receive out-of-equilibrium corrections from derivatives of the charge density $n$. In these frames, the causality conditions take the simpler form \eqref{causality-conditions-simpler}, in agreement with Ref.~\cite{Hoult:2020eho}. We found that if the equation of state obeys $(\partial P/\partial n)_\ve = 0$, then the stability conditions in these special frames become drastically simpler; see Eq.~\eqref{stability-simpler} and the surrounding text.

While BDNK first-order hydrodynamics is parametrized by the $14$ transport parameters appearing in Eqs.~\eqref{constitutive-kovtun}, causality is determined by just three combinations, namely $\lambda_1/\lambda_0$, $\lambda_2/\lambda_0$, and $\lambda_3/\lambda_0$ [see Eqs.~\eqref{lambdai}], as depicted in Fig.~\ref{fig:amplituhedron}. This is reminiscent of the hydrohedron construction discussed in \cite{Heller:2023jtd}. Causality and stability together are decided by the values of $12$ combinations: $\eta$, $\theta_2$, $\lambda_0$, $\lambda_1$, $\lambda_2$, $\lambda_3$, $A$, $B$, $C$, $D$, $E$ and $F$ [see Eqs.\eqref{ABCDEF}], along with the equation of state. Two of these combinations, $\eta$ and $F$, are invariant under frame transformations. Any nine of the remaining ten can be varied independently under frame transformations, with the last one being determined.

In Sec.~\ref{causality-subsec}, we pointed out some nongeneric cases of interest which fall outside the scope of our analysis and are obtained by imposing relations among the transport parameters. These can be treated on a case-by-case basis using techniques similar to those employed for the generic frames. Some have already been considered in the literature in detail \cite{Bemfica:2020zjp, Hoult:2020eho}, but not all. Notably, there are scenarios in which the conditions for nonlinear causality differ from those for linear causality. One such scenario [see Eq.~\eqref{example-outside}] has already arisen in the literature \cite{Rocha:2023hts}, but its full causality and stability properties are not yet known.

We wrote, for the first time our knowledge, the BDNK first-order theory with a $U(1)_A$ anomaly, and found necessary and sufficient conditions for its causality. We found that causality requires $\theta_\omega=0$; that is, vorticity-induced heat flux must be absent from the theory. Apart from this, the generic-frame conditions for causality---even in the presence of an external gauge field---and stability are identical to Eqs.~\eqref{causality-conditions} and \eqref{stability-conditions} from the $U(1)_V$ case. Causality and stability conditions for the combined $U(1)_V\times U(1)_A$ case were derived in Sec.~\ref{VA}. It is notable that, as a consequence of the vanishing of $\theta_\omega$, one cannot formulate first-order chiral hydrodynamics in the thermodynamic frame. The same was shown to be true of the zeroth-order truncation, i.e., ideal chiral hydrodynamics \cite{Speranza:2021bxf}. Results such as these fit into the wider research area which aims to understand how fundamental principles such as thermodynamic consistency, causality, and stability constrain the macroscopic behavior emerging from inherently quantum-mechanical microscopic effects such as those due to quantum anomalies \cite{Son:2009tf}. 
Looking forward, in this respect, it will be important to investigate in detail the causality and stability properties of hydrodynamics with dynamical spin degrees of freedom, see, e.g., \cite{Florkowski:2017ruc,Hattori:2019lfp,Bhadury:2020puc,Weickgenannt:2020aaf,Gallegos:2021bzp,Hongo:2021ona,Weickgenannt:2022zxs,Daher:2022wzf,Xie:2023gbo, Sarwar:2022yzs}. The first step toward a BDNK formulation of spin hydrodynamics was recently taken in \cite{Weickgenannt:2023btk}.

The results of our paper are crucial for numerical implementations of first-order relativistic hydrodynamics for studies of heavy-ion collisions and astrophysical systems such as neutron star mergers \cite{Pandya:2021ief, Pandya:2022pif,  Pandya:2022sff, Bantilan:2022ech}. In addition, numerical simulations aiming to evaluate the role of chirality and quantum anomalies in heavy-ion collisions \cite{Shi:2017cpu,Shi:2019wzi,Buzzegoli:2022kqx} can now track such effects throughout the hydrodynamic evolution of the quark-gluon plasma using the self-contained causal and stable hydrodynamic theory presented in this work.

\section*{Acknowledgments}
We thank M.~M.~Disconzi, G.~S.~Denicol, I.~A.~Shovkovy, P.~Kovtun, and R.~Hoult for useful discussions. J.N.~is partially supported by the U.S. Department of Energy, Office of Science, Office for Nuclear Physics under Awards No. DE-SC0021301 and DE-SC0023861. J.N.~and E.S.~thank KITP - UC Santa Barbara for its hospitality during ``The Many Faces of Relativistic Fluid Dynamics" program, where this work's last stages were completed. This research was partly supported by the National Science Foundation under Grant No. NSF PHY-1748958.

\appendix

\section{}

The following are explicit expressions for the $\D{i,j}$ defined in Eqs.~\eqref{Deltai}:

\begin{subequations} \label{Deltaij}
\begin{align}
    \D{1,0} &= \bar A\, \bar B-\bar \lambda _0, \\
    \D{1,2} &= \bar \lambda_1\bar A -\bar \lambda_0\bar C,  \\
    \D{2,0} &= \D{1, 0}, \\
    \D{2,2} &= \bar C\D{1,0} + \D{1,2} - \bar A(\bar A\, \bar D - \bar\lambda_0 c_s^2), \\
    \D{2,4} &= \bar C\D{1,2} - \bar A(\bar{\lambda}_2\bar{A} - \bar{\lambda }_0\bar{E}), \\
    \D{3,2} &= (\bar D - \bar B c_s^2)\D{1,0},\\
    \D{3,4} &= (\bar A\, \bar F - \bar B\, \bar E  - c_s^2 \bar\lambda_1 +\bar \lambda_2 )\D{1,0} - \bar B c_s^2 \D{1,2} +\bar D \D{2,2} + c_s^2 \bar \lambda_0( \bar A\, \bar D - c_s^2 \bar \lambda_0), \\
    \D{3,6} &= (\bar\lambda_3\bar A-\bar\lambda_1 \bar E)\D{1,0} +(\bar A\, \bar F - \bar B\,\bar E-c_s^2\bar\lambda_1)\D{1,2} + \bar \lambda_2 \D{2,2} + \bar D\D{2,4} \\
    &\qquad+ \bar\lambda_0[ \bar E(\bar A\, \bar D - \bar \lambda_0 c_s^2) + c_s^2(\bar \lambda_2 \bar A - \bar \lambda_0 \bar E)], \nonumber \\
    \D{3,8} &= (\bar \lambda_3\bar A- \bar \lambda_1 \bar E)\D{1,2} + \bar \lambda_2\D{2,4} + \bar\lambda_0 \bar E(\bar \lambda_2 \bar A - \bar\lambda_0 \bar E), \\
    \D{4,4} &= -\bar F \D{1,0} + c_s^2 \D{3,2}, \\
    \D{4,6} &= [\bar F(\bar A c_s^2 - \bar C)-\bar\lambda_3]\D{1,0} - \bar F \D{2,2} + \bar E \D{3,2} + c_s^2 \D{3,4}, \\
    \D{4,8} &= [\bar \lambda_3(\bar A c_s^2 - \bar C) + \bar A \,\bar E\,\bar F]\D{1,0} + \bar A\,\bar F c_s^2 \D{1,2} - (\bar\lambda_3 + \bar C\,\bar F) \D{2,2} - \bar F \D{2,4}, \\
    &\qquad+ \bar E \D{3,4} + c_s^2 \D{3,6} - \bar A^3 \bar F^2, \nonumber \\
    \D{4,10} &= \bar \lambda_3 \bar A\,\bar E \D{1,0} + \bar A (\bar E\,\bar F + \bar \lambda_3 c_s^2) \D{1,2} - \bar\lambda_3 \bar C \D{2,2} - (\bar\lambda_3 + \bar C\, \bar F) \D{2,4} \\
    &\qquad+ \bar E \D{3,6} + c_s^2 \D{3,8} - 2 \bar\lambda_3 \bar A^3 \bar F, \nonumber \\
    \D{4,12} &= \bar \lambda_3 \bar A\, \bar E \D{1,2} - \bar \lambda_3 \bar C \D{2,4} + \bar E \D{3,8} - \bar \lambda_3^2 \bar A^3.
\end{align}
\end{subequations}

\bibliography{references}

\def\cprime{$'$}
\begin{thebibliography}{115}%
\makeatletter
\providecommand \@ifxundefined [1]{%
 \@ifx{#1\undefined}
}%
\providecommand \@ifnum [1]{%
 \ifnum #1\expandafter \@firstoftwo
 \else \expandafter \@secondoftwo
 \fi
}%
\providecommand \@ifx [1]{%
 \ifx #1\expandafter \@firstoftwo
 \else \expandafter \@secondoftwo
 \fi
}%
\providecommand \natexlab [1]{#1}%
\providecommand \enquote  [1]{``#1''}%
\providecommand \bibnamefont  [1]{#1}%
\providecommand \bibfnamefont [1]{#1}%
\providecommand \citenamefont [1]{#1}%
\providecommand \href@noop [0]{\@secondoftwo}%
\providecommand \href [0]{\begingroup \@sanitize@url \@href}%
\providecommand \@href[1]{\@@startlink{#1}\@@href}%
\providecommand \@@href[1]{\endgroup#1\@@endlink}%
\providecommand \@sanitize@url [0]{\catcode `\\12\catcode `\$12\catcode
  `\&12\catcode `\#12\catcode `\^12\catcode `\_12\catcode `\%12\relax}%
\providecommand \@@startlink[1]{}%
\providecommand \@@endlink[0]{}%
\providecommand \url  [0]{\begingroup\@sanitize@url \@url }%
\providecommand \@url [1]{\endgroup\@href {#1}{\urlprefix }}%
\providecommand \urlprefix  [0]{URL }%
\providecommand \Eprint [0]{\href }%
\providecommand \doibase [0]{https://doi.org/}%
\providecommand \selectlanguage [0]{\@gobble}%
\providecommand \bibinfo  [0]{\@secondoftwo}%
\providecommand \bibfield  [0]{\@secondoftwo}%
\providecommand \translation [1]{[#1]}%
\providecommand \BibitemOpen [0]{}%
\providecommand \bibitemStop [0]{}%
\providecommand \bibitemNoStop [0]{.\EOS\space}%
\providecommand \EOS [0]{\spacefactor3000\relax}%
\providecommand \BibitemShut  [1]{\csname bibitem#1\endcsname}%
\let\auto@bib@innerbib\@empty
\bibitem [{\citenamefont {Weinberg}(2008)}]{WeinbergCosmology}%
  \BibitemOpen
  \bibfield  {author} {\bibinfo {author} {\bibfnamefont {S.}~\bibnamefont
  {Weinberg}},\ }\href@noop {} {\emph {\bibinfo {title} {Cosmology}}}\
  (\bibinfo  {publisher} {Oxford University Press},\ \bibinfo {year} {2008})\
  p.\ \bibinfo {pages} {593}\BibitemShut {NoStop}%
\bibitem [{\citenamefont {Rezzolla}\ and\ \citenamefont
  {Zanotti}(2013)}]{Rezzolla_Zanotti_book}%
  \BibitemOpen
  \bibfield  {author} {\bibinfo {author} {\bibfnamefont {L.}~\bibnamefont
  {Rezzolla}}\ and\ \bibinfo {author} {\bibfnamefont {O.}~\bibnamefont
  {Zanotti}},\ }\href@noop {} {\emph {\bibinfo {title} {Relativistic
  hydrodynamics}}}\ (\bibinfo  {publisher} {Oxford University Press},\ \bibinfo
  {address} {New York},\ \bibinfo {year} {2013})\BibitemShut {NoStop}%
\bibitem [{\citenamefont {Romatschke}\ and\ \citenamefont
  {Romatschke}(2019)}]{Romatschke:2017ejr}%
  \BibitemOpen
  \bibfield  {author} {\bibinfo {author} {\bibfnamefont {P.}~\bibnamefont
  {Romatschke}}\ and\ \bibinfo {author} {\bibfnamefont {U.}~\bibnamefont
  {Romatschke}},\ }\href@noop {} {\emph {\bibinfo {title} {Relativistic Fluid
  Dynamics In and Out of Equilibrium}}},\ Cambridge Monographs on Mathematical
  Physics\ (\bibinfo  {publisher} {Cambridge University Press},\ \bibinfo
  {year} {2019})\ \Eprint {https://arxiv.org/abs/1712.05815} {arXiv:1712.05815
  [nucl-th]} \BibitemShut {NoStop}%
\bibitem [{\citenamefont {Kharzeev}\ \emph {et~al.}(2016)\citenamefont
  {Kharzeev}, \citenamefont {Liao}, \citenamefont {Voloshin},\ and\
  \citenamefont {Wang}}]{Kharzeev:2015znc}%
  \BibitemOpen
  \bibfield  {author} {\bibinfo {author} {\bibfnamefont {D.~E.}\ \bibnamefont
  {Kharzeev}}, \bibinfo {author} {\bibfnamefont {J.}~\bibnamefont {Liao}},
  \bibinfo {author} {\bibfnamefont {S.~A.}\ \bibnamefont {Voloshin}},\ and\
  \bibinfo {author} {\bibfnamefont {G.}~\bibnamefont {Wang}},\ }\bibfield
  {title} {\bibinfo {title} {{Chiral magnetic and vortical effects in
  high-energy nuclear collisions\textemdash{}A status report}},\ }\href
  {https://doi.org/10.1016/j.ppnp.2016.01.001} {\bibfield  {journal} {\bibinfo
  {journal} {Prog. Part. Nucl. Phys.}\ }\textbf {\bibinfo {volume} {88}},\
  \bibinfo {pages} {1} (\bibinfo {year} {2016})},\ \Eprint
  {https://arxiv.org/abs/1511.04050} {arXiv:1511.04050 [hep-ph]} \BibitemShut
  {NoStop}%
\bibitem [{\citenamefont {Huang}(2016)}]{Huang:2015oca}%
  \BibitemOpen
  \bibfield  {author} {\bibinfo {author} {\bibfnamefont {X.-G.}\ \bibnamefont
  {Huang}},\ }\bibfield  {title} {\bibinfo {title} {{Electromagnetic fields and
  anomalous transports in heavy-ion collisions --- A pedagogical review}},\
  }\href {https://doi.org/10.1088/0034-4885/79/7/076302} {\bibfield  {journal}
  {\bibinfo  {journal} {Rept. Prog. Phys.}\ }\textbf {\bibinfo {volume} {79}},\
  \bibinfo {pages} {076302} (\bibinfo {year} {2016})},\ \Eprint
  {https://arxiv.org/abs/1509.04073} {arXiv:1509.04073 [nucl-th]} \BibitemShut
  {NoStop}%
\bibitem [{\citenamefont {Hosur}\ and\ \citenamefont
  {Qi}(2013)}]{Hosur:2013kxa}%
  \BibitemOpen
  \bibfield  {author} {\bibinfo {author} {\bibfnamefont {P.}~\bibnamefont
  {Hosur}}\ and\ \bibinfo {author} {\bibfnamefont {X.}~\bibnamefont {Qi}},\
  }\bibfield  {title} {\bibinfo {title} {{Recent developments in transport
  phenomena in Weyl semimetals}},\ }\href
  {https://doi.org/10.1016/j.crhy.2013.10.010} {\bibfield  {journal} {\bibinfo
  {journal} {Comptes Rendus Physique}\ }\textbf {\bibinfo {volume} {14}},\
  \bibinfo {pages} {857} (\bibinfo {year} {2013})},\ \Eprint
  {https://arxiv.org/abs/1309.4464} {arXiv:1309.4464 [cond-mat.str-el]}
  \BibitemShut {NoStop}%
\bibitem [{\citenamefont {Landau}\ and\ \citenamefont
  {Lifshitz}(1987)}]{LandauLifshitzFluids}%
  \BibitemOpen
  \bibfield  {author} {\bibinfo {author} {\bibfnamefont {L.~D.}\ \bibnamefont
  {Landau}}\ and\ \bibinfo {author} {\bibfnamefont {E.~M.}\ \bibnamefont
  {Lifshitz}},\ }\href@noop {} {\emph {\bibinfo {title} {Fluid Mechanics -
  Volume 6 (Course of Theoretical Physics)}}},\ \bibinfo {edition} {2nd}\ ed.\
  (\bibinfo  {publisher} {Butterworth-Heinemann},\ \bibinfo {year} {1987})\ p.\
  \bibinfo {pages} {552}\BibitemShut {NoStop}%
\bibitem [{\citenamefont {Heinz}\ and\ \citenamefont
  {Snellings}(2013)}]{Heinz:2013th}%
  \BibitemOpen
  \bibfield  {author} {\bibinfo {author} {\bibfnamefont {U.}~\bibnamefont
  {Heinz}}\ and\ \bibinfo {author} {\bibfnamefont {R.}~\bibnamefont
  {Snellings}},\ }\bibfield  {title} {\bibinfo {title} {Collective flow and
  viscosity in relativistic heavy-ion collisions},\ }\href
  {https://doi.org/10.1146/annurev-nucl-102212-170540} {\bibfield  {journal}
  {\bibinfo  {journal} {Ann. Rev. Nucl. Part. Sci.}\ }\textbf {\bibinfo
  {volume} {63}},\ \bibinfo {pages} {123} (\bibinfo {year} {2013})},\ \Eprint
  {https://arxiv.org/abs/1301.2826} {arXiv:1301.2826 [nucl-th]} \BibitemShut
  {NoStop}%
\bibitem [{\citenamefont {Baiotti}\ and\ \citenamefont
  {Rezzolla}(2017)}]{Baiotti:2016qnr}%
  \BibitemOpen
  \bibfield  {author} {\bibinfo {author} {\bibfnamefont {L.}~\bibnamefont
  {Baiotti}}\ and\ \bibinfo {author} {\bibfnamefont {L.}~\bibnamefont
  {Rezzolla}},\ }\bibfield  {title} {\bibinfo {title} {Binary neutron star
  mergers: a review of einstein's richest laboratory},\ }\href
  {https://doi.org/10.1088/1361-6633/aa67bb} {\bibfield  {journal} {\bibinfo
  {journal} {Rept. Prog. Phys.}\ }\textbf {\bibinfo {volume} {80}},\ \bibinfo
  {pages} {096901} (\bibinfo {year} {2017})},\ \Eprint
  {https://arxiv.org/abs/1607.03540} {arXiv:1607.03540 [gr-qc]} \BibitemShut
  {NoStop}%
\bibitem [{\citenamefont {Alford}\ \emph {et~al.}(2018)\citenamefont {Alford},
  \citenamefont {Bovard}, \citenamefont {Hanauske}, \citenamefont {Rezzolla},\
  and\ \citenamefont {Schwenzer}}]{Alford:2017rxf}%
  \BibitemOpen
  \bibfield  {author} {\bibinfo {author} {\bibfnamefont {M.~G.}\ \bibnamefont
  {Alford}}, \bibinfo {author} {\bibfnamefont {L.}~\bibnamefont {Bovard}},
  \bibinfo {author} {\bibfnamefont {M.}~\bibnamefont {Hanauske}}, \bibinfo
  {author} {\bibfnamefont {L.}~\bibnamefont {Rezzolla}},\ and\ \bibinfo
  {author} {\bibfnamefont {K.}~\bibnamefont {Schwenzer}},\ }\bibfield  {title}
  {\bibinfo {title} {Viscous dissipation and heat conduction in binary
  neutron-star mergers},\ }\href
  {https://doi.org/10.1103/PhysRevLett.120.041101} {\bibfield  {journal}
  {\bibinfo  {journal} {Phys. Rev. Lett.}\ }\textbf {\bibinfo {volume} {120}},\
  \bibinfo {pages} {041101} (\bibinfo {year} {2018})},\ \Eprint
  {https://arxiv.org/abs/1707.09475} {arXiv:1707.09475 [gr-qc]} \BibitemShut
  {NoStop}%
\bibitem [{\citenamefont {Most}\ \emph {et~al.}(2021)\citenamefont {Most},
  \citenamefont {Harris}, \citenamefont {Plumberg}, \citenamefont {Alford},
  \citenamefont {Noronha}, \citenamefont {Noronha-Hostler}, \citenamefont
  {Pretorius}, \citenamefont {Witek},\ and\ \citenamefont
  {Yunes}}]{Most:2021zvc}%
  \BibitemOpen
  \bibfield  {author} {\bibinfo {author} {\bibfnamefont {E.~R.}\ \bibnamefont
  {Most}}, \bibinfo {author} {\bibfnamefont {S.~P.}\ \bibnamefont {Harris}},
  \bibinfo {author} {\bibfnamefont {C.}~\bibnamefont {Plumberg}}, \bibinfo
  {author} {\bibfnamefont {M.~G.}\ \bibnamefont {Alford}}, \bibinfo {author}
  {\bibfnamefont {J.}~\bibnamefont {Noronha}}, \bibinfo {author} {\bibfnamefont
  {J.}~\bibnamefont {Noronha-Hostler}}, \bibinfo {author} {\bibfnamefont
  {F.}~\bibnamefont {Pretorius}}, \bibinfo {author} {\bibfnamefont
  {H.}~\bibnamefont {Witek}},\ and\ \bibinfo {author} {\bibfnamefont
  {N.}~\bibnamefont {Yunes}},\ }\bibfield  {title} {\bibinfo {title}
  {{Projecting the likely importance of weak-interaction-driven bulk viscosity
  in neutron star mergers}},\ }\href {https://doi.org/10.1093/mnras/stab2793}
  {\bibfield  {journal} {\bibinfo  {journal} {Mon. Not. Roy. Astron. Soc.}\
  }\textbf {\bibinfo {volume} {509}},\ \bibinfo {pages} {1096} (\bibinfo {year}
  {2021})},\ \Eprint {https://arxiv.org/abs/2107.05094} {arXiv:2107.05094
  [astro-ph.HE]} \BibitemShut {NoStop}%
\bibitem [{\citenamefont {Most}\ \emph {et~al.}(2022)\citenamefont {Most},
  \citenamefont {Haber}, \citenamefont {Harris}, \citenamefont {Zhang},
  \citenamefont {Alford},\ and\ \citenamefont {Noronha}}]{Most:2022yhe}%
  \BibitemOpen
  \bibfield  {author} {\bibinfo {author} {\bibfnamefont {E.~R.}\ \bibnamefont
  {Most}}, \bibinfo {author} {\bibfnamefont {A.}~\bibnamefont {Haber}},
  \bibinfo {author} {\bibfnamefont {S.~P.}\ \bibnamefont {Harris}}, \bibinfo
  {author} {\bibfnamefont {Z.}~\bibnamefont {Zhang}}, \bibinfo {author}
  {\bibfnamefont {M.~G.}\ \bibnamefont {Alford}},\ and\ \bibinfo {author}
  {\bibfnamefont {J.}~\bibnamefont {Noronha}},\ }\bibfield  {title} {\bibinfo
  {title} {{Emergence of microphysical viscosity in binary neutron star
  post-merger dynamics}},\ }\href@noop {} {\  (\bibinfo {year} {2022})},\
  \Eprint {https://arxiv.org/abs/2207.00442} {arXiv:2207.00442 [astro-ph.HE]}
  \BibitemShut {NoStop}%
\bibitem [{\citenamefont {Choquet-Bruhat}(2009)}]{ChoquetBruhatGRBook}%
  \BibitemOpen
  \bibfield  {author} {\bibinfo {author} {\bibfnamefont {Y.}~\bibnamefont
  {Choquet-Bruhat}},\ }\href@noop {} {\emph {\bibinfo {title} {General
  Relativity and the Einstein Equations}}}\ (\bibinfo  {publisher} {Oxford
  University Press},\ \bibinfo {address} {New York},\ \bibinfo {year}
  {2009})\BibitemShut {NoStop}%
\bibitem [{\citenamefont {Eckart}(1940)}]{EckartViscous}%
  \BibitemOpen
  \bibfield  {author} {\bibinfo {author} {\bibfnamefont {C.}~\bibnamefont
  {Eckart}},\ }\bibfield  {title} {\bibinfo {title} {{The Thermodynamics of
  irreversible processes. 3.. Relativistic theory of the simple fluid}},\
  }\href {https://doi.org/10.1103/PhysRev.58.919} {\bibfield  {journal}
  {\bibinfo  {journal} {Phys. Rev.}\ }\textbf {\bibinfo {volume} {58}},\
  \bibinfo {pages} {919} (\bibinfo {year} {1940})}\BibitemShut {NoStop}%
\bibitem [{\citenamefont {M{\"u}ller}(1967)}]{Muller:1967zza}%
  \BibitemOpen
  \bibfield  {author} {\bibinfo {author} {\bibfnamefont {I.}~\bibnamefont
  {M{\"u}ller}},\ }\bibfield  {title} {\bibinfo {title} {{Zum Paradoxon der
  Warmeleitungstheorie}},\ }\href {https://doi.org/10.1007/BF01326412}
  {\bibfield  {journal} {\bibinfo  {journal} {Z. Phys.}\ }\textbf {\bibinfo
  {volume} {198}},\ \bibinfo {pages} {329} (\bibinfo {year}
  {1967})}\BibitemShut {NoStop}%
\bibitem [{\citenamefont {Israel}\ and\ \citenamefont {Stewart}(1979)}]{MIS-6}%
  \BibitemOpen
  \bibfield  {author} {\bibinfo {author} {\bibfnamefont {W.}~\bibnamefont
  {Israel}}\ and\ \bibinfo {author} {\bibfnamefont {J.~M.}\ \bibnamefont
  {Stewart}},\ }\bibfield  {title} {\bibinfo {title} {{Transient relativistic
  thermodynamics and kinetic theory}},\ }\href
  {https://doi.org/10.1016/0003-4916(79)90130-1} {\bibfield  {journal}
  {\bibinfo  {journal} {Annals Phys.}\ }\textbf {\bibinfo {volume} {118}},\
  \bibinfo {pages} {341} (\bibinfo {year} {1979})}\BibitemShut {NoStop}%
\bibitem [{\citenamefont {Israel}(1976{\natexlab{a}})}]{MIS-2}%
  \BibitemOpen
  \bibfield  {author} {\bibinfo {author} {\bibfnamefont {W.}~\bibnamefont
  {Israel}},\ }\bibfield  {title} {\bibinfo {title} {{Nonstationary
  irreversible thermodynamics: A Causal relativistic theory}},\ }\href
  {https://doi.org/10.1016/0003-4916(76)90064-6} {\bibfield  {journal}
  {\bibinfo  {journal} {Annals Phys.}\ }\textbf {\bibinfo {volume} {100}},\
  \bibinfo {pages} {310} (\bibinfo {year} {1976}{\natexlab{a}})}\BibitemShut
  {NoStop}%
\bibitem [{\citenamefont {Hiscock}\ and\ \citenamefont
  {Lindblom}(1983)}]{Hiscock_Lindblom_stability_1983}%
  \BibitemOpen
  \bibfield  {author} {\bibinfo {author} {\bibfnamefont {W.~A.}\ \bibnamefont
  {Hiscock}}\ and\ \bibinfo {author} {\bibfnamefont {L.}~\bibnamefont
  {Lindblom}},\ }\bibfield  {title} {\bibinfo {title} {{Stability and causality
  in dissipative relativistic fluids}},\ }\href
  {https://doi.org/10.1016/0003-4916(83)90288-9} {\bibfield  {journal}
  {\bibinfo  {journal} {Annals Phys.}\ }\textbf {\bibinfo {volume} {151}},\
  \bibinfo {pages} {466} (\bibinfo {year} {1983})}\BibitemShut {NoStop}%
\bibitem [{\citenamefont {Olson}(1990)}]{Olson:1989ey}%
  \BibitemOpen
  \bibfield  {author} {\bibinfo {author} {\bibfnamefont {T.~S.}\ \bibnamefont
  {Olson}},\ }\bibfield  {title} {\bibinfo {title} {Stability and causality in
  the {I}srael-{S}tewart energy frame theory},\ }\href
  {https://doi.org/10.1016/0003-4916(90)90366-V} {\bibfield  {journal}
  {\bibinfo  {journal} {Annals Phys.}\ }\textbf {\bibinfo {volume} {199}},\
  \bibinfo {pages} {18} (\bibinfo {year} {1990})}\BibitemShut {NoStop}%
\bibitem [{\citenamefont {Floerchinger}\ and\ \citenamefont
  {Grossi}(2018)}]{Floerchinger:2017cii}%
  \BibitemOpen
  \bibfield  {author} {\bibinfo {author} {\bibfnamefont {S.}~\bibnamefont
  {Floerchinger}}\ and\ \bibinfo {author} {\bibfnamefont {E.}~\bibnamefont
  {Grossi}},\ }\bibfield  {title} {\bibinfo {title} {{Causality of fluid
  dynamics for high-energy nuclear collisions}},\ }\href
  {https://doi.org/10.1007/JHEP08(2018)186} {\bibfield  {journal} {\bibinfo
  {journal} {JHEP}\ }\textbf {\bibinfo {volume} {08}},\ \bibinfo {pages}
  {186}},\ \Eprint {https://arxiv.org/abs/1711.06687} {arXiv:1711.06687
  [nucl-th]} \BibitemShut {NoStop}%
\bibitem [{\citenamefont {Bemfica}\ \emph
  {et~al.}(2019{\natexlab{a}})\citenamefont {Bemfica}, \citenamefont
  {Disconzi},\ and\ \citenamefont {Noronha}}]{Bemfica:2019cop}%
  \BibitemOpen
  \bibfield  {author} {\bibinfo {author} {\bibfnamefont {F.~S.}\ \bibnamefont
  {Bemfica}}, \bibinfo {author} {\bibfnamefont {M.~M.}\ \bibnamefont
  {Disconzi}},\ and\ \bibinfo {author} {\bibfnamefont {J.}~\bibnamefont
  {Noronha}},\ }\bibfield  {title} {\bibinfo {title} {{Causality of the
  Einstein-Israel-Stewart Theory with Bulk Viscosity}},\ }\href
  {https://doi.org/10.1103/PhysRevLett.122.221602} {\bibfield  {journal}
  {\bibinfo  {journal} {Phys. Rev. Lett.}\ }\textbf {\bibinfo {volume} {122}},\
  \bibinfo {pages} {221602} (\bibinfo {year} {2019}{\natexlab{a}})},\ \Eprint
  {https://arxiv.org/abs/1901.06701} {arXiv:1901.06701 [gr-qc]} \BibitemShut
  {NoStop}%
\bibitem [{\citenamefont {Bemfica}\ \emph {et~al.}(2021)\citenamefont
  {Bemfica}, \citenamefont {Disconzi}, \citenamefont {Hoang}, \citenamefont
  {Noronha},\ and\ \citenamefont {Radosz}}]{Bemfica:2020xym}%
  \BibitemOpen
  \bibfield  {author} {\bibinfo {author} {\bibfnamefont {F.~S.}\ \bibnamefont
  {Bemfica}}, \bibinfo {author} {\bibfnamefont {M.~M.}\ \bibnamefont
  {Disconzi}}, \bibinfo {author} {\bibfnamefont {V.}~\bibnamefont {Hoang}},
  \bibinfo {author} {\bibfnamefont {J.}~\bibnamefont {Noronha}},\ and\ \bibinfo
  {author} {\bibfnamefont {M.}~\bibnamefont {Radosz}},\ }\bibfield  {title}
  {\bibinfo {title} {{Nonlinear Constraints on Relativistic Fluids Far from
  Equilibrium}},\ }\href {https://doi.org/10.1103/PhysRevLett.126.222301}
  {\bibfield  {journal} {\bibinfo  {journal} {Phys. Rev. Lett.}\ }\textbf
  {\bibinfo {volume} {126}},\ \bibinfo {pages} {222301} (\bibinfo {year}
  {2021})},\ \Eprint {https://arxiv.org/abs/2005.11632} {arXiv:2005.11632
  [hep-th]} \BibitemShut {NoStop}%
\bibitem [{\citenamefont {Bemfica}\ \emph {et~al.}(2018)\citenamefont
  {Bemfica}, \citenamefont {Disconzi},\ and\ \citenamefont
  {Noronha}}]{Bemfica:2017wps}%
  \BibitemOpen
  \bibfield  {author} {\bibinfo {author} {\bibfnamefont {F.~S.}\ \bibnamefont
  {Bemfica}}, \bibinfo {author} {\bibfnamefont {M.~M.}\ \bibnamefont
  {Disconzi}},\ and\ \bibinfo {author} {\bibfnamefont {J.}~\bibnamefont
  {Noronha}},\ }\bibfield  {title} {\bibinfo {title} {{Causality and existence
  of solutions of relativistic viscous fluid dynamics with gravity}},\ }\href
  {https://doi.org/10.1103/PhysRevD.98.104064} {\bibfield  {journal} {\bibinfo
  {journal} {Phys. Rev. D}\ }\textbf {\bibinfo {volume} {98}},\ \bibinfo
  {pages} {104064} (\bibinfo {year} {2018})},\ \Eprint
  {https://arxiv.org/abs/1708.06255} {arXiv:1708.06255 [gr-qc]} \BibitemShut
  {NoStop}%
\bibitem [{\citenamefont {Kovtun}(2019)}]{Kovtun:2019hdm}%
  \BibitemOpen
  \bibfield  {author} {\bibinfo {author} {\bibfnamefont {P.}~\bibnamefont
  {Kovtun}},\ }\bibfield  {title} {\bibinfo {title} {{First-order relativistic
  hydrodynamics is stable}},\ }\href {https://doi.org/10.1007/JHEP10(2019)034}
  {\bibfield  {journal} {\bibinfo  {journal} {JHEP}\ }\textbf {\bibinfo
  {volume} {10}},\ \bibinfo {pages} {034}},\ \Eprint
  {https://arxiv.org/abs/1907.08191} {arXiv:1907.08191 [hep-th]} \BibitemShut
  {NoStop}%
\bibitem [{\citenamefont {Bemfica}\ \emph
  {et~al.}(2019{\natexlab{b}})\citenamefont {Bemfica}, \citenamefont
  {Disconzi},\ and\ \citenamefont {Noronha}}]{Bemfica:2019knx}%
  \BibitemOpen
  \bibfield  {author} {\bibinfo {author} {\bibfnamefont {F.~S.}\ \bibnamefont
  {Bemfica}}, \bibinfo {author} {\bibfnamefont {M.~M.}\ \bibnamefont
  {Disconzi}},\ and\ \bibinfo {author} {\bibfnamefont {J.}~\bibnamefont
  {Noronha}},\ }\bibfield  {title} {\bibinfo {title} {{Nonlinear Causality of
  General First-Order Relativistic Viscous Hydrodynamics}},\ }\href
  {https://doi.org/10.1103/PhysRevD.100.104020} {\bibfield  {journal} {\bibinfo
   {journal} {Phys. Rev. D}\ }\textbf {\bibinfo {volume} {100}},\ \bibinfo
  {pages} {104020} (\bibinfo {year} {2019}{\natexlab{b}})},\ \Eprint
  {https://arxiv.org/abs/1907.12695} {arXiv:1907.12695 [gr-qc]} \BibitemShut
  {NoStop}%
\bibitem [{\citenamefont {Hoult}\ and\ \citenamefont
  {Kovtun}(2020)}]{Hoult:2020eho}%
  \BibitemOpen
  \bibfield  {author} {\bibinfo {author} {\bibfnamefont {R.~E.}\ \bibnamefont
  {Hoult}}\ and\ \bibinfo {author} {\bibfnamefont {P.}~\bibnamefont {Kovtun}},\
  }\bibfield  {title} {\bibinfo {title} {{Stable and causal relativistic
  Navier-Stokes equations}},\ }\href {https://doi.org/10.1007/JHEP06(2020)067}
  {\bibfield  {journal} {\bibinfo  {journal} {JHEP}\ }\textbf {\bibinfo
  {volume} {06}},\ \bibinfo {pages} {067}},\ \Eprint
  {https://arxiv.org/abs/2004.04102} {arXiv:2004.04102 [hep-th]} \BibitemShut
  {NoStop}%
\bibitem [{\citenamefont {Bemfica}\ \emph {et~al.}(2022)\citenamefont
  {Bemfica}, \citenamefont {Disconzi},\ and\ \citenamefont
  {Noronha}}]{Bemfica:2020zjp}%
  \BibitemOpen
  \bibfield  {author} {\bibinfo {author} {\bibfnamefont {F.~S.}\ \bibnamefont
  {Bemfica}}, \bibinfo {author} {\bibfnamefont {M.~M.}\ \bibnamefont
  {Disconzi}},\ and\ \bibinfo {author} {\bibfnamefont {J.}~\bibnamefont
  {Noronha}},\ }\bibfield  {title} {\bibinfo {title} {{First-Order
  General-Relativistic Viscous Fluid Dynamics}},\ }\href
  {https://doi.org/10.1103/PhysRevX.12.021044} {\bibfield  {journal} {\bibinfo
  {journal} {Phys. Rev. X}\ }\textbf {\bibinfo {volume} {12}},\ \bibinfo
  {pages} {021044} (\bibinfo {year} {2022})},\ \Eprint
  {https://arxiv.org/abs/2009.11388} {arXiv:2009.11388 [gr-qc]} \BibitemShut
  {NoStop}%
\bibitem [{\citenamefont {Rocha}\ \emph {et~al.}(2022)\citenamefont {Rocha},
  \citenamefont {Denicol},\ and\ \citenamefont {Noronha}}]{Rocha:2022ind}%
  \BibitemOpen
  \bibfield  {author} {\bibinfo {author} {\bibfnamefont {G.~S.}\ \bibnamefont
  {Rocha}}, \bibinfo {author} {\bibfnamefont {G.~S.}\ \bibnamefont {Denicol}},\
  and\ \bibinfo {author} {\bibfnamefont {J.}~\bibnamefont {Noronha}},\
  }\bibfield  {title} {\bibinfo {title} {{Perturbative approaches in
  relativistic kinetic theory and the emergence of first-order
  hydrodynamics}},\ }\href {https://doi.org/10.1103/PhysRevD.106.036010}
  {\bibfield  {journal} {\bibinfo  {journal} {Phys. Rev. D}\ }\textbf {\bibinfo
  {volume} {106}},\ \bibinfo {pages} {036010} (\bibinfo {year} {2022})},\
  \Eprint {https://arxiv.org/abs/2205.00078} {arXiv:2205.00078 [nucl-th]}
  \BibitemShut {NoStop}%
\bibitem [{\citenamefont {Hoult}\ and\ \citenamefont
  {Kovtun}(2022)}]{Hoult:2021gnb}%
  \BibitemOpen
  \bibfield  {author} {\bibinfo {author} {\bibfnamefont {R.~E.}\ \bibnamefont
  {Hoult}}\ and\ \bibinfo {author} {\bibfnamefont {P.}~\bibnamefont {Kovtun}},\
  }\bibfield  {title} {\bibinfo {title} {{Causal first-order hydrodynamics from
  kinetic theory and holography}},\ }\href
  {https://doi.org/10.1103/PhysRevD.106.066023} {\bibfield  {journal} {\bibinfo
   {journal} {Phys. Rev. D}\ }\textbf {\bibinfo {volume} {106}},\ \bibinfo
  {pages} {066023} (\bibinfo {year} {2022})},\ \Eprint
  {https://arxiv.org/abs/2112.14042} {arXiv:2112.14042 [hep-th]} \BibitemShut
  {NoStop}%
\bibitem [{\citenamefont {Biswas}\ \emph {et~al.}(2022)\citenamefont {Biswas},
  \citenamefont {Mitra},\ and\ \citenamefont {Roy}}]{Biswas:2022cla}%
  \BibitemOpen
  \bibfield  {author} {\bibinfo {author} {\bibfnamefont {R.}~\bibnamefont
  {Biswas}}, \bibinfo {author} {\bibfnamefont {S.}~\bibnamefont {Mitra}},\ and\
  \bibinfo {author} {\bibfnamefont {V.}~\bibnamefont {Roy}},\ }\bibfield
  {title} {\bibinfo {title} {{Is first-order relativistic hydrodynamics in a
  general frame stable and causal for arbitrary interactions?}},\ }\href
  {https://doi.org/10.1103/PhysRevD.106.L011501} {\bibfield  {journal}
  {\bibinfo  {journal} {Phys. Rev. D}\ }\textbf {\bibinfo {volume} {106}},\
  \bibinfo {pages} {L011501} (\bibinfo {year} {2022})},\ \Eprint
  {https://arxiv.org/abs/2202.08685} {arXiv:2202.08685 [nucl-th]} \BibitemShut
  {NoStop}%
\bibitem [{\citenamefont {Biswas}\ \emph {et~al.}(2023)\citenamefont {Biswas},
  \citenamefont {Mitra},\ and\ \citenamefont {Roy}}]{Biswas:2022hiv}%
  \BibitemOpen
  \bibfield  {author} {\bibinfo {author} {\bibfnamefont {R.}~\bibnamefont
  {Biswas}}, \bibinfo {author} {\bibfnamefont {S.}~\bibnamefont {Mitra}},\ and\
  \bibinfo {author} {\bibfnamefont {V.}~\bibnamefont {Roy}},\ }\bibfield
  {title} {\bibinfo {title} {{An expedition to the islands of stability in the
  first-order causal hydrodynamics}},\ }\href
  {https://doi.org/10.1016/j.physletb.2023.137725} {\bibfield  {journal}
  {\bibinfo  {journal} {Phys. Lett. B}\ }\textbf {\bibinfo {volume} {838}},\
  \bibinfo {pages} {137725} (\bibinfo {year} {2023})},\ \Eprint
  {https://arxiv.org/abs/2211.11358} {arXiv:2211.11358 [nucl-th]} \BibitemShut
  {NoStop}%
\bibitem [{\citenamefont {Noronha}\ \emph {et~al.}(2022)\citenamefont
  {Noronha}, \citenamefont {Spali\'nski},\ and\ \citenamefont
  {Speranza}}]{Noronha:2021syv}%
  \BibitemOpen
  \bibfield  {author} {\bibinfo {author} {\bibfnamefont {J.}~\bibnamefont
  {Noronha}}, \bibinfo {author} {\bibfnamefont {M.}~\bibnamefont
  {Spali\'nski}},\ and\ \bibinfo {author} {\bibfnamefont {E.}~\bibnamefont
  {Speranza}},\ }\bibfield  {title} {\bibinfo {title} {{Transient Relativistic
  Fluid Dynamics in a General Hydrodynamic Frame}},\ }\href
  {https://doi.org/10.1103/PhysRevLett.128.252302} {\bibfield  {journal}
  {\bibinfo  {journal} {Phys. Rev. Lett.}\ }\textbf {\bibinfo {volume} {128}},\
  \bibinfo {pages} {252302} (\bibinfo {year} {2022})},\ \Eprint
  {https://arxiv.org/abs/2105.01034} {arXiv:2105.01034 [nucl-th]} \BibitemShut
  {NoStop}%
\bibitem [{\citenamefont {Rodino}(1993)}]{RodinoGevreyBook}%
  \BibitemOpen
  \bibfield  {author} {\bibinfo {author} {\bibfnamefont {L.}~\bibnamefont
  {Rodino}},\ }\href@noop {} {\emph {\bibinfo {title} {Linear partial
  differential operators in Gevrey spaces}}}\ (\bibinfo  {publisher} {World
  Scientific},\ \bibinfo {address} {Singapore},\ \bibinfo {year}
  {1993})\BibitemShut {NoStop}%
\bibitem [{\citenamefont {Kharzeev}\ \emph {et~al.}(2008)\citenamefont
  {Kharzeev}, \citenamefont {McLerran},\ and\ \citenamefont
  {Warringa}}]{Kharzeev:2007jp}%
  \BibitemOpen
  \bibfield  {author} {\bibinfo {author} {\bibfnamefont {D.~E.}\ \bibnamefont
  {Kharzeev}}, \bibinfo {author} {\bibfnamefont {L.~D.}\ \bibnamefont
  {McLerran}},\ and\ \bibinfo {author} {\bibfnamefont {H.~J.}\ \bibnamefont
  {Warringa}},\ }\bibfield  {title} {\bibinfo {title} {{The Effects of
  topological charge change in heavy ion collisions: 'Event by event P and CP
  violation'}},\ }\href {https://doi.org/10.1016/j.nuclphysa.2008.02.298}
  {\bibfield  {journal} {\bibinfo  {journal} {Nucl. Phys. A}\ }\textbf
  {\bibinfo {volume} {803}},\ \bibinfo {pages} {227} (\bibinfo {year}
  {2008})},\ \Eprint {https://arxiv.org/abs/0711.0950} {arXiv:0711.0950
  [hep-ph]} \BibitemShut {NoStop}%
\bibitem [{\citenamefont {Fukushima}\ \emph {et~al.}(2008)\citenamefont
  {Fukushima}, \citenamefont {Kharzeev},\ and\ \citenamefont
  {Warringa}}]{Fukushima:2008xe}%
  \BibitemOpen
  \bibfield  {author} {\bibinfo {author} {\bibfnamefont {K.}~\bibnamefont
  {Fukushima}}, \bibinfo {author} {\bibfnamefont {D.~E.}\ \bibnamefont
  {Kharzeev}},\ and\ \bibinfo {author} {\bibfnamefont {H.~J.}\ \bibnamefont
  {Warringa}},\ }\bibfield  {title} {\bibinfo {title} {{The Chiral Magnetic
  Effect}},\ }\href {https://doi.org/10.1103/PhysRevD.78.074033} {\bibfield
  {journal} {\bibinfo  {journal} {Phys. Rev. D}\ }\textbf {\bibinfo {volume}
  {78}},\ \bibinfo {pages} {074033} (\bibinfo {year} {2008})},\ \Eprint
  {https://arxiv.org/abs/0808.3382} {arXiv:0808.3382 [hep-ph]} \BibitemShut
  {NoStop}%
\bibitem [{\citenamefont {Vilenkin}(1979)}]{vilenkin1979macroscopic}%
  \BibitemOpen
  \bibfield  {author} {\bibinfo {author} {\bibfnamefont {A.}~\bibnamefont
  {Vilenkin}},\ }\bibfield  {title} {\bibinfo {title} {Macroscopic
  parity-violating effects: Neutrino fluxes from rotating black holes and in
  rotating thermal radiation},\ }\href
  {https://doi.org/10.1103/PhysRevD.20.1807} {\bibfield  {journal} {\bibinfo
  {journal} {Phys. Rev. D}\ }\textbf {\bibinfo {volume} {20}},\ \bibinfo
  {pages} {1807} (\bibinfo {year} {1979})}\BibitemShut {NoStop}%
\bibitem [{\citenamefont {Vilenkin}(1980)}]{vilenkin1980quantum}%
  \BibitemOpen
  \bibfield  {author} {\bibinfo {author} {\bibfnamefont {A.}~\bibnamefont
  {Vilenkin}},\ }\bibfield  {title} {\bibinfo {title} {Quantum field theory at
  finite temperature in a rotating system},\ }\href
  {https://doi.org/10.1103/PhysRevD.21.2260} {\bibfield  {journal} {\bibinfo
  {journal} {Phys. Rev. D}\ }\textbf {\bibinfo {volume} {21}},\ \bibinfo
  {pages} {2260} (\bibinfo {year} {1980})}\BibitemShut {NoStop}%
\bibitem [{\citenamefont {Erdmenger}\ \emph {et~al.}(2009)\citenamefont
  {Erdmenger}, \citenamefont {Haack}, \citenamefont {Kaminski},\ and\
  \citenamefont {Yarom}}]{Erdmenger:2008rm}%
  \BibitemOpen
  \bibfield  {author} {\bibinfo {author} {\bibfnamefont {J.}~\bibnamefont
  {Erdmenger}}, \bibinfo {author} {\bibfnamefont {M.}~\bibnamefont {Haack}},
  \bibinfo {author} {\bibfnamefont {M.}~\bibnamefont {Kaminski}},\ and\
  \bibinfo {author} {\bibfnamefont {A.}~\bibnamefont {Yarom}},\ }\bibfield
  {title} {\bibinfo {title} {{Fluid dynamics of R-charged black holes}},\
  }\href {https://doi.org/10.1088/1126-6708/2009/01/055} {\bibfield  {journal}
  {\bibinfo  {journal} {JHEP}\ }\textbf {\bibinfo {volume} {01}},\ \bibinfo
  {pages} {055}},\ \Eprint {https://arxiv.org/abs/0809.2488} {arXiv:0809.2488
  [hep-th]} \BibitemShut {NoStop}%
\bibitem [{\citenamefont {Banerjee}\ \emph {et~al.}(2011)\citenamefont
  {Banerjee}, \citenamefont {Bhattacharya}, \citenamefont {Bhattacharyya},
  \citenamefont {Dutta}, \citenamefont {Loganayagam},\ and\ \citenamefont
  {Surowka}}]{Banerjee:2008th}%
  \BibitemOpen
  \bibfield  {author} {\bibinfo {author} {\bibfnamefont {N.}~\bibnamefont
  {Banerjee}}, \bibinfo {author} {\bibfnamefont {J.}~\bibnamefont
  {Bhattacharya}}, \bibinfo {author} {\bibfnamefont {S.}~\bibnamefont
  {Bhattacharyya}}, \bibinfo {author} {\bibfnamefont {S.}~\bibnamefont
  {Dutta}}, \bibinfo {author} {\bibfnamefont {R.}~\bibnamefont {Loganayagam}},\
  and\ \bibinfo {author} {\bibfnamefont {P.}~\bibnamefont {Surowka}},\
  }\bibfield  {title} {\bibinfo {title} {{Hydrodynamics from charged black
  branes}},\ }\href {https://doi.org/10.1007/JHEP01(2011)094} {\bibfield
  {journal} {\bibinfo  {journal} {JHEP}\ }\textbf {\bibinfo {volume} {01}},\
  \bibinfo {pages} {094}},\ \Eprint {https://arxiv.org/abs/0809.2596}
  {arXiv:0809.2596 [hep-th]} \BibitemShut {NoStop}%
\bibitem [{\citenamefont {Son}\ and\ \citenamefont
  {Surowka}(2009)}]{Son:2009tf}%
  \BibitemOpen
  \bibfield  {author} {\bibinfo {author} {\bibfnamefont {D.~T.}\ \bibnamefont
  {Son}}\ and\ \bibinfo {author} {\bibfnamefont {P.}~\bibnamefont {Surowka}},\
  }\bibfield  {title} {\bibinfo {title} {{Hydrodynamics with Triangle
  Anomalies}},\ }\href {https://doi.org/10.1103/PhysRevLett.103.191601}
  {\bibfield  {journal} {\bibinfo  {journal} {Phys. Rev. Lett.}\ }\textbf
  {\bibinfo {volume} {103}},\ \bibinfo {pages} {191601} (\bibinfo {year}
  {2009})},\ \Eprint {https://arxiv.org/abs/0906.5044} {arXiv:0906.5044
  [hep-th]} \BibitemShut {NoStop}%
\bibitem [{\citenamefont {Landsteiner}\ \emph
  {et~al.}(2011{\natexlab{a}})\citenamefont {Landsteiner}, \citenamefont
  {Megias},\ and\ \citenamefont {Pena-Benitez}}]{Landsteiner:2011cp}%
  \BibitemOpen
  \bibfield  {author} {\bibinfo {author} {\bibfnamefont {K.}~\bibnamefont
  {Landsteiner}}, \bibinfo {author} {\bibfnamefont {E.}~\bibnamefont
  {Megias}},\ and\ \bibinfo {author} {\bibfnamefont {F.}~\bibnamefont
  {Pena-Benitez}},\ }\bibfield  {title} {\bibinfo {title} {{Gravitational
  Anomaly and Transport}},\ }\href
  {https://doi.org/10.1103/PhysRevLett.107.021601} {\bibfield  {journal}
  {\bibinfo  {journal} {Phys. Rev. Lett.}\ }\textbf {\bibinfo {volume} {107}},\
  \bibinfo {pages} {021601} (\bibinfo {year} {2011}{\natexlab{a}})},\ \Eprint
  {https://arxiv.org/abs/1103.5006} {arXiv:1103.5006 [hep-ph]} \BibitemShut
  {NoStop}%
\bibitem [{\citenamefont {Landsteiner}\ \emph
  {et~al.}(2011{\natexlab{b}})\citenamefont {Landsteiner}, \citenamefont
  {Megias}, \citenamefont {Melgar},\ and\ \citenamefont
  {Pena-Benitez}}]{Landsteiner:2011iq}%
  \BibitemOpen
  \bibfield  {author} {\bibinfo {author} {\bibfnamefont {K.}~\bibnamefont
  {Landsteiner}}, \bibinfo {author} {\bibfnamefont {E.}~\bibnamefont {Megias}},
  \bibinfo {author} {\bibfnamefont {L.}~\bibnamefont {Melgar}},\ and\ \bibinfo
  {author} {\bibfnamefont {F.}~\bibnamefont {Pena-Benitez}},\ }\bibfield
  {title} {\bibinfo {title} {{Holographic Gravitational Anomaly and Chiral
  Vortical Effect}},\ }\href {https://doi.org/10.1007/JHEP09(2011)121}
  {\bibfield  {journal} {\bibinfo  {journal} {JHEP}\ }\textbf {\bibinfo
  {volume} {09}},\ \bibinfo {pages} {121}},\ \Eprint
  {https://arxiv.org/abs/1107.0368} {arXiv:1107.0368 [hep-th]} \BibitemShut
  {NoStop}%
\bibitem [{\citenamefont {Amado}\ \emph {et~al.}(2011)\citenamefont {Amado},
  \citenamefont {Landsteiner},\ and\ \citenamefont
  {Pena-Benitez}}]{Amado:2011zx}%
  \BibitemOpen
  \bibfield  {author} {\bibinfo {author} {\bibfnamefont {I.}~\bibnamefont
  {Amado}}, \bibinfo {author} {\bibfnamefont {K.}~\bibnamefont {Landsteiner}},\
  and\ \bibinfo {author} {\bibfnamefont {F.}~\bibnamefont {Pena-Benitez}},\
  }\bibfield  {title} {\bibinfo {title} {{Anomalous transport coefficients from
  Kubo formulas in Holography}},\ }\href
  {https://doi.org/10.1007/JHEP05(2011)081} {\bibfield  {journal} {\bibinfo
  {journal} {JHEP}\ }\textbf {\bibinfo {volume} {05}},\ \bibinfo {pages}
  {081}},\ \Eprint {https://arxiv.org/abs/1102.4577} {arXiv:1102.4577 [hep-th]}
  \BibitemShut {NoStop}%
\bibitem [{\citenamefont {Glorioso}\ \emph {et~al.}(2019)\citenamefont
  {Glorioso}, \citenamefont {Liu},\ and\ \citenamefont
  {Rajagopal}}]{Glorioso:2017lcn}%
  \BibitemOpen
  \bibfield  {author} {\bibinfo {author} {\bibfnamefont {P.}~\bibnamefont
  {Glorioso}}, \bibinfo {author} {\bibfnamefont {H.}~\bibnamefont {Liu}},\ and\
  \bibinfo {author} {\bibfnamefont {S.}~\bibnamefont {Rajagopal}},\ }\bibfield
  {title} {\bibinfo {title} {{Global Anomalies, Discrete Symmetries, and
  Hydrodynamic Effective Actions}},\ }\href
  {https://doi.org/10.1007/JHEP01(2019)043} {\bibfield  {journal} {\bibinfo
  {journal} {JHEP}\ }\textbf {\bibinfo {volume} {01}},\ \bibinfo {pages}
  {043}},\ \Eprint {https://arxiv.org/abs/1710.03768} {arXiv:1710.03768
  [hep-th]} \BibitemShut {NoStop}%
\bibitem [{\citenamefont {Flachi}\ and\ \citenamefont
  {Fukushima}(2018)}]{Flachi:2017vlp}%
  \BibitemOpen
  \bibfield  {author} {\bibinfo {author} {\bibfnamefont {A.}~\bibnamefont
  {Flachi}}\ and\ \bibinfo {author} {\bibfnamefont {K.}~\bibnamefont
  {Fukushima}},\ }\bibfield  {title} {\bibinfo {title} {{Chiral vortical effect
  with finite rotation, temperature, and curvature}},\ }\href
  {https://doi.org/10.1103/PhysRevD.98.096011} {\bibfield  {journal} {\bibinfo
  {journal} {Phys. Rev. D}\ }\textbf {\bibinfo {volume} {98}},\ \bibinfo
  {pages} {096011} (\bibinfo {year} {2018})},\ \Eprint
  {https://arxiv.org/abs/1702.04753} {arXiv:1702.04753 [hep-th]} \BibitemShut
  {NoStop}%
\bibitem [{\citenamefont {Avkhadiev}\ and\ \citenamefont
  {Sadofyev}(2017)}]{Avkhadiev:2017fxj}%
  \BibitemOpen
  \bibfield  {author} {\bibinfo {author} {\bibfnamefont {A.}~\bibnamefont
  {Avkhadiev}}\ and\ \bibinfo {author} {\bibfnamefont {A.~V.}\ \bibnamefont
  {Sadofyev}},\ }\bibfield  {title} {\bibinfo {title} {{Chiral Vortical Effect
  for Bosons}},\ }\href {https://doi.org/10.1103/PhysRevD.96.045015} {\bibfield
   {journal} {\bibinfo  {journal} {Phys. Rev. D}\ }\textbf {\bibinfo {volume}
  {96}},\ \bibinfo {pages} {045015} (\bibinfo {year} {2017})},\ \Eprint
  {https://arxiv.org/abs/1702.07340} {arXiv:1702.07340 [hep-th]} \BibitemShut
  {NoStop}%
\bibitem [{\citenamefont {Stone}\ and\ \citenamefont
  {Kim}(2018)}]{Stone:2018zel}%
  \BibitemOpen
  \bibfield  {author} {\bibinfo {author} {\bibfnamefont {M.}~\bibnamefont
  {Stone}}\ and\ \bibinfo {author} {\bibfnamefont {J.}~\bibnamefont {Kim}},\
  }\bibfield  {title} {\bibinfo {title} {{Mixed Anomalies: Chiral Vortical
  Effect and the Sommerfeld Expansion}},\ }\href
  {https://doi.org/10.1103/PhysRevD.98.025012} {\bibfield  {journal} {\bibinfo
  {journal} {Phys. Rev. D}\ }\textbf {\bibinfo {volume} {98}},\ \bibinfo
  {pages} {025012} (\bibinfo {year} {2018})},\ \Eprint
  {https://arxiv.org/abs/1804.08668} {arXiv:1804.08668 [cond-mat.mes-hall]}
  \BibitemShut {NoStop}%
\bibitem [{\citenamefont {Buzzegoli}\ and\ \citenamefont
  {Becattini}(2018)}]{Buzzegoli:2018wpy}%
  \BibitemOpen
  \bibfield  {author} {\bibinfo {author} {\bibfnamefont {M.}~\bibnamefont
  {Buzzegoli}}\ and\ \bibinfo {author} {\bibfnamefont {F.}~\bibnamefont
  {Becattini}},\ }\bibfield  {title} {\bibinfo {title} {{General thermodynamic
  equilibrium with axial chemical potential for the free Dirac field}},\ }\href
  {https://doi.org/10.1007/JHEP12(2018)002} {\bibfield  {journal} {\bibinfo
  {journal} {JHEP}\ }\textbf {\bibinfo {volume} {12}},\ \bibinfo {pages}
  {002}},\ \Eprint {https://arxiv.org/abs/1807.02071} {arXiv:1807.02071
  [hep-th]} \BibitemShut {NoStop}%
\bibitem [{\citenamefont {Prokhorov}\ \emph {et~al.}(2020)\citenamefont
  {Prokhorov}, \citenamefont {Teryaev},\ and\ \citenamefont
  {Zakharov}}]{Prokhorov:2020okl}%
  \BibitemOpen
  \bibfield  {author} {\bibinfo {author} {\bibfnamefont {G.~Y.}\ \bibnamefont
  {Prokhorov}}, \bibinfo {author} {\bibfnamefont {O.~V.}\ \bibnamefont
  {Teryaev}},\ and\ \bibinfo {author} {\bibfnamefont {V.~I.}\ \bibnamefont
  {Zakharov}},\ }\bibfield  {title} {\bibinfo {title} {{Chiral vortical effect:
  Black-hole versus flat-space derivation}},\ }\href
  {https://doi.org/10.1103/PhysRevD.102.121702} {\bibfield  {journal} {\bibinfo
   {journal} {Phys. Rev. D}\ }\textbf {\bibinfo {volume} {102}},\ \bibinfo
  {pages} {121702} (\bibinfo {year} {2020})},\ \Eprint
  {https://arxiv.org/abs/2003.11119} {arXiv:2003.11119 [hep-th]} \BibitemShut
  {NoStop}%
\bibitem [{\citenamefont {Huang}\ \emph {et~al.}(2020)\citenamefont {Huang},
  \citenamefont {Mitkin}, \citenamefont {Sadofyev},\ and\ \citenamefont
  {Speranza}}]{Huang:2020kik}%
  \BibitemOpen
  \bibfield  {author} {\bibinfo {author} {\bibfnamefont {X.-G.}\ \bibnamefont
  {Huang}}, \bibinfo {author} {\bibfnamefont {P.}~\bibnamefont {Mitkin}},
  \bibinfo {author} {\bibfnamefont {A.~V.}\ \bibnamefont {Sadofyev}},\ and\
  \bibinfo {author} {\bibfnamefont {E.}~\bibnamefont {Speranza}},\ }\bibfield
  {title} {\bibinfo {title} {{Zilch Vortical Effect, Berry Phase, and Kinetic
  Theory}},\ }\href {https://doi.org/10.1007/JHEP10(2020)117} {\bibfield
  {journal} {\bibinfo  {journal} {JHEP}\ }\textbf {\bibinfo {volume} {10}},\
  \bibinfo {pages} {117}},\ \Eprint {https://arxiv.org/abs/2006.03591}
  {arXiv:2006.03591 [hep-th]} \BibitemShut {NoStop}%
\bibitem [{\citenamefont {Sadofyev}\ and\ \citenamefont
  {Isachenkov}(2011)}]{Sadofyev:2010pr}%
  \BibitemOpen
  \bibfield  {author} {\bibinfo {author} {\bibfnamefont {A.~V.}\ \bibnamefont
  {Sadofyev}}\ and\ \bibinfo {author} {\bibfnamefont {M.~V.}\ \bibnamefont
  {Isachenkov}},\ }\bibfield  {title} {\bibinfo {title} {{The Chiral magnetic
  effect in hydrodynamical approach}},\ }\href
  {https://doi.org/10.1016/j.physletb.2011.02.041} {\bibfield  {journal}
  {\bibinfo  {journal} {Phys. Lett. B}\ }\textbf {\bibinfo {volume} {697}},\
  \bibinfo {pages} {404} (\bibinfo {year} {2011})},\ \Eprint
  {https://arxiv.org/abs/1010.1550} {arXiv:1010.1550 [hep-th]} \BibitemShut
  {NoStop}%
\bibitem [{\citenamefont {Neiman}\ and\ \citenamefont
  {Oz}(2011)}]{Neiman:2010zi}%
  \BibitemOpen
  \bibfield  {author} {\bibinfo {author} {\bibfnamefont {Y.}~\bibnamefont
  {Neiman}}\ and\ \bibinfo {author} {\bibfnamefont {Y.}~\bibnamefont {Oz}},\
  }\bibfield  {title} {\bibinfo {title} {{Relativistic Hydrodynamics with
  General Anomalous Charges}},\ }\href
  {https://doi.org/10.1007/JHEP03(2011)023} {\bibfield  {journal} {\bibinfo
  {journal} {JHEP}\ }\textbf {\bibinfo {volume} {03}},\ \bibinfo {pages}
  {023}},\ \Eprint {https://arxiv.org/abs/1011.5107} {arXiv:1011.5107 [hep-th]}
  \BibitemShut {NoStop}%
\bibitem [{\citenamefont {Landsteiner}\ \emph {et~al.}(2013)\citenamefont
  {Landsteiner}, \citenamefont {Megias},\ and\ \citenamefont
  {Pena-Benitez}}]{Landsteiner:2012kd}%
  \BibitemOpen
  \bibfield  {author} {\bibinfo {author} {\bibfnamefont {K.}~\bibnamefont
  {Landsteiner}}, \bibinfo {author} {\bibfnamefont {E.}~\bibnamefont
  {Megias}},\ and\ \bibinfo {author} {\bibfnamefont {F.}~\bibnamefont
  {Pena-Benitez}},\ }\bibfield  {title} {\bibinfo {title} {{Anomalous Transport
  from Kubo Formulae}},\ }\href {https://doi.org/10.1007/978-3-642-37305-3_17}
  {\bibfield  {journal} {\bibinfo  {journal} {Lect. Notes Phys.}\ }\textbf
  {\bibinfo {volume} {871}},\ \bibinfo {pages} {433} (\bibinfo {year}
  {2013})},\ \Eprint {https://arxiv.org/abs/1207.5808} {arXiv:1207.5808
  [hep-th]} \BibitemShut {NoStop}%
\bibitem [{\citenamefont {Speranza}\ \emph {et~al.}(2023)\citenamefont
  {Speranza}, \citenamefont {Bemfica}, \citenamefont {Disconzi},\ and\
  \citenamefont {Noronha}}]{Speranza:2021bxf}%
  \BibitemOpen
  \bibfield  {author} {\bibinfo {author} {\bibfnamefont {E.}~\bibnamefont
  {Speranza}}, \bibinfo {author} {\bibfnamefont {F.~S.}\ \bibnamefont
  {Bemfica}}, \bibinfo {author} {\bibfnamefont {M.~M.}\ \bibnamefont
  {Disconzi}},\ and\ \bibinfo {author} {\bibfnamefont {J.}~\bibnamefont
  {Noronha}},\ }\bibfield  {title} {\bibinfo {title} {{Challenges in solving
  chiral hydrodynamics}},\ }\href {https://doi.org/10.1103/PhysRevD.107.054029}
  {\bibfield  {journal} {\bibinfo  {journal} {Phys. Rev. D}\ }\textbf {\bibinfo
  {volume} {107}},\ \bibinfo {pages} {054029} (\bibinfo {year} {2023})},\
  \Eprint {https://arxiv.org/abs/2104.02110} {arXiv:2104.02110 [hep-th]}
  \BibitemShut {NoStop}%
\bibitem [{\citenamefont {Chen}\ \emph {et~al.}(2015)\citenamefont {Chen},
  \citenamefont {Son},\ and\ \citenamefont {Stephanov}}]{Chen:2015gta}%
  \BibitemOpen
  \bibfield  {author} {\bibinfo {author} {\bibfnamefont {J.-Y.}\ \bibnamefont
  {Chen}}, \bibinfo {author} {\bibfnamefont {D.~T.}\ \bibnamefont {Son}},\ and\
  \bibinfo {author} {\bibfnamefont {M.~A.}\ \bibnamefont {Stephanov}},\
  }\bibfield  {title} {\bibinfo {title} {{Collisions in Chiral Kinetic
  Theory}},\ }\href {https://doi.org/10.1103/PhysRevLett.115.021601} {\bibfield
   {journal} {\bibinfo  {journal} {Phys. Rev. Lett.}\ }\textbf {\bibinfo
  {volume} {115}},\ \bibinfo {pages} {021601} (\bibinfo {year} {2015})},\
  \Eprint {https://arxiv.org/abs/1502.06966} {arXiv:1502.06966 [hep-th]}
  \BibitemShut {NoStop}%
\bibitem [{\citenamefont {Son}\ and\ \citenamefont
  {Yamamoto}(2012)}]{Son:2012wh}%
  \BibitemOpen
  \bibfield  {author} {\bibinfo {author} {\bibfnamefont {D.~T.}\ \bibnamefont
  {Son}}\ and\ \bibinfo {author} {\bibfnamefont {N.}~\bibnamefont {Yamamoto}},\
  }\bibfield  {title} {\bibinfo {title} {{Berry Curvature, Triangle Anomalies,
  and the Chiral Magnetic Effect in Fermi Liquids}},\ }\href
  {https://doi.org/10.1103/PhysRevLett.109.181602} {\bibfield  {journal}
  {\bibinfo  {journal} {Phys. Rev. Lett.}\ }\textbf {\bibinfo {volume} {109}},\
  \bibinfo {pages} {181602} (\bibinfo {year} {2012})},\ \Eprint
  {https://arxiv.org/abs/1203.2697} {arXiv:1203.2697 [cond-mat.mes-hall]}
  \BibitemShut {NoStop}%
\bibitem [{\citenamefont {Stephanov}\ and\ \citenamefont
  {Yin}(2012)}]{Stephanov:2012ki}%
  \BibitemOpen
  \bibfield  {author} {\bibinfo {author} {\bibfnamefont {M.~A.}\ \bibnamefont
  {Stephanov}}\ and\ \bibinfo {author} {\bibfnamefont {Y.}~\bibnamefont
  {Yin}},\ }\bibfield  {title} {\bibinfo {title} {{Chiral Kinetic Theory}},\
  }\href {https://doi.org/10.1103/PhysRevLett.109.162001} {\bibfield  {journal}
  {\bibinfo  {journal} {Phys. Rev. Lett.}\ }\textbf {\bibinfo {volume} {109}},\
  \bibinfo {pages} {162001} (\bibinfo {year} {2012})},\ \Eprint
  {https://arxiv.org/abs/1207.0747} {arXiv:1207.0747 [hep-th]} \BibitemShut
  {NoStop}%
\bibitem [{\citenamefont {Chen}\ \emph {et~al.}(2013)\citenamefont {Chen},
  \citenamefont {Pu}, \citenamefont {Wang},\ and\ \citenamefont
  {Wang}}]{Chen:2012ca}%
  \BibitemOpen
  \bibfield  {author} {\bibinfo {author} {\bibfnamefont {J.-W.}\ \bibnamefont
  {Chen}}, \bibinfo {author} {\bibfnamefont {S.}~\bibnamefont {Pu}}, \bibinfo
  {author} {\bibfnamefont {Q.}~\bibnamefont {Wang}},\ and\ \bibinfo {author}
  {\bibfnamefont {X.-N.}\ \bibnamefont {Wang}},\ }\bibfield  {title} {\bibinfo
  {title} {{Berry Curvature and Four-Dimensional Monopoles in the Relativistic
  Chiral Kinetic Equation}},\ }\href
  {https://doi.org/10.1103/PhysRevLett.110.262301} {\bibfield  {journal}
  {\bibinfo  {journal} {Phys. Rev. Lett.}\ }\textbf {\bibinfo {volume} {110}},\
  \bibinfo {pages} {262301} (\bibinfo {year} {2013})},\ \Eprint
  {https://arxiv.org/abs/1210.8312} {arXiv:1210.8312 [hep-th]} \BibitemShut
  {NoStop}%
\bibitem [{\citenamefont {Son}\ and\ \citenamefont
  {Yamamoto}(2013)}]{Son:2012zy}%
  \BibitemOpen
  \bibfield  {author} {\bibinfo {author} {\bibfnamefont {D.~T.}\ \bibnamefont
  {Son}}\ and\ \bibinfo {author} {\bibfnamefont {N.}~\bibnamefont {Yamamoto}},\
  }\bibfield  {title} {\bibinfo {title} {{Kinetic theory with Berry curvature
  from quantum field theories}},\ }\href
  {https://doi.org/10.1103/PhysRevD.87.085016} {\bibfield  {journal} {\bibinfo
  {journal} {Phys. Rev. D}\ }\textbf {\bibinfo {volume} {87}},\ \bibinfo
  {pages} {085016} (\bibinfo {year} {2013})},\ \Eprint
  {https://arxiv.org/abs/1210.8158} {arXiv:1210.8158 [hep-th]} \BibitemShut
  {NoStop}%
\bibitem [{\citenamefont {Manuel}\ and\ \citenamefont
  {Torres-Rincon}(2014)}]{Manuel:2013zaa}%
  \BibitemOpen
  \bibfield  {author} {\bibinfo {author} {\bibfnamefont {C.}~\bibnamefont
  {Manuel}}\ and\ \bibinfo {author} {\bibfnamefont {J.~M.}\ \bibnamefont
  {Torres-Rincon}},\ }\bibfield  {title} {\bibinfo {title} {{Kinetic theory of
  chiral relativistic plasmas and energy density of their gauge collective
  excitations}},\ }\href {https://doi.org/10.1103/PhysRevD.89.096002}
  {\bibfield  {journal} {\bibinfo  {journal} {Phys. Rev. D}\ }\textbf {\bibinfo
  {volume} {89}},\ \bibinfo {pages} {096002} (\bibinfo {year} {2014})},\
  \Eprint {https://arxiv.org/abs/1312.1158} {arXiv:1312.1158 [hep-ph]}
  \BibitemShut {NoStop}%
\bibitem [{\citenamefont {Hidaka}\ \emph {et~al.}(2017)\citenamefont {Hidaka},
  \citenamefont {Pu},\ and\ \citenamefont {Yang}}]{Hidaka:2016yjf}%
  \BibitemOpen
  \bibfield  {author} {\bibinfo {author} {\bibfnamefont {Y.}~\bibnamefont
  {Hidaka}}, \bibinfo {author} {\bibfnamefont {S.}~\bibnamefont {Pu}},\ and\
  \bibinfo {author} {\bibfnamefont {D.-L.}\ \bibnamefont {Yang}},\ }\bibfield
  {title} {\bibinfo {title} {{Relativistic Chiral Kinetic Theory from Quantum
  Field Theories}},\ }\href {https://doi.org/10.1103/PhysRevD.95.091901}
  {\bibfield  {journal} {\bibinfo  {journal} {Phys. Rev. D}\ }\textbf {\bibinfo
  {volume} {95}},\ \bibinfo {pages} {091901} (\bibinfo {year} {2017})},\
  \Eprint {https://arxiv.org/abs/1612.04630} {arXiv:1612.04630 [hep-th]}
  \BibitemShut {NoStop}%
\bibitem [{\citenamefont {Huang}\ \emph {et~al.}(2018)\citenamefont {Huang},
  \citenamefont {Shi}, \citenamefont {Jiang}, \citenamefont {Liao},\ and\
  \citenamefont {Zhuang}}]{Huang:2018wdl}%
  \BibitemOpen
  \bibfield  {author} {\bibinfo {author} {\bibfnamefont {A.}~\bibnamefont
  {Huang}}, \bibinfo {author} {\bibfnamefont {S.}~\bibnamefont {Shi}}, \bibinfo
  {author} {\bibfnamefont {Y.}~\bibnamefont {Jiang}}, \bibinfo {author}
  {\bibfnamefont {J.}~\bibnamefont {Liao}},\ and\ \bibinfo {author}
  {\bibfnamefont {P.}~\bibnamefont {Zhuang}},\ }\bibfield  {title} {\bibinfo
  {title} {{Complete and Consistent Chiral Transport from Wigner Function
  Formalism}},\ }\href {https://doi.org/10.1103/PhysRevD.98.036010} {\bibfield
  {journal} {\bibinfo  {journal} {Phys. Rev. D}\ }\textbf {\bibinfo {volume}
  {98}},\ \bibinfo {pages} {036010} (\bibinfo {year} {2018})},\ \Eprint
  {https://arxiv.org/abs/1801.03640} {arXiv:1801.03640 [hep-th]} \BibitemShut
  {NoStop}%
\bibitem [{\citenamefont {Yang}(2018)}]{Yang:2018lew}%
  \BibitemOpen
  \bibfield  {author} {\bibinfo {author} {\bibfnamefont {D.-L.}\ \bibnamefont
  {Yang}},\ }\bibfield  {title} {\bibinfo {title} {{Side-Jump Induced
  Spin-Orbit Interaction of Chiral Fluids from Kinetic Theory}},\ }\href
  {https://doi.org/10.1103/PhysRevD.98.076019} {\bibfield  {journal} {\bibinfo
  {journal} {Phys. Rev. D}\ }\textbf {\bibinfo {volume} {98}},\ \bibinfo
  {pages} {076019} (\bibinfo {year} {2018})},\ \Eprint
  {https://arxiv.org/abs/1807.02395} {arXiv:1807.02395 [nucl-th]} \BibitemShut
  {NoStop}%
\bibitem [{\citenamefont {Jensen}(2012)}]{Jensen:2012jy}%
  \BibitemOpen
  \bibfield  {author} {\bibinfo {author} {\bibfnamefont {K.}~\bibnamefont
  {Jensen}},\ }\bibfield  {title} {\bibinfo {title} {{Triangle Anomalies,
  Thermodynamics, and Hydrodynamics}},\ }\href
  {https://doi.org/10.1103/PhysRevD.85.125017} {\bibfield  {journal} {\bibinfo
  {journal} {Phys. Rev. D}\ }\textbf {\bibinfo {volume} {85}},\ \bibinfo
  {pages} {125017} (\bibinfo {year} {2012})},\ \Eprint
  {https://arxiv.org/abs/1203.3599} {arXiv:1203.3599 [hep-th]} \BibitemShut
  {NoStop}%
\bibitem [{\citenamefont {Jensen}\ \emph {et~al.}(2012)\citenamefont {Jensen},
  \citenamefont {Kaminski}, \citenamefont {Kovtun}, \citenamefont {Meyer},
  \citenamefont {Ritz},\ and\ \citenamefont {Yarom}}]{Jensen:2012jh}%
  \BibitemOpen
  \bibfield  {author} {\bibinfo {author} {\bibfnamefont {K.}~\bibnamefont
  {Jensen}}, \bibinfo {author} {\bibfnamefont {M.}~\bibnamefont {Kaminski}},
  \bibinfo {author} {\bibfnamefont {P.}~\bibnamefont {Kovtun}}, \bibinfo
  {author} {\bibfnamefont {R.}~\bibnamefont {Meyer}}, \bibinfo {author}
  {\bibfnamefont {A.}~\bibnamefont {Ritz}},\ and\ \bibinfo {author}
  {\bibfnamefont {A.}~\bibnamefont {Yarom}},\ }\bibfield  {title} {\bibinfo
  {title} {{Towards hydrodynamics without an entropy current}},\ }\href
  {https://doi.org/10.1103/PhysRevLett.109.101601} {\bibfield  {journal}
  {\bibinfo  {journal} {Phys. Rev. Lett.}\ }\textbf {\bibinfo {volume} {109}},\
  \bibinfo {pages} {101601} (\bibinfo {year} {2012})},\ \Eprint
  {https://arxiv.org/abs/1203.3556} {arXiv:1203.3556 [hep-th]} \BibitemShut
  {NoStop}%
\bibitem [{\citenamefont {Banerjee}\ \emph {et~al.}(2012)\citenamefont
  {Banerjee}, \citenamefont {Bhattacharya}, \citenamefont {Bhattacharyya},
  \citenamefont {Jain}, \citenamefont {Minwalla},\ and\ \citenamefont
  {Sharma}}]{Banerjee:2012iz}%
  \BibitemOpen
  \bibfield  {author} {\bibinfo {author} {\bibfnamefont {N.}~\bibnamefont
  {Banerjee}}, \bibinfo {author} {\bibfnamefont {J.}~\bibnamefont
  {Bhattacharya}}, \bibinfo {author} {\bibfnamefont {S.}~\bibnamefont
  {Bhattacharyya}}, \bibinfo {author} {\bibfnamefont {S.}~\bibnamefont {Jain}},
  \bibinfo {author} {\bibfnamefont {S.}~\bibnamefont {Minwalla}},\ and\
  \bibinfo {author} {\bibfnamefont {T.}~\bibnamefont {Sharma}},\ }\bibfield
  {title} {\bibinfo {title} {{Constraints on Fluid Dynamics from Equilibrium
  Partition Functions}},\ }\href {https://doi.org/10.1007/JHEP09(2012)046}
  {\bibfield  {journal} {\bibinfo  {journal} {JHEP}\ }\textbf {\bibinfo
  {volume} {09}},\ \bibinfo {pages} {046}},\ \Eprint
  {https://arxiv.org/abs/1203.3544} {arXiv:1203.3544 [hep-th]} \BibitemShut
  {NoStop}%
\bibitem [{\citenamefont {Ammon}\ \emph {et~al.}(2021)\citenamefont {Ammon},
  \citenamefont {Grieninger}, \citenamefont {Hernandez}, \citenamefont
  {Kaminski}, \citenamefont {Koirala}, \citenamefont {Leiber},\ and\
  \citenamefont {Wu}}]{Ammon:2020rvg}%
  \BibitemOpen
  \bibfield  {author} {\bibinfo {author} {\bibfnamefont {M.}~\bibnamefont
  {Ammon}}, \bibinfo {author} {\bibfnamefont {S.}~\bibnamefont {Grieninger}},
  \bibinfo {author} {\bibfnamefont {J.}~\bibnamefont {Hernandez}}, \bibinfo
  {author} {\bibfnamefont {M.}~\bibnamefont {Kaminski}}, \bibinfo {author}
  {\bibfnamefont {R.}~\bibnamefont {Koirala}}, \bibinfo {author} {\bibfnamefont
  {J.}~\bibnamefont {Leiber}},\ and\ \bibinfo {author} {\bibfnamefont
  {J.}~\bibnamefont {Wu}},\ }\bibfield  {title} {\bibinfo {title} {{Chiral
  hydrodynamics in strong external magnetic fields}},\ }\href
  {https://doi.org/10.1007/JHEP04(2021)078} {\bibfield  {journal} {\bibinfo
  {journal} {JHEP}\ }\textbf {\bibinfo {volume} {04}},\ \bibinfo {pages}
  {078}},\ \Eprint {https://arxiv.org/abs/2012.09183} {arXiv:2012.09183
  [hep-th]} \BibitemShut {NoStop}%
\bibitem [{\citenamefont {Gorbar}\ \emph {et~al.}(2017)\citenamefont {Gorbar},
  \citenamefont {Rybalka},\ and\ \citenamefont {Shovkovy}}]{Gorbar:2017toh}%
  \BibitemOpen
  \bibfield  {author} {\bibinfo {author} {\bibfnamefont {E.~V.}\ \bibnamefont
  {Gorbar}}, \bibinfo {author} {\bibfnamefont {D.~O.}\ \bibnamefont
  {Rybalka}},\ and\ \bibinfo {author} {\bibfnamefont {I.~A.}\ \bibnamefont
  {Shovkovy}},\ }\bibfield  {title} {\bibinfo {title} {{Second-order
  dissipative hydrodynamics for plasma with chiral asymmetry and vorticity}},\
  }\href {https://doi.org/10.1103/PhysRevD.95.096010} {\bibfield  {journal}
  {\bibinfo  {journal} {Phys. Rev. D}\ }\textbf {\bibinfo {volume} {95}},\
  \bibinfo {pages} {096010} (\bibinfo {year} {2017})},\ \Eprint
  {https://arxiv.org/abs/1702.07791} {arXiv:1702.07791 [hep-th]} \BibitemShut
  {NoStop}%
\bibitem [{\citenamefont {Kovtun}(2023)}]{Kovtun:2022vas}%
  \BibitemOpen
  \bibfield  {author} {\bibinfo {author} {\bibfnamefont {P.}~\bibnamefont
  {Kovtun}},\ }\bibfield  {title} {\bibinfo {title} {{Temperature in
  relativistic fluids}},\ }\href {https://doi.org/10.1103/PhysRevD.107.086012}
  {\bibfield  {journal} {\bibinfo  {journal} {Phys. Rev. D}\ }\textbf {\bibinfo
  {volume} {107}},\ \bibinfo {pages} {086012} (\bibinfo {year} {2023})},\
  \Eprint {https://arxiv.org/abs/2210.15605} {arXiv:2210.15605 [gr-qc]}
  \BibitemShut {NoStop}%
\bibitem [{\citenamefont {Van}\ and\ \citenamefont {Biro}(2012)}]{Van:2011yn}%
  \BibitemOpen
  \bibfield  {author} {\bibinfo {author} {\bibfnamefont {P.}~\bibnamefont
  {Van}}\ and\ \bibinfo {author} {\bibfnamefont {T.~S.}\ \bibnamefont {Biro}},\
  }\bibfield  {title} {\bibinfo {title} {First order and stable relativistic
  dissipative hydrodynamics},\ }\href
  {https://doi.org/10.1016/j.physletb.2012.02.006} {\bibfield  {journal}
  {\bibinfo  {journal} {Phys. Lett.}\ }\textbf {\bibinfo {volume} {B709}},\
  \bibinfo {pages} {106} (\bibinfo {year} {2012})},\ \Eprint
  {https://arxiv.org/abs/1109.0985} {arXiv:1109.0985 [nucl-th]} \BibitemShut
  {NoStop}%
\bibitem [{\citenamefont {V\'an}\ and\ \citenamefont
  {Bir\'o}(2013)}]{Van:2013sma}%
  \BibitemOpen
  \bibfield  {author} {\bibinfo {author} {\bibfnamefont {P.}~\bibnamefont
  {V\'an}}\ and\ \bibinfo {author} {\bibfnamefont {T.~S.}\ \bibnamefont
  {Bir\'o}},\ }\bibfield  {title} {\bibinfo {title} {{Dissipation flow-frames:
  particle, energy, thermometer}},\ }in\ \href@noop {} {\emph {\bibinfo
  {booktitle} {{12th Joint European Thermodynamics Conference}}}}\ (\bibinfo
  {year} {2013})\ \Eprint {https://arxiv.org/abs/1305.3190} {arXiv:1305.3190
  [gr-qc]} \BibitemShut {NoStop}%
\bibitem [{\citenamefont {Becattini}\ \emph {et~al.}(2015)\citenamefont
  {Becattini}, \citenamefont {Bucciantini}, \citenamefont {Grossi},\ and\
  \citenamefont {Tinti}}]{Becattini:2014yxa}%
  \BibitemOpen
  \bibfield  {author} {\bibinfo {author} {\bibfnamefont {F.}~\bibnamefont
  {Becattini}}, \bibinfo {author} {\bibfnamefont {L.}~\bibnamefont
  {Bucciantini}}, \bibinfo {author} {\bibfnamefont {E.}~\bibnamefont
  {Grossi}},\ and\ \bibinfo {author} {\bibfnamefont {L.}~\bibnamefont
  {Tinti}},\ }\bibfield  {title} {\bibinfo {title} {{Local thermodynamical
  equilibrium and the beta frame for a quantum relativistic fluid}},\ }\href
  {https://doi.org/10.1140/epjc/s10052-015-3384-y} {\bibfield  {journal}
  {\bibinfo  {journal} {Eur. Phys. J. C}\ }\textbf {\bibinfo {volume} {75}},\
  \bibinfo {pages} {191} (\bibinfo {year} {2015})},\ \Eprint
  {https://arxiv.org/abs/1403.6265} {arXiv:1403.6265 [hep-th]} \BibitemShut
  {NoStop}%
\bibitem [{\citenamefont {Kondepudi}\ and\ \citenamefont
  {I.~Prigogine}(2014)}]{prigogine2014modern}%
  \BibitemOpen
  \bibfield  {author} {\bibinfo {author} {\bibfnamefont {D.}~\bibnamefont
  {Kondepudi}}\ and\ \bibinfo {author} {\bibfnamefont {D.}~\bibnamefont
  {I.~Prigogine}},\ }\href@noop {} {\emph {\bibinfo {title} {Modern
  Thermodynamics}}}\ (\bibinfo  {publisher} {John Wiley \& Sons},\ \bibinfo
  {year} {2014})\BibitemShut {NoStop}%
\bibitem [{\citenamefont {Denicol}\ \emph {et~al.}(2012)\citenamefont
  {Denicol}, \citenamefont {Niemi}, \citenamefont {Molnar},\ and\ \citenamefont
  {Rischke}}]{Denicol:2012cn}%
  \BibitemOpen
  \bibfield  {author} {\bibinfo {author} {\bibfnamefont {G.~S.}\ \bibnamefont
  {Denicol}}, \bibinfo {author} {\bibfnamefont {H.}~\bibnamefont {Niemi}},
  \bibinfo {author} {\bibfnamefont {E.}~\bibnamefont {Molnar}},\ and\ \bibinfo
  {author} {\bibfnamefont {D.~H.}\ \bibnamefont {Rischke}},\ }\bibfield
  {title} {\bibinfo {title} {Derivation of transient relativistic fluid
  dynamics from the {B}oltzmann equation},\ }\href
  {https://doi.org/10.1103/PhysRevD.85.114047, 10.1103/PhysRevD.91.039902}
  {\bibfield  {journal} {\bibinfo  {journal} {Phys. Rev.}\ }\textbf {\bibinfo
  {volume} {D85}},\ \bibinfo {pages} {114047} (\bibinfo {year} {2012})},\
  \bibinfo {note} {[Erratum: Phys. Rev.D91,no.3,039902(2015)]},\ \Eprint
  {https://arxiv.org/abs/1202.4551} {arXiv:1202.4551 [nucl-th]} \BibitemShut
  {NoStop}%
\bibitem [{\citenamefont {Israel}(1976{\natexlab{b}})}]{Israel:1976tn}%
  \BibitemOpen
  \bibfield  {author} {\bibinfo {author} {\bibfnamefont {W.}~\bibnamefont
  {Israel}},\ }\bibfield  {title} {\bibinfo {title} {{Nonstationary
  irreversible thermodynamics: A Causal relativistic theory}},\ }\href
  {https://doi.org/10.1016/0003-4916(76)90064-6} {\bibfield  {journal}
  {\bibinfo  {journal} {Annals Phys.}\ }\textbf {\bibinfo {volume} {100}},\
  \bibinfo {pages} {310} (\bibinfo {year} {1976}{\natexlab{b}})}\BibitemShut
  {NoStop}%
\bibitem [{\citenamefont {Kovtun}(2012)}]{Kovtun:2012rj}%
  \BibitemOpen
  \bibfield  {author} {\bibinfo {author} {\bibfnamefont {K.}~\bibnamefont
  {Kovtun}},\ }\bibfield  {title} {\bibinfo {title} {Lectures on hydrodynamic
  fluctuations in relativistic theories},\ }\bibfield  {booktitle} {\emph
  {\bibinfo {booktitle} {INT Summer School on Applications of String Theory
  Seattle, Washington, USA, July 18-29, 2011}},\ }\href
  {https://doi.org/10.1088/1751-8113/45/47/473001} {\bibfield  {journal}
  {\bibinfo  {journal} {J. Phys.}\ }\textbf {\bibinfo {volume} {A45}},\
  \bibinfo {pages} {473001} (\bibinfo {year} {2012})},\ \Eprint
  {https://arxiv.org/abs/1205.5040} {arXiv:1205.5040 [hep-th]} \BibitemShut
  {NoStop}%
\bibitem [{\citenamefont {Courant}\ and\ \citenamefont
  {Hilbert}(1991)}]{Courant_and_Hilbert_book_2}%
  \BibitemOpen
  \bibfield  {author} {\bibinfo {author} {\bibfnamefont {C.}~\bibnamefont
  {Courant}}\ and\ \bibinfo {author} {\bibfnamefont {D.}~\bibnamefont
  {Hilbert}},\ }\href@noop {} {\emph {\bibinfo {title} {Methods of Mathematical
  Physics}}},\ \bibinfo {edition} {1st}\ ed.,\ Vol.~\bibinfo {volume} {2}\
  (\bibinfo  {publisher} {John Wiley \& Sons, Inc.},\ \bibinfo {year} {1991})\
  p.\ \bibinfo {pages} {852}\BibitemShut {NoStop}%
\bibitem [{\citenamefont {Pichon}(1965)}]{PichonViscous}%
  \BibitemOpen
  \bibfield  {author} {\bibinfo {author} {\bibfnamefont {G.}~\bibnamefont
  {Pichon}},\ }\bibfield  {title} {\bibinfo {title} {{\'E}tude relativiste de
  fluides visqueux et charg{\'e}s},\ }\href
  {http://www.numdam.org/item/AIHPA_1965__2_1_21_0} {\bibfield  {journal}
  {\bibinfo  {journal} {Annales de l'I.H.P. Physique th{\'e}orique}\ }\textbf
  {\bibinfo {volume} {2}},\ \bibinfo {pages} {21} (\bibinfo {year}
  {1965})}\BibitemShut {NoStop}%
\bibitem [{\citenamefont {Hiscock}\ and\ \citenamefont
  {Lindblom}(1985)}]{Hiscock_Lindblom_instability_1985}%
  \BibitemOpen
  \bibfield  {author} {\bibinfo {author} {\bibfnamefont {W.~A.}\ \bibnamefont
  {Hiscock}}\ and\ \bibinfo {author} {\bibfnamefont {L.}~\bibnamefont
  {Lindblom}},\ }\bibfield  {title} {\bibinfo {title} {{Generic instabilities
  in first-order dissipative relativistic fluid theories}},\ }\href
  {https://doi.org/10.1103/PhysRevD.31.725} {\bibfield  {journal} {\bibinfo
  {journal} {Phys. Rev. D}\ }\textbf {\bibinfo {volume} {31}},\ \bibinfo
  {pages} {725} (\bibinfo {year} {1985})}\BibitemShut {NoStop}%
\bibitem [{\citenamefont {Denicol}\ \emph {et~al.}(2008)\citenamefont
  {Denicol}, \citenamefont {Kodama}, \citenamefont {Koide},\ and\ \citenamefont
  {Mota}}]{Denicol:2008ha}%
  \BibitemOpen
  \bibfield  {author} {\bibinfo {author} {\bibfnamefont {G.~S.}\ \bibnamefont
  {Denicol}}, \bibinfo {author} {\bibfnamefont {T.}~\bibnamefont {Kodama}},
  \bibinfo {author} {\bibfnamefont {T.}~\bibnamefont {Koide}},\ and\ \bibinfo
  {author} {\bibfnamefont {P.}~\bibnamefont {Mota}},\ }\bibfield  {title}
  {\bibinfo {title} {Stability and causality in relativistic dissipative
  hydrodynamics},\ }\href {https://doi.org/10.1088/0954-3899/35/11/115102}
  {\bibfield  {journal} {\bibinfo  {journal} {J. Phys.}\ }\textbf {\bibinfo
  {volume} {G35}},\ \bibinfo {pages} {115102} (\bibinfo {year} {2008})},\
  \Eprint {https://arxiv.org/abs/0807.3120} {arXiv:0807.3120 [hep-ph]}
  \BibitemShut {NoStop}%
\bibitem [{\citenamefont {Disconzi}(2019)}]{DisconziFollowupBemficaNoronha}%
  \BibitemOpen
  \bibfield  {author} {\bibinfo {author} {\bibfnamefont {M.~M.}\ \bibnamefont
  {Disconzi}},\ }\bibfield  {title} {\bibinfo {title} {On the existence of
  solutions and causality for relativistic viscous conformal fluids},\ }\href
  {https://doi.org/10.3934/cpaa.2019075} {\bibfield  {journal} {\bibinfo
  {journal} {Communications in Pure and Applied Analysis}\ }\textbf {\bibinfo
  {volume} {18}},\ \bibinfo {pages} {1567} (\bibinfo {year} {2019})},\ \Eprint
  {https://arxiv.org/abs/1708.06572} {arXiv:1708.06572 [math.AP]} \BibitemShut
  {NoStop}%
\bibitem [{\citenamefont {Disconzi}(2023)}]{Disconzi:2023rtt}%
  \BibitemOpen
  \bibfield  {author} {\bibinfo {author} {\bibfnamefont {M.~M.}\ \bibnamefont
  {Disconzi}},\ }\bibfield  {title} {\bibinfo {title} {{Recent developments in
  mathematical aspects of relativistic fluids}},\ }\href@noop {} {\  (\bibinfo
  {year} {2023})},\ \Eprint {https://arxiv.org/abs/2308.09844}
  {arXiv:2308.09844 [math.AP]} \BibitemShut {NoStop}%
\bibitem [{\citenamefont {Rocha}\ \emph {et~al.}(2023)\citenamefont {Rocha},
  \citenamefont {de~Brito},\ and\ \citenamefont {Denicol}}]{Rocha:2023hts}%
  \BibitemOpen
  \bibfield  {author} {\bibinfo {author} {\bibfnamefont {G.~S.}\ \bibnamefont
  {Rocha}}, \bibinfo {author} {\bibfnamefont {C.~V.~P.}\ \bibnamefont
  {de~Brito}},\ and\ \bibinfo {author} {\bibfnamefont {G.~S.}\ \bibnamefont
  {Denicol}},\ }\bibfield  {title} {\bibinfo {title} {{Hydrodynamic theories
  for a system of weakly self-interacting classical ultra-relativistic scalar
  particles: microscopic derivations and attractors}},\ }\href@noop {} {\
  (\bibinfo {year} {2023})},\ \Eprint {https://arxiv.org/abs/2306.07423}
  {arXiv:2306.07423 [nucl-th]} \BibitemShut {NoStop}%
\bibitem [{\citenamefont {Gavassino}(2022)}]{Gavassino:2021owo}%
  \BibitemOpen
  \bibfield  {author} {\bibinfo {author} {\bibfnamefont {L.}~\bibnamefont
  {Gavassino}},\ }\bibfield  {title} {\bibinfo {title} {{Can We Make Sense of
  Dissipation without Causality?}},\ }\href
  {https://doi.org/10.1103/PhysRevX.12.041001} {\bibfield  {journal} {\bibinfo
  {journal} {Phys. Rev. X}\ }\textbf {\bibinfo {volume} {12}},\ \bibinfo
  {pages} {041001} (\bibinfo {year} {2022})},\ \Eprint
  {https://arxiv.org/abs/2111.05254} {arXiv:2111.05254 [gr-qc]} \BibitemShut
  {NoStop}%
\bibitem [{\citenamefont {Gavassino}\ \emph {et~al.}(2022)\citenamefont
  {Gavassino}, \citenamefont {Antonelli},\ and\ \citenamefont
  {Haskell}}]{Gavassino:2021kjm}%
  \BibitemOpen
  \bibfield  {author} {\bibinfo {author} {\bibfnamefont {L.}~\bibnamefont
  {Gavassino}}, \bibinfo {author} {\bibfnamefont {M.}~\bibnamefont
  {Antonelli}},\ and\ \bibinfo {author} {\bibfnamefont {B.}~\bibnamefont
  {Haskell}},\ }\bibfield  {title} {\bibinfo {title} {{Thermodynamic Stability
  Implies Causality}},\ }\href {https://doi.org/10.1103/PhysRevLett.128.010606}
  {\bibfield  {journal} {\bibinfo  {journal} {Phys. Rev. Lett.}\ }\textbf
  {\bibinfo {volume} {128}},\ \bibinfo {pages} {010606} (\bibinfo {year}
  {2022})},\ \Eprint {https://arxiv.org/abs/2105.14621} {arXiv:2105.14621
  [gr-qc]} \BibitemShut {NoStop}%
\bibitem [{\citenamefont {Gavassino}(2023)}]{Gavassino:2023myj}%
  \BibitemOpen
  \bibfield  {author} {\bibinfo {author} {\bibfnamefont {L.}~\bibnamefont
  {Gavassino}},\ }\bibfield  {title} {\bibinfo {title} {{Bounds on transport
  from hydrodynamic stability}},\ }\href
  {https://doi.org/10.1016/j.physletb.2023.137854} {\bibfield  {journal}
  {\bibinfo  {journal} {Phys. Lett. B}\ }\textbf {\bibinfo {volume} {840}},\
  \bibinfo {pages} {137854} (\bibinfo {year} {2023})},\ \Eprint
  {https://arxiv.org/abs/2301.06651} {arXiv:2301.06651 [hep-th]} \BibitemShut
  {NoStop}%
\bibitem [{\citenamefont {Gavassino}\ \emph {et~al.}(2023)\citenamefont
  {Gavassino}, \citenamefont {Disconzi},\ and\ \citenamefont
  {Noronha}}]{Gavassino:2023mad}%
  \BibitemOpen
  \bibfield  {author} {\bibinfo {author} {\bibfnamefont {L.}~\bibnamefont
  {Gavassino}}, \bibinfo {author} {\bibfnamefont {M.~M.}\ \bibnamefont
  {Disconzi}},\ and\ \bibinfo {author} {\bibfnamefont {J.}~\bibnamefont
  {Noronha}},\ }\bibfield  {title} {\bibinfo {title} {{Dispersion relations
  alone cannot guarantee causality}},\ }\href@noop {} {\  (\bibinfo {year}
  {2023})},\ \Eprint {https://arxiv.org/abs/2307.05987} {arXiv:2307.05987
  [hep-th]} \BibitemShut {NoStop}%
\bibitem [{\citenamefont {Gradshteyn}\ and\ \citenamefont
  {Ryzhik}(2007)}]{gradshteyn2007}%
  \BibitemOpen
  \bibfield  {author} {\bibinfo {author} {\bibfnamefont {I.~S.}\ \bibnamefont
  {Gradshteyn}}\ and\ \bibinfo {author} {\bibfnamefont {I.~M.}\ \bibnamefont
  {Ryzhik}},\ }\href@noop {} {\emph {\bibinfo {title} {Table of integrals,
  series, and products}}},\ \bibinfo {edition} {seventh}\ ed.\ (\bibinfo
  {publisher} {Elsevier/Academic Press, Amsterdam},\ \bibinfo {year} {2007})\
  pp.\ \bibinfo {pages} {xlviii+1171},\ \bibinfo {note} {translated from the
  Russian, Translation edited and with a preface by Alan Jeffrey and Daniel
  Zwillinger, With one CD-ROM (Windows, Macintosh and UNIX)}\BibitemShut
  {NoStop}%
\bibitem [{\citenamefont {Ulrich}\ and\ \citenamefont
  {Watson}(1994)}]{doi:10.1137/0915035}%
  \BibitemOpen
  \bibfield  {author} {\bibinfo {author} {\bibfnamefont {G.}~\bibnamefont
  {Ulrich}}\ and\ \bibinfo {author} {\bibfnamefont {L.~T.}\ \bibnamefont
  {Watson}},\ }\bibfield  {title} {\bibinfo {title} {Positivity conditions for
  quartic polynomials},\ }\href {https://doi.org/10.1137/0915035} {\bibfield
  {journal} {\bibinfo  {journal} {SIAM Journal on Scientific Computing}\
  }\textbf {\bibinfo {volume} {15}},\ \bibinfo {pages} {528} (\bibinfo {year}
  {1994})}\BibitemShut {NoStop}%
\bibitem [{\citenamefont {Pandya}\ \emph
  {et~al.}(2022{\natexlab{a}})\citenamefont {Pandya}, \citenamefont {Most},\
  and\ \citenamefont {Pretorius}}]{Pandya:2022sff}%
  \BibitemOpen
  \bibfield  {author} {\bibinfo {author} {\bibfnamefont {A.}~\bibnamefont
  {Pandya}}, \bibinfo {author} {\bibfnamefont {E.~R.}\ \bibnamefont {Most}},\
  and\ \bibinfo {author} {\bibfnamefont {F.}~\bibnamefont {Pretorius}},\
  }\bibfield  {title} {\bibinfo {title} {{Causal, stable first-order viscous
  relativistic hydrodynamics with ideal gas microphysics}},\ }\href
  {https://doi.org/10.1103/PhysRevD.106.123036} {\bibfield  {journal} {\bibinfo
   {journal} {Phys. Rev. D}\ }\textbf {\bibinfo {volume} {106}},\ \bibinfo
  {pages} {123036} (\bibinfo {year} {2022}{\natexlab{a}})},\ \Eprint
  {https://arxiv.org/abs/2209.09265} {arXiv:2209.09265 [gr-qc]} \BibitemShut
  {NoStop}%
\bibitem [{\citenamefont {Bhattacharya}\ \emph {et~al.}(2014)\citenamefont
  {Bhattacharya}, \citenamefont {Bhattacharyya}, \citenamefont {Minwalla},\
  and\ \citenamefont {Yarom}}]{Bhattacharya:2011tra}%
  \BibitemOpen
  \bibfield  {author} {\bibinfo {author} {\bibfnamefont {J.}~\bibnamefont
  {Bhattacharya}}, \bibinfo {author} {\bibfnamefont {S.}~\bibnamefont
  {Bhattacharyya}}, \bibinfo {author} {\bibfnamefont {S.}~\bibnamefont
  {Minwalla}},\ and\ \bibinfo {author} {\bibfnamefont {A.}~\bibnamefont
  {Yarom}},\ }\bibfield  {title} {\bibinfo {title} {{A Theory of first order
  dissipative superfluid dynamics}},\ }\href
  {https://doi.org/10.1007/JHEP05(2014)147} {\bibfield  {journal} {\bibinfo
  {journal} {JHEP}\ }\textbf {\bibinfo {volume} {05}},\ \bibinfo {pages}
  {147}},\ \Eprint {https://arxiv.org/abs/1105.3733} {arXiv:1105.3733 [hep-th]}
  \BibitemShut {NoStop}%
\bibitem [{\citenamefont {Jensen}\ \emph {et~al.}(2014)\citenamefont {Jensen},
  \citenamefont {Loganayagam},\ and\ \citenamefont {Yarom}}]{Jensen:2013kka}%
  \BibitemOpen
  \bibfield  {author} {\bibinfo {author} {\bibfnamefont {K.}~\bibnamefont
  {Jensen}}, \bibinfo {author} {\bibfnamefont {R.}~\bibnamefont
  {Loganayagam}},\ and\ \bibinfo {author} {\bibfnamefont {A.}~\bibnamefont
  {Yarom}},\ }\bibfield  {title} {\bibinfo {title} {{Anomaly inflow and thermal
  equilibrium}},\ }\href {https://doi.org/10.1007/JHEP05(2014)134} {\bibfield
  {journal} {\bibinfo  {journal} {JHEP}\ }\textbf {\bibinfo {volume} {05}},\
  \bibinfo {pages} {134}},\ \Eprint {https://arxiv.org/abs/1310.7024}
  {arXiv:1310.7024 [hep-th]} \BibitemShut {NoStop}%
\bibitem [{\citenamefont {Stephanov}\ and\ \citenamefont
  {Yee}(2016)}]{Stephanov:2015roa}%
  \BibitemOpen
  \bibfield  {author} {\bibinfo {author} {\bibfnamefont {M.~A.}\ \bibnamefont
  {Stephanov}}\ and\ \bibinfo {author} {\bibfnamefont {H.-U.}\ \bibnamefont
  {Yee}},\ }\bibfield  {title} {\bibinfo {title} {{No-Drag Frame for Anomalous
  Chiral Fluid}},\ }\href {https://doi.org/10.1103/PhysRevLett.116.122302}
  {\bibfield  {journal} {\bibinfo  {journal} {Phys. Rev. Lett.}\ }\textbf
  {\bibinfo {volume} {116}},\ \bibinfo {pages} {122302} (\bibinfo {year}
  {2016})},\ \Eprint {https://arxiv.org/abs/1508.02396} {arXiv:1508.02396
  [hep-th]} \BibitemShut {NoStop}%
\bibitem [{\citenamefont {Rajagopal}\ and\ \citenamefont
  {Sadofyev}(2015)}]{Rajagopal:2015roa}%
  \BibitemOpen
  \bibfield  {author} {\bibinfo {author} {\bibfnamefont {K.}~\bibnamefont
  {Rajagopal}}\ and\ \bibinfo {author} {\bibfnamefont {A.~V.}\ \bibnamefont
  {Sadofyev}},\ }\bibfield  {title} {\bibinfo {title} {{Chiral drag force}},\
  }\href {https://doi.org/10.1007/JHEP10(2015)018} {\bibfield  {journal}
  {\bibinfo  {journal} {JHEP}\ }\textbf {\bibinfo {volume} {10}},\ \bibinfo
  {pages} {018}},\ \Eprint {https://arxiv.org/abs/1505.07379} {arXiv:1505.07379
  [hep-th]} \BibitemShut {NoStop}%
\bibitem [{\citenamefont {Akamatsu}\ and\ \citenamefont
  {Yamamoto}(2013)}]{Akamatsu:2013pjd}%
  \BibitemOpen
  \bibfield  {author} {\bibinfo {author} {\bibfnamefont {Y.}~\bibnamefont
  {Akamatsu}}\ and\ \bibinfo {author} {\bibfnamefont {N.}~\bibnamefont
  {Yamamoto}},\ }\bibfield  {title} {\bibinfo {title} {{Chiral Plasma
  Instabilities}},\ }\href {https://doi.org/10.1103/PhysRevLett.111.052002}
  {\bibfield  {journal} {\bibinfo  {journal} {Phys. Rev. Lett.}\ }\textbf
  {\bibinfo {volume} {111}},\ \bibinfo {pages} {052002} (\bibinfo {year}
  {2013})},\ \Eprint {https://arxiv.org/abs/1302.2125} {arXiv:1302.2125
  [nucl-th]} \BibitemShut {NoStop}%
\bibitem [{\citenamefont {Hirono}\ \emph {et~al.}(2015)\citenamefont {Hirono},
  \citenamefont {Kharzeev},\ and\ \citenamefont {Yin}}]{Hirono:2015rla}%
  \BibitemOpen
  \bibfield  {author} {\bibinfo {author} {\bibfnamefont {Y.}~\bibnamefont
  {Hirono}}, \bibinfo {author} {\bibfnamefont {D.}~\bibnamefont {Kharzeev}},\
  and\ \bibinfo {author} {\bibfnamefont {Y.}~\bibnamefont {Yin}},\ }\bibfield
  {title} {\bibinfo {title} {{Self-similar inverse cascade of magnetic helicity
  driven by the chiral anomaly}},\ }\href
  {https://doi.org/10.1103/PhysRevD.92.125031} {\bibfield  {journal} {\bibinfo
  {journal} {Phys. Rev. D}\ }\textbf {\bibinfo {volume} {92}},\ \bibinfo
  {pages} {125031} (\bibinfo {year} {2015})},\ \Eprint
  {https://arxiv.org/abs/1509.07790} {arXiv:1509.07790 [hep-th]} \BibitemShut
  {NoStop}%
\bibitem [{\citenamefont {Manuel}\ and\ \citenamefont
  {Torres-Rincon}(2015)}]{Manuel:2015zpa}%
  \BibitemOpen
  \bibfield  {author} {\bibinfo {author} {\bibfnamefont {C.}~\bibnamefont
  {Manuel}}\ and\ \bibinfo {author} {\bibfnamefont {J.~M.}\ \bibnamefont
  {Torres-Rincon}},\ }\bibfield  {title} {\bibinfo {title} {{Dynamical
  evolution of the chiral magnetic effect: Applications to the quark-gluon
  plasma}},\ }\href {https://doi.org/10.1103/PhysRevD.92.074018} {\bibfield
  {journal} {\bibinfo  {journal} {Phys. Rev. D}\ }\textbf {\bibinfo {volume}
  {92}},\ \bibinfo {pages} {074018} (\bibinfo {year} {2015})},\ \Eprint
  {https://arxiv.org/abs/1501.07608} {arXiv:1501.07608 [hep-ph]} \BibitemShut
  {NoStop}%
\bibitem [{\citenamefont {Heller}\ \emph {et~al.}(2023)\citenamefont {Heller},
  \citenamefont {Serantes}, \citenamefont {Spali\'nski},\ and\ \citenamefont
  {Withers}}]{Heller:2023jtd}%
  \BibitemOpen
  \bibfield  {author} {\bibinfo {author} {\bibfnamefont {M.~P.}\ \bibnamefont
  {Heller}}, \bibinfo {author} {\bibfnamefont {A.}~\bibnamefont {Serantes}},
  \bibinfo {author} {\bibfnamefont {M.}~\bibnamefont {Spali\'nski}},\ and\
  \bibinfo {author} {\bibfnamefont {B.}~\bibnamefont {Withers}},\ }\bibfield
  {title} {\bibinfo {title} {{The Hydrohedron: Bootstrapping Relativistic
  Hydrodynamics}},\ }\href@noop {} {\  (\bibinfo {year} {2023})},\ \Eprint
  {https://arxiv.org/abs/2305.07703} {arXiv:2305.07703 [hep-th]} \BibitemShut
  {NoStop}%
\bibitem [{\citenamefont {Florkowski}\ \emph {et~al.}(2018)\citenamefont
  {Florkowski}, \citenamefont {Friman}, \citenamefont {Jaiswal},\ and\
  \citenamefont {Speranza}}]{Florkowski:2017ruc}%
  \BibitemOpen
  \bibfield  {author} {\bibinfo {author} {\bibfnamefont {W.}~\bibnamefont
  {Florkowski}}, \bibinfo {author} {\bibfnamefont {B.}~\bibnamefont {Friman}},
  \bibinfo {author} {\bibfnamefont {A.}~\bibnamefont {Jaiswal}},\ and\ \bibinfo
  {author} {\bibfnamefont {E.}~\bibnamefont {Speranza}},\ }\bibfield  {title}
  {\bibinfo {title} {{Relativistic fluid dynamics with spin}},\ }\href
  {https://doi.org/10.1103/PhysRevC.97.041901} {\bibfield  {journal} {\bibinfo
  {journal} {Phys. Rev. C}\ }\textbf {\bibinfo {volume} {97}},\ \bibinfo
  {pages} {041901} (\bibinfo {year} {2018})},\ \Eprint
  {https://arxiv.org/abs/1705.00587} {arXiv:1705.00587 [nucl-th]} \BibitemShut
  {NoStop}%
\bibitem [{\citenamefont {Hattori}\ \emph {et~al.}(2019)\citenamefont
  {Hattori}, \citenamefont {Hongo}, \citenamefont {Huang}, \citenamefont
  {Matsuo},\ and\ \citenamefont {Taya}}]{Hattori:2019lfp}%
  \BibitemOpen
  \bibfield  {author} {\bibinfo {author} {\bibfnamefont {K.}~\bibnamefont
  {Hattori}}, \bibinfo {author} {\bibfnamefont {M.}~\bibnamefont {Hongo}},
  \bibinfo {author} {\bibfnamefont {X.-G.}\ \bibnamefont {Huang}}, \bibinfo
  {author} {\bibfnamefont {M.}~\bibnamefont {Matsuo}},\ and\ \bibinfo {author}
  {\bibfnamefont {H.}~\bibnamefont {Taya}},\ }\bibfield  {title} {\bibinfo
  {title} {{Fate of spin polarization in a relativistic fluid: An
  entropy-current analysis}},\ }\href
  {https://doi.org/10.1016/j.physletb.2019.05.040} {\bibfield  {journal}
  {\bibinfo  {journal} {Phys. Lett. B}\ }\textbf {\bibinfo {volume} {795}},\
  \bibinfo {pages} {100} (\bibinfo {year} {2019})},\ \Eprint
  {https://arxiv.org/abs/1901.06615} {arXiv:1901.06615 [hep-th]} \BibitemShut
  {NoStop}%
\bibitem [{\citenamefont {Bhadury}\ \emph {et~al.}(2021)\citenamefont
  {Bhadury}, \citenamefont {Florkowski}, \citenamefont {Jaiswal}, \citenamefont
  {Kumar},\ and\ \citenamefont {Ryblewski}}]{Bhadury:2020puc}%
  \BibitemOpen
  \bibfield  {author} {\bibinfo {author} {\bibfnamefont {S.}~\bibnamefont
  {Bhadury}}, \bibinfo {author} {\bibfnamefont {W.}~\bibnamefont {Florkowski}},
  \bibinfo {author} {\bibfnamefont {A.}~\bibnamefont {Jaiswal}}, \bibinfo
  {author} {\bibfnamefont {A.}~\bibnamefont {Kumar}},\ and\ \bibinfo {author}
  {\bibfnamefont {R.}~\bibnamefont {Ryblewski}},\ }\bibfield  {title} {\bibinfo
  {title} {{Relativistic dissipative spin dynamics in the relaxation time
  approximation}},\ }\href {https://doi.org/10.1016/j.physletb.2021.136096}
  {\bibfield  {journal} {\bibinfo  {journal} {Phys. Lett. B}\ }\textbf
  {\bibinfo {volume} {814}},\ \bibinfo {pages} {136096} (\bibinfo {year}
  {2021})},\ \Eprint {https://arxiv.org/abs/2002.03937} {arXiv:2002.03937
  [hep-ph]} \BibitemShut {NoStop}%
\bibitem [{\citenamefont {Weickgenannt}\ \emph {et~al.}(2021)\citenamefont
  {Weickgenannt}, \citenamefont {Speranza}, \citenamefont {Sheng},
  \citenamefont {Wang},\ and\ \citenamefont {Rischke}}]{Weickgenannt:2020aaf}%
  \BibitemOpen
  \bibfield  {author} {\bibinfo {author} {\bibfnamefont {N.}~\bibnamefont
  {Weickgenannt}}, \bibinfo {author} {\bibfnamefont {E.}~\bibnamefont
  {Speranza}}, \bibinfo {author} {\bibfnamefont {X.-l.}\ \bibnamefont {Sheng}},
  \bibinfo {author} {\bibfnamefont {Q.}~\bibnamefont {Wang}},\ and\ \bibinfo
  {author} {\bibfnamefont {D.~H.}\ \bibnamefont {Rischke}},\ }\bibfield
  {title} {\bibinfo {title} {{Generating Spin Polarization from Vorticity
  through Nonlocal Collisions}},\ }\href
  {https://doi.org/10.1103/PhysRevLett.127.052301} {\bibfield  {journal}
  {\bibinfo  {journal} {Phys. Rev. Lett.}\ }\textbf {\bibinfo {volume} {127}},\
  \bibinfo {pages} {052301} (\bibinfo {year} {2021})},\ \Eprint
  {https://arxiv.org/abs/2005.01506} {arXiv:2005.01506 [hep-ph]} \BibitemShut
  {NoStop}%
\bibitem [{\citenamefont {Gallegos}\ \emph {et~al.}(2021)\citenamefont
  {Gallegos}, \citenamefont {G\"ursoy},\ and\ \citenamefont
  {Yarom}}]{Gallegos:2021bzp}%
  \BibitemOpen
  \bibfield  {author} {\bibinfo {author} {\bibfnamefont {A.~D.}\ \bibnamefont
  {Gallegos}}, \bibinfo {author} {\bibfnamefont {U.}~\bibnamefont {G\"ursoy}},\
  and\ \bibinfo {author} {\bibfnamefont {A.}~\bibnamefont {Yarom}},\ }\bibfield
   {title} {\bibinfo {title} {{Hydrodynamics of spin currents}},\ }\href
  {https://doi.org/10.21468/SciPostPhys.11.2.041} {\bibfield  {journal}
  {\bibinfo  {journal} {SciPost Phys.}\ }\textbf {\bibinfo {volume} {11}},\
  \bibinfo {pages} {041} (\bibinfo {year} {2021})},\ \Eprint
  {https://arxiv.org/abs/2101.04759} {arXiv:2101.04759 [hep-th]} \BibitemShut
  {NoStop}%
\bibitem [{\citenamefont {Hongo}\ \emph {et~al.}(2021)\citenamefont {Hongo},
  \citenamefont {Huang}, \citenamefont {Kaminski}, \citenamefont {Stephanov},\
  and\ \citenamefont {Yee}}]{Hongo:2021ona}%
  \BibitemOpen
  \bibfield  {author} {\bibinfo {author} {\bibfnamefont {M.}~\bibnamefont
  {Hongo}}, \bibinfo {author} {\bibfnamefont {X.-G.}\ \bibnamefont {Huang}},
  \bibinfo {author} {\bibfnamefont {M.}~\bibnamefont {Kaminski}}, \bibinfo
  {author} {\bibfnamefont {M.}~\bibnamefont {Stephanov}},\ and\ \bibinfo
  {author} {\bibfnamefont {H.-U.}\ \bibnamefont {Yee}},\ }\bibfield  {title}
  {\bibinfo {title} {{Relativistic spin hydrodynamics with torsion and linear
  response theory for spin relaxation}},\ }\href
  {https://doi.org/10.1007/JHEP11(2021)150} {\bibfield  {journal} {\bibinfo
  {journal} {JHEP}\ }\textbf {\bibinfo {volume} {11}},\ \bibinfo {pages}
  {150}},\ \Eprint {https://arxiv.org/abs/2107.14231} {arXiv:2107.14231
  [hep-th]} \BibitemShut {NoStop}%
\bibitem [{\citenamefont {Weickgenannt}\ \emph {et~al.}(2022)\citenamefont
  {Weickgenannt}, \citenamefont {Wagner}, \citenamefont {Speranza},\ and\
  \citenamefont {Rischke}}]{Weickgenannt:2022zxs}%
  \BibitemOpen
  \bibfield  {author} {\bibinfo {author} {\bibfnamefont {N.}~\bibnamefont
  {Weickgenannt}}, \bibinfo {author} {\bibfnamefont {D.}~\bibnamefont
  {Wagner}}, \bibinfo {author} {\bibfnamefont {E.}~\bibnamefont {Speranza}},\
  and\ \bibinfo {author} {\bibfnamefont {D.~H.}\ \bibnamefont {Rischke}},\
  }\bibfield  {title} {\bibinfo {title} {{Relativistic second-order dissipative
  spin hydrodynamics from the method of moments}},\ }\href
  {https://doi.org/10.1103/PhysRevD.106.096014} {\bibfield  {journal} {\bibinfo
   {journal} {Phys. Rev. D}\ }\textbf {\bibinfo {volume} {106}},\ \bibinfo
  {pages} {096014} (\bibinfo {year} {2022})},\ \Eprint
  {https://arxiv.org/abs/2203.04766} {arXiv:2203.04766 [nucl-th]} \BibitemShut
  {NoStop}%
\bibitem [{\citenamefont {Daher}\ \emph {et~al.}(2023)\citenamefont {Daher},
  \citenamefont {Das},\ and\ \citenamefont {Ryblewski}}]{Daher:2022wzf}%
  \BibitemOpen
  \bibfield  {author} {\bibinfo {author} {\bibfnamefont {A.}~\bibnamefont
  {Daher}}, \bibinfo {author} {\bibfnamefont {A.}~\bibnamefont {Das}},\ and\
  \bibinfo {author} {\bibfnamefont {R.}~\bibnamefont {Ryblewski}},\ }\bibfield
  {title} {\bibinfo {title} {{Stability studies of first-order
  spin-hydrodynamic frameworks}},\ }\href
  {https://doi.org/10.1103/PhysRevD.107.054043} {\bibfield  {journal} {\bibinfo
   {journal} {Phys. Rev. D}\ }\textbf {\bibinfo {volume} {107}},\ \bibinfo
  {pages} {054043} (\bibinfo {year} {2023})},\ \Eprint
  {https://arxiv.org/abs/2209.10460} {arXiv:2209.10460 [nucl-th]} \BibitemShut
  {NoStop}%
\bibitem [{\citenamefont {Xie}\ \emph {et~al.}(2023)\citenamefont {Xie},
  \citenamefont {Wang}, \citenamefont {Yang},\ and\ \citenamefont
  {Pu}}]{Xie:2023gbo}%
  \BibitemOpen
  \bibfield  {author} {\bibinfo {author} {\bibfnamefont {X.-Q.}\ \bibnamefont
  {Xie}}, \bibinfo {author} {\bibfnamefont {D.-L.}\ \bibnamefont {Wang}},
  \bibinfo {author} {\bibfnamefont {C.}~\bibnamefont {Yang}},\ and\ \bibinfo
  {author} {\bibfnamefont {S.}~\bibnamefont {Pu}},\ }\bibfield  {title}
  {\bibinfo {title} {{Causality and stability analysis for the minimal causal
  spin hydrodynamics}},\ }\href@noop {} {\  (\bibinfo {year} {2023})},\ \Eprint
  {https://arxiv.org/abs/2306.13880} {arXiv:2306.13880 [hep-ph]} \BibitemShut
  {NoStop}%
\bibitem [{\citenamefont {Sarwar}\ \emph {et~al.}(2023)\citenamefont {Sarwar},
  \citenamefont {Hasanujjaman}, \citenamefont {Bhatt}, \citenamefont {Mishra},\
  and\ \citenamefont {Alam}}]{Sarwar:2022yzs}%
  \BibitemOpen
  \bibfield  {author} {\bibinfo {author} {\bibfnamefont {G.}~\bibnamefont
  {Sarwar}}, \bibinfo {author} {\bibfnamefont {M.}~\bibnamefont
  {Hasanujjaman}}, \bibinfo {author} {\bibfnamefont {J.~R.}\ \bibnamefont
  {Bhatt}}, \bibinfo {author} {\bibfnamefont {H.}~\bibnamefont {Mishra}},\ and\
  \bibinfo {author} {\bibfnamefont {J.-e.}\ \bibnamefont {Alam}},\ }\bibfield
  {title} {\bibinfo {title} {{Causality and stability of relativistic spin
  hydrodynamics}},\ }\href {https://doi.org/10.1103/PhysRevD.107.054031}
  {\bibfield  {journal} {\bibinfo  {journal} {Phys. Rev. D}\ }\textbf {\bibinfo
  {volume} {107}},\ \bibinfo {pages} {054031} (\bibinfo {year} {2023})},\
  \Eprint {https://arxiv.org/abs/2209.08652} {arXiv:2209.08652 [nucl-th]}
  \BibitemShut {NoStop}%
\bibitem [{\citenamefont {Weickgenannt}(2023)}]{Weickgenannt:2023btk}%
  \BibitemOpen
  \bibfield  {author} {\bibinfo {author} {\bibfnamefont {N.}~\bibnamefont
  {Weickgenannt}},\ }\bibfield  {title} {\bibinfo {title} {{Linearly stable and
  causal relativistic first-order spin hydrodynamics}},\ }\href@noop {} {\
  (\bibinfo {year} {2023})},\ \Eprint {https://arxiv.org/abs/2307.13561}
  {arXiv:2307.13561 [nucl-th]} \BibitemShut {NoStop}%
\bibitem [{\citenamefont {Pandya}\ and\ \citenamefont
  {Pretorius}(2021)}]{Pandya:2021ief}%
  \BibitemOpen
  \bibfield  {author} {\bibinfo {author} {\bibfnamefont {A.}~\bibnamefont
  {Pandya}}\ and\ \bibinfo {author} {\bibfnamefont {F.}~\bibnamefont
  {Pretorius}},\ }\bibfield  {title} {\bibinfo {title} {{Numerical exploration
  of first-order relativistic hydrodynamics}},\ }\href
  {https://doi.org/10.1103/PhysRevD.104.023015} {\bibfield  {journal} {\bibinfo
   {journal} {Phys. Rev. D}\ }\textbf {\bibinfo {volume} {104}},\ \bibinfo
  {pages} {023015} (\bibinfo {year} {2021})},\ \Eprint
  {https://arxiv.org/abs/2104.00804} {arXiv:2104.00804 [gr-qc]} \BibitemShut
  {NoStop}%
\bibitem [{\citenamefont {Pandya}\ \emph
  {et~al.}(2022{\natexlab{b}})\citenamefont {Pandya}, \citenamefont {Most},\
  and\ \citenamefont {Pretorius}}]{Pandya:2022pif}%
  \BibitemOpen
  \bibfield  {author} {\bibinfo {author} {\bibfnamefont {A.}~\bibnamefont
  {Pandya}}, \bibinfo {author} {\bibfnamefont {E.~R.}\ \bibnamefont {Most}},\
  and\ \bibinfo {author} {\bibfnamefont {F.}~\bibnamefont {Pretorius}},\
  }\bibfield  {title} {\bibinfo {title} {{Conservative finite volume scheme for
  first-order viscous relativistic hydrodynamics}},\ }\href
  {https://doi.org/10.1103/PhysRevD.105.123001} {\bibfield  {journal} {\bibinfo
   {journal} {Phys. Rev. D}\ }\textbf {\bibinfo {volume} {105}},\ \bibinfo
  {pages} {123001} (\bibinfo {year} {2022}{\natexlab{b}})},\ \Eprint
  {https://arxiv.org/abs/2201.12317} {arXiv:2201.12317 [gr-qc]} \BibitemShut
  {NoStop}%
\bibitem [{\citenamefont {Bantilan}\ \emph {et~al.}(2022)\citenamefont
  {Bantilan}, \citenamefont {Bea},\ and\ \citenamefont
  {Figueras}}]{Bantilan:2022ech}%
  \BibitemOpen
  \bibfield  {author} {\bibinfo {author} {\bibfnamefont {H.}~\bibnamefont
  {Bantilan}}, \bibinfo {author} {\bibfnamefont {Y.}~\bibnamefont {Bea}},\ and\
  \bibinfo {author} {\bibfnamefont {P.}~\bibnamefont {Figueras}},\ }\bibfield
  {title} {\bibinfo {title} {{Evolutions in first-order viscous
  hydrodynamics}},\ }\href {https://doi.org/10.1007/JHEP08(2022)298} {\bibfield
   {journal} {\bibinfo  {journal} {JHEP}\ }\textbf {\bibinfo {volume} {08}},\
  \bibinfo {pages} {298}},\ \Eprint {https://arxiv.org/abs/2201.13359}
  {arXiv:2201.13359 [hep-th]} \BibitemShut {NoStop}%
\bibitem [{\citenamefont {Shi}\ \emph {et~al.}(2018)\citenamefont {Shi},
  \citenamefont {Jiang}, \citenamefont {Lilleskov},\ and\ \citenamefont
  {Liao}}]{Shi:2017cpu}%
  \BibitemOpen
  \bibfield  {author} {\bibinfo {author} {\bibfnamefont {S.}~\bibnamefont
  {Shi}}, \bibinfo {author} {\bibfnamefont {Y.}~\bibnamefont {Jiang}}, \bibinfo
  {author} {\bibfnamefont {E.}~\bibnamefont {Lilleskov}},\ and\ \bibinfo
  {author} {\bibfnamefont {J.}~\bibnamefont {Liao}},\ }\bibfield  {title}
  {\bibinfo {title} {{Anomalous Chiral Transport in Heavy Ion Collisions from
  Anomalous-Viscous Fluid Dynamics}},\ }\href
  {https://doi.org/10.1016/j.aop.2018.04.026} {\bibfield  {journal} {\bibinfo
  {journal} {Annals Phys.}\ }\textbf {\bibinfo {volume} {394}},\ \bibinfo
  {pages} {50} (\bibinfo {year} {2018})},\ \Eprint
  {https://arxiv.org/abs/1711.02496} {arXiv:1711.02496 [nucl-th]} \BibitemShut
  {NoStop}%
\bibitem [{\citenamefont {Shi}\ \emph {et~al.}(2020)\citenamefont {Shi},
  \citenamefont {Zhang}, \citenamefont {Hou},\ and\ \citenamefont
  {Liao}}]{Shi:2019wzi}%
  \BibitemOpen
  \bibfield  {author} {\bibinfo {author} {\bibfnamefont {S.}~\bibnamefont
  {Shi}}, \bibinfo {author} {\bibfnamefont {H.}~\bibnamefont {Zhang}}, \bibinfo
  {author} {\bibfnamefont {D.}~\bibnamefont {Hou}},\ and\ \bibinfo {author}
  {\bibfnamefont {J.}~\bibnamefont {Liao}},\ }\bibfield  {title} {\bibinfo
  {title} {{Signatures of Chiral Magnetic Effect in the Collisions of
  Isobars}},\ }\href {https://doi.org/10.1103/PhysRevLett.125.242301}
  {\bibfield  {journal} {\bibinfo  {journal} {Phys. Rev. Lett.}\ }\textbf
  {\bibinfo {volume} {125}},\ \bibinfo {pages} {242301} (\bibinfo {year}
  {2020})},\ \Eprint {https://arxiv.org/abs/1910.14010} {arXiv:1910.14010
  [nucl-th]} \BibitemShut {NoStop}%
\bibitem [{\citenamefont {Buzzegoli}\ \emph {et~al.}(2022)\citenamefont
  {Buzzegoli}, \citenamefont {Kharzeev}, \citenamefont {Liu}, \citenamefont
  {Shi}, \citenamefont {Voloshin},\ and\ \citenamefont
  {Yee}}]{Buzzegoli:2022kqx}%
  \BibitemOpen
  \bibfield  {author} {\bibinfo {author} {\bibfnamefont {M.}~\bibnamefont
  {Buzzegoli}}, \bibinfo {author} {\bibfnamefont {D.~E.}\ \bibnamefont
  {Kharzeev}}, \bibinfo {author} {\bibfnamefont {Y.-C.}\ \bibnamefont {Liu}},
  \bibinfo {author} {\bibfnamefont {S.}~\bibnamefont {Shi}}, \bibinfo {author}
  {\bibfnamefont {S.~A.}\ \bibnamefont {Voloshin}},\ and\ \bibinfo {author}
  {\bibfnamefont {H.-U.}\ \bibnamefont {Yee}},\ }\bibfield  {title} {\bibinfo
  {title} {{Shear-induced anomalous transport and charge asymmetry of
  triangular flow in heavy-ion collisions}},\ }\href
  {https://doi.org/10.1103/PhysRevC.106.L051902} {\bibfield  {journal}
  {\bibinfo  {journal} {Phys. Rev. C}\ }\textbf {\bibinfo {volume} {106}},\
  \bibinfo {pages} {L051902} (\bibinfo {year} {2022})},\ \Eprint
  {https://arxiv.org/abs/2206.11382} {arXiv:2206.11382 [hep-ph]} \BibitemShut
  {NoStop}%
\end{thebibliography}%
\end{document}